\documentclass[11pt,a4paper]{article}
\pdfoutput=1
\usepackage{jcappub}

\newcommand{\bi}{\begin{itemize}}
\newcommand{\ei}{\end{itemize}}
\newcommand{\beq}{\begin{equation}}
\newcommand{\eeq}{\end{equation}}
\newcommand{\bea}{\begin{eqnarray}}
\newcommand{\eea}{\end{eqnarray}}

\def\d{{\delta}}

\def\bx{{\bf x}}

\def\l{\langle}

\def\N{{\cal N}}

\def\f{f}

\def\k{{\bf k}}

\def\curv{\sigma}
\def\f{f_c}

\def\n{{\bf \hat n}}
\def\tf{\tilde f}

\def\d{\delta}
\def\l{l}

\begin{document}
\title{Isocurvature modes in the CMB bispectrum}
\author[a,b]{David Langlois}
\author[c]{and Bartjan van Tent}
\affiliation[a]{APC, Astroparticules et Cosmologie, Universit\'e Paris Diderot, CNRS/IN2P3, CEA/Irfu, Observatoire de Paris, Sorbonne Paris Cit\'e, 10 rue Alice Domon et L\'eonie Duquet, 75205 Paris Cedex 13, France}
\affiliation[b]{IAP, 98bis Boulevard Arago, 75014 Paris, France}
\affiliation[c]{Laboratoire de Physique Th\'eorique, Universit\'e Paris-Sud 11 
and CNRS, B\^atiment 210, 91405 Orsay Cedex, France}
\emailAdd{langlois@apc.univ-paris7.fr}
\emailAdd{vantent@th.u-psud.fr}

\abstract{
We study the angular bispectrum of local type arising from the  (possibly correlated) combination of a primordial adiabatic mode with an isocurvature one. Generically, this bispectrum can be decomposed into six elementary bispectra. We estimate how precisely  CMB data, including polarization,
 can enable us to measure or constrain  the six corresponding amplitudes, 
considering separately the four types of isocurvature modes (CDM, baryon, neutrino density, neutrino velocity).  
Finally, we discuss how the model-independent constraints on the bispectrum can be combined to get constraints on the parameters of multiple-field inflation models.}

\arxivnumber{1204.5042}

\begin{flushright}
LPT-12-35
\end{flushright}

\maketitle

\section{Introduction}

Inflation is currently  the best candidate to explain the generation of primordial perturbations, but many of its realizations remain compatible with the present data. One can hope that future data will enable us to find additional 
information in the primordial perturbations that could help to discriminate between  the
various mechanisms that can have taken place in the very early Universe. 

In this respect, it is important 
to test the adiabatic nature of the primordial perturbations. 
Since single-field inflation predicts only adiabatic perturbations, the detection of a fraction of an isocurvature mode in the cosmological data would rule out the simplest models of inflation. By contrast, multiple-field inflation could easily account for the presence of isocurvature modes~\cite{LEBEDEV-85-65}, which can even be correlated with the adiabatic component~\cite{Langlois:1999dw,Langlois:2000ar}. 

As shown in \cite{Bucher:1999re}, the most  general primordial perturbation is a priori a linear combination of the usual adiabatic mode with four types of isocurvature modes, respectively the Cold Dark Matter (CDM), baryon, neutrino density  and neutrino velocity isocurvature modes. The existence, and amplitude, of  isocurvature modes depends on the details of the thermal history of the Universe. Various scenarios that  can lead to observable isocurvature modes have been discussed in the literature (double inflation \cite{Silk:1986vc,Polarski:1994rz,Langlois:1999dw}, axions \cite{Seckel:1985tj}, curvatons 
\cite{Linde:1996gt,Lyth:2001nq,Moroi:2001ct,Moroi:2002rd,Lyth:2003ip}).

In parallel to the possible presence of isocurvature modes, another property that could distinguish multiple-field models from single-field models is  a detectable primordial non-Gaussianity of the local type. 
 So far, the   WMAP measurements  of the CMB anisotropies~\cite{Komatsu:2010fb}  have set the present limit $f_{\rm
  NL}^{\rm local} =32\pm 21$ (68\,\% CL) [and $-10 < f_{\rm NL}^{\rm
  local} < 74$ (95\,\% CL)] on the parameter $f_{\rm
  NL}^{\rm local} $ that characterizes the amplitude of the simplest type of non-Gaussianity, namely the local shape.   
Similarly to isocurvature modes,  a detection of local primordial non-Gaussianity   would rule out all inflation models based on a single scalar field, since they generate only 
unobservably small local non-Gaussianities~\cite{Creminelli:2004yq}. 
Scenarios with additional scalar fields, 
 such as another inflaton (see e.g.\ \cite{Byrnes:2010em,Tzavara:2010ge}),  a curvaton~\cite{Lyth:2002my} or a modulaton~\cite{dgz,kofman,Langlois:2009jp}, which can 
 produce  detectable local non-Gaussianity, would then move to the front stage.

Isocurvature modes are usually investigated by constraining the  power spectrum of primordial perturbations with CMB or large-scale structure data (see e.g.\ \cite{Bean:2006qz,Sollom:2009vd,Mangilli:2010ut,Li:2010yb,Kawasaki:2011ze,Kasanda:2011np,DiValentino:2011sv,Valiviita:2012ub}).
However, isocurvature modes could also contribute to non-Gaussianities as discussed in several works~\cite{Bartolo:2001cw,Kawasaki:2008sn,Langlois:2008vk,Kawasaki:2008pa,Hikage:2008sk,Kawakami:2009iu, Langlois:2011zz,Langlois:2010fe,LvT}. Moreover, there exist models 
\cite{Langlois:2011zz}  where isocurvature modes, while remaining a small fraction at the linear level,  would dominate the non-Gaussianity. As shown in \cite{LvT}, these CDM isocurvature modes would be potentially detectable via their non-Gaussianity in the CMB data such as collected by Planck. Non-Gaussianity can thus be considered as a complementary probe of isocurvature modes. 

In the present work, we refine and extend our  previous analysis~\cite{LvT} by considering all types of isocurvature modes, not only the CDM isocurvature mode. We analyse the bispectrum generated by the  adiabatic mode together with   one of the four isocurvature modes. The total  angular bispectrum can be decomposed  into six distinct components: the usual purely adiabatic bispectrum, a purely isocurvature bispectrum, and four other bispectra that arise from the possible 
 correlations between the  adiabatic and isocurvature mode. 
 Because these six bispectra have  different shapes in $l$-space, their amplitude can  in principle be measured  in the CMB data and  we have computed, for each type of isocurvature mode,  the associated $6\times 6$  Fisher matrix to estimate what precision on these six parameters could be reached with the Planck data. 
 We also show that the inclusion of polarization measurements improves the predicted precision of some
isocurvature non-Gaussianity parameters significantly. 
 
The elementary bispectra discussed above depend only on the adiabatic and isocurvature transfer functions  and are thus independent of the details of the generation mechanism. Now, by assuming a specific class of inflationary models, one obtains particular relations between the six bispectra, which can be used as consistency relations for the model or to constrain  the model parameters. 
We illustrate this in the context of curvaton-type models, generating adiabatic and CDM  isocurvature perturbations. 

The outline of the paper is the following. In the next section, we present the various isocurvature perturbations and discuss their impact on the CMB angular power spectrum. The following section is devoted to the angular bispectrum and its decomposition into six elementary bispectra. We then discuss the observational prospects to detect these elementary local bispectra in the future data, distinguishing the various isocurvature modes. Finally, we consider  models where primordial perturbations are generated by an inflaton and a curvaton, 
and show that the amplitudes of all six bispectra depend on only two coefficients, which can be constrained from the data. The last section contains our conclusions.

\section{Isocurvature perturbations}
In this section, we recall  the definition of isocurvature modes in the context of {\it linear} cosmological perturbations. At the time of last scattering, the main components in the Universe are the CDM (c), the baryons (b), the photons ($\gamma$) and the neutrinos ($\nu$). All these components are characterized by their individual energy density perturbation $\delta\rho_i$ and their velocities ${\cal V}_i$ (as well as higher momenta of their phase space distribution functions, which we do not  discuss here; see \cite{ma_bertschinger} for details). The ``primordial'' perturbations for each Fourier mode $k$ are usually defined on super-Hubble scales, i.e.\ when $k\ll aH$, deep in the radiation dominated era. 

The most common type of perturbation is the adiabatic mode, 
characterized by
\beq
\frac{\delta n_c}{n_c}=\frac{\delta n_b}{n_b}=\frac{\delta n_\nu}{n_\nu}=\frac{\delta n_\gamma}{n_\gamma}\,,
\eeq
which means that the number of photons (or neutrinos, or CDM particles) per baryon is not fluctuating.
 In terms of the energy density contrasts ($\delta\equiv \delta\rho/\rho$), the above condition is expressed as
\beq
\label{adiabatic}
\delta_c=\delta_b= \frac34\delta_\nu=\frac34\delta_\gamma\,,
\eeq
where the $3/4$ factor, for photons and neutrinos, comes from the relation $\rho\propto n^{4/3}$ for relativistic species. 

Assuming adiabatic initial conditions is  natural if all particles have been created by the decay of a single degree of freedom, such as a single inflaton, and,  so far, the CMB data are fully compatible with purely adiabatic perturbations.
However, other types of perturbations can be included in a more general framework. In addition to the adiabatic mode, one can consider  four distinct types of so-called isocurvature modes~\cite{Bucher:1999re}: the CDM isocurvature mode $S_c$, the baryon isocurvature mode $S_b$, the neutrino density  isocurvature mode $S_{\nu d}$ and the neutrino velocity  isocurvature mode $S_{\nu v}$. 
At zeroth order in $k\tau$, where $\tau$ is the conformal time, the  first three are  characterized, respectively,  by 
\begin{eqnarray}
\delta_c&=&S_c+\frac34 \d_\gamma, \qquad 
\delta_b= \frac34\delta_\nu=\frac34\delta_\gamma \qquad {\rm (CDM\ isocurvature)}
\\
\delta_b &=& S_b +\frac34 \d_\gamma, \qquad 
\delta_c= \frac34\delta_\nu=\frac34\delta_\gamma \qquad {\rm (baryon \ isocurvature)}
\\
\frac34\delta_\nu &=& S_{\nu d} +\frac34 \d_\gamma, \qquad 
\d_b=\delta_c=\frac34\delta_\gamma \qquad {\rm (neutrino \ density \ isocurvature)},
\end{eqnarray}
while  the corresponding velocities tend to zero. 
As for the neutrino velocity isocurvature mode, it  is characterized by non vanishing ''initial velocities'',
\beq
{\cal V}_\nu= S_{\nu v}, \qquad  {\cal V}_{\gamma b}=-\frac78 N_\nu \left(\frac{4}{11}\right)^{4/3} S_{\nu v}
\qquad  {\rm (neutrino \ velocity \ isocurvature)},
\eeq
where ${\cal V}_{\gamma b}$ is the common velocity of the photon-baryon plasma (the photons and baryons are initially tightly coupled via the Thomson scattering off free electons) and $N_\nu$ is the number of species of massless neutrinos. The above relation between the two velocities ensures that they exactly cancel in the momentum density, 
while the energy densities satisfy the adiabatic condition (\ref{adiabatic}). 

In the following, each mode will be characterized by its amplitude: the curvature perturbation on constant energy  hypersurfaces,  $\zeta$,  for the adiabatic mode, and $S_c$, $S_b$, $S_{\nu d}$ and $S_{\nu v}$ for the four isocurvature modes. These variables will be  denoted collectively as $X^I$. 
In the context of inflation, a necessary condition for these isocurvature modes to be created is that several light degrees of freedom exist during inflation. Moreover, since the adiabatic and isocurvature modes can be related in various ways to these degrees of freedom during inflation, one can envisage the existence of correlations between these modes.

These various modes lead to {\it different} predictions for the CMB temperature and polarization.  
Let us consider for instance the temperature anisotropies, which can be decomposed into spherical harmonics:
\beq
\frac{\Delta T}{T}=\sum_{lm} a_{lm} Y_{lm}\,.
\eeq
The multipole coefficients $a_{lm}$ can be related linearly to  any of the primordial modes. The precise correspondance can be computed numerically and written in the form
\beq
a_{lm}^{I}=4\pi (-i)^l \int \frac{d^3\k}{(2\pi)^3} X^I(\k)\,  g^I_l(k)\, Y^*_{lm}(\hat\k)\,,
\eeq
where  $g^I_l(k)$ is the transfer function associated with the corresponding primordial perturbation ($g^I_l(k)$ depends also on the various cosmological parameters). 

For each type of perturbation, the angular power spectrum  is thus given by
\beq
C_l^{I}=\langle a_{lm}^{I}a_{lm}^{I*}\rangle=\frac{2}{\pi}\int_0^\infty dk\, k^2 \left[g_l^I(k)\right]^2  P_I(k),
\eeq
where we have introduced the primordial power spectrum $P_I(k)$ defined by 
\beq
\langle X^{I}(\k_1) X^I(\k_2)  \rangle \equiv  (2 \pi)^3 \delta ( \k_1+\k_2) P_{I}(k_1)\,.
 \eeq
For our purposes, the crucial point is that the transfer functions associated with isocurvature perturbations are very different from the adiabatic transfer function. 
Moreover, each isocurvature mode leads to a specific signature that enables one to distinguish it from the other isocurvature modes. The only exception are  the CDM and baryon isocurvature modes  which give  exactly  the same pattern, up to the rescaling:
\beq
\label{omega_bc}
S_b=\omega_{bc} \, S_c\,,\qquad \omega_{bc}\equiv\frac{\Omega_b}{\Omega_c}\,,
\eeq
where the parameters  $\Omega_b$ and $\Omega_c$ denote, as usual, the present energy density fractions, respectively for baryons and CDM (note however that these two modes can in principle be discriminated via other effects, see e.g.\ \cite{
Holder:2009gd,Gordon:2009wx, Kawasaki:2011ze,Grin:2011tf}).

More generally, when we also allow for possible correlations between the modes
and include E-polarization, the angular power spectra are given by
\beq
C_l^{IJ;\, \alpha \alpha'}=\frac{2}{\pi}\int_0^\infty dk\, k^2 g_l^{I; \alpha}(k)  
g_l^{J;\alpha'}(k) P_{IJ}(k),
\eeq
where $I$ and $J$ label the isocurvature mode and $\alpha$ and $\alpha'$ the polarization
(i.e.\ either T (temperature) or E (polarization)). The primordial power spectrum
$P_{IJ}(k)$ is now defined by
\beq
\langle X^{I}(\k_1) X^J(\k_2)  \rangle \equiv  (2 \pi)^3 \delta ( \k_1+\k_2) P_{IJ}(k_1)\,,
 \eeq
which generalizes the previous definition to include the presence of 
correlations between the modes, which corresponds to the situation
where $P_{IJ}$ with different $I$ and $J$ does not vanish. 

\begin{figure}
\centering
\includegraphics[width=0.49\textwidth, clip=true]{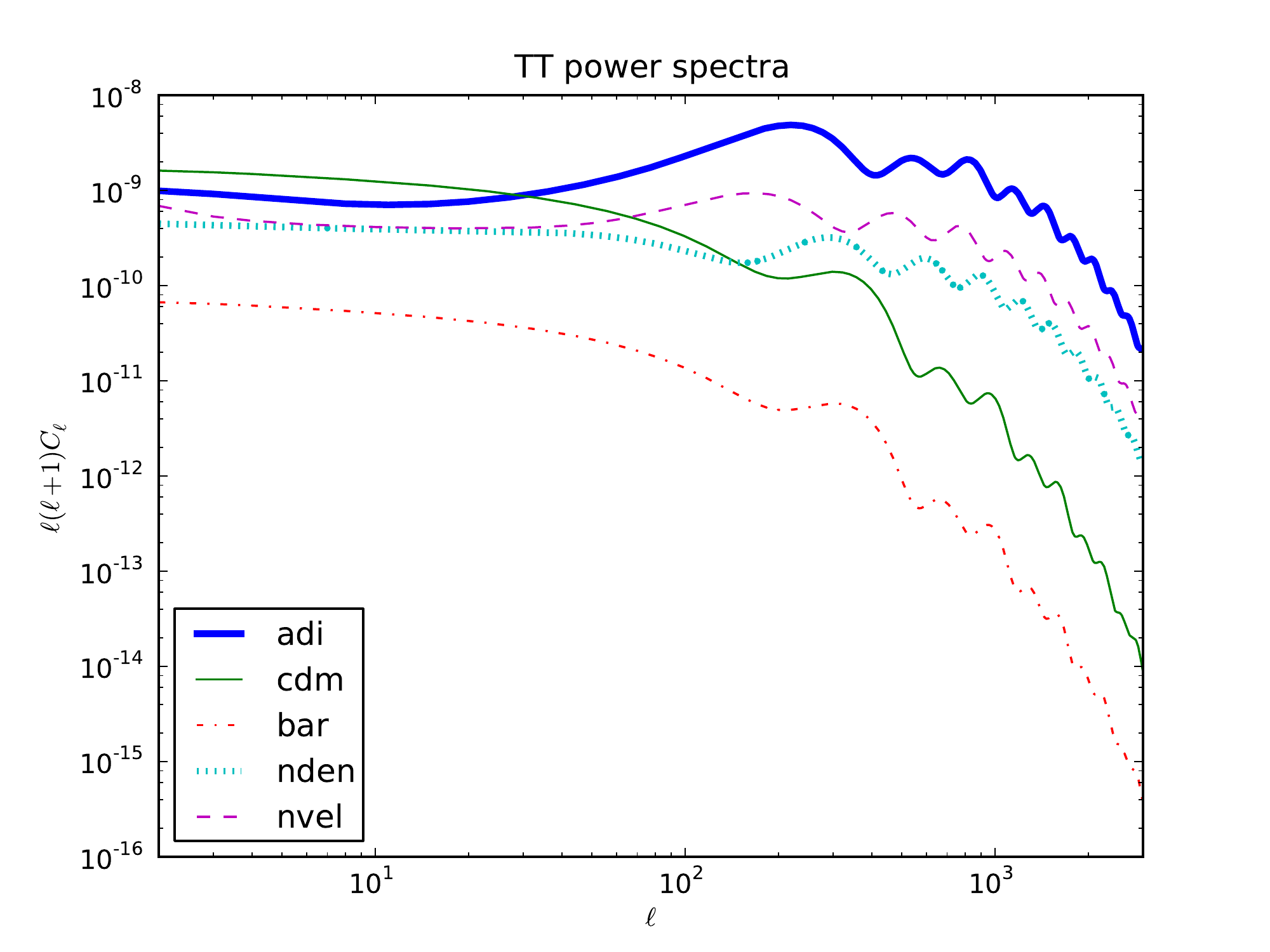}
\includegraphics[width=0.49\textwidth, clip=true]{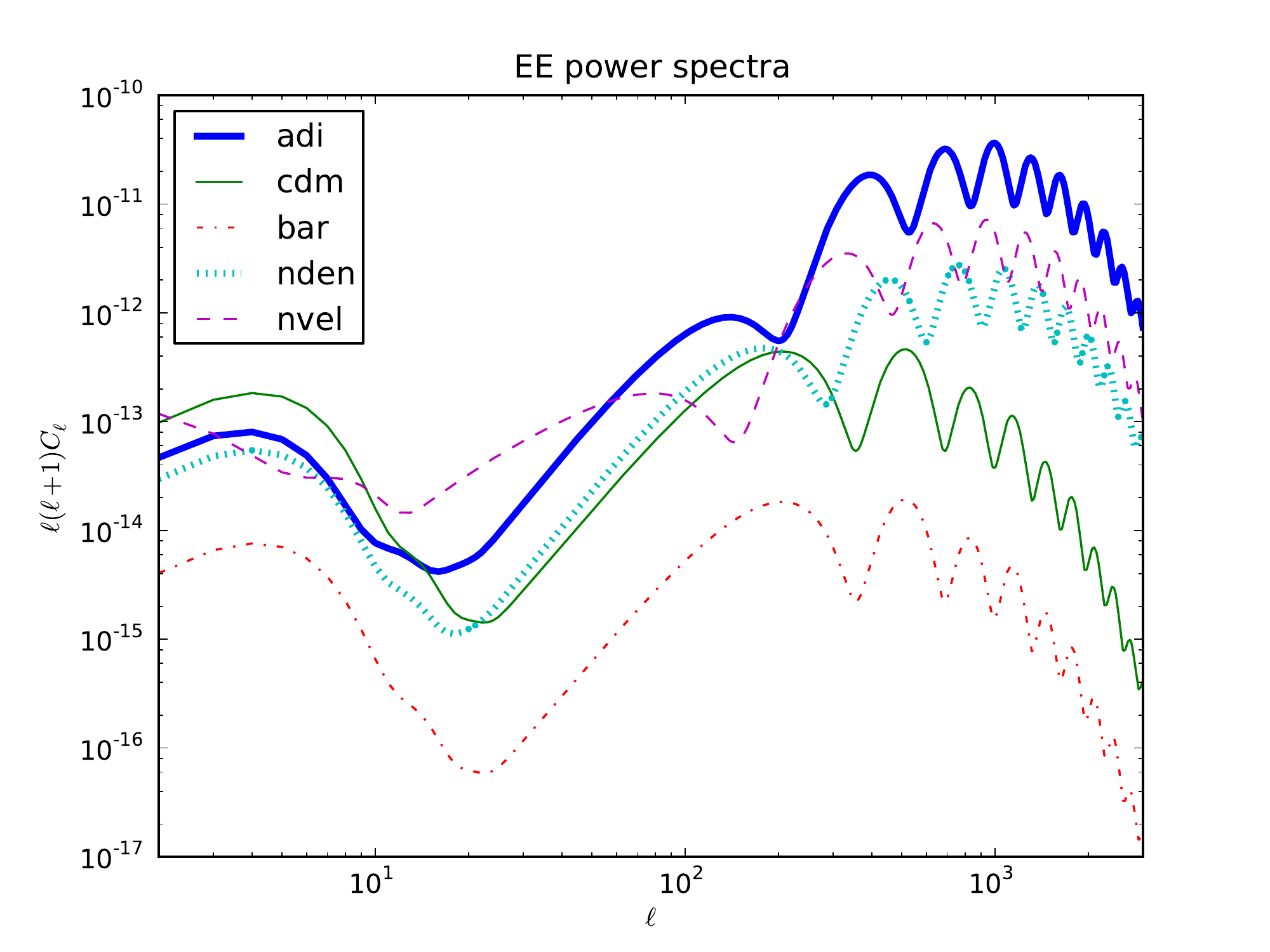}
\caption{Angular power spectra (multiplied by $l(l+1)$) for the temperature 
(left) and polarization (right) obtained from
purely adiabatic  or purely isocurvature initial conditions.  The amplitude 
and spectral index of the primordial power spectrum, as well as the 
cosmological parameters, on which the transfer functions depend, correspond 
to the WMAP7-only best-fit parameters 
(WMAP7-only best-fit parameters are used for all the figures and explicit
computations in this paper).}
\label{spectra_fig}
\end{figure}

\begin{figure}
\centering
\includegraphics[width=0.49\textwidth, clip=true]{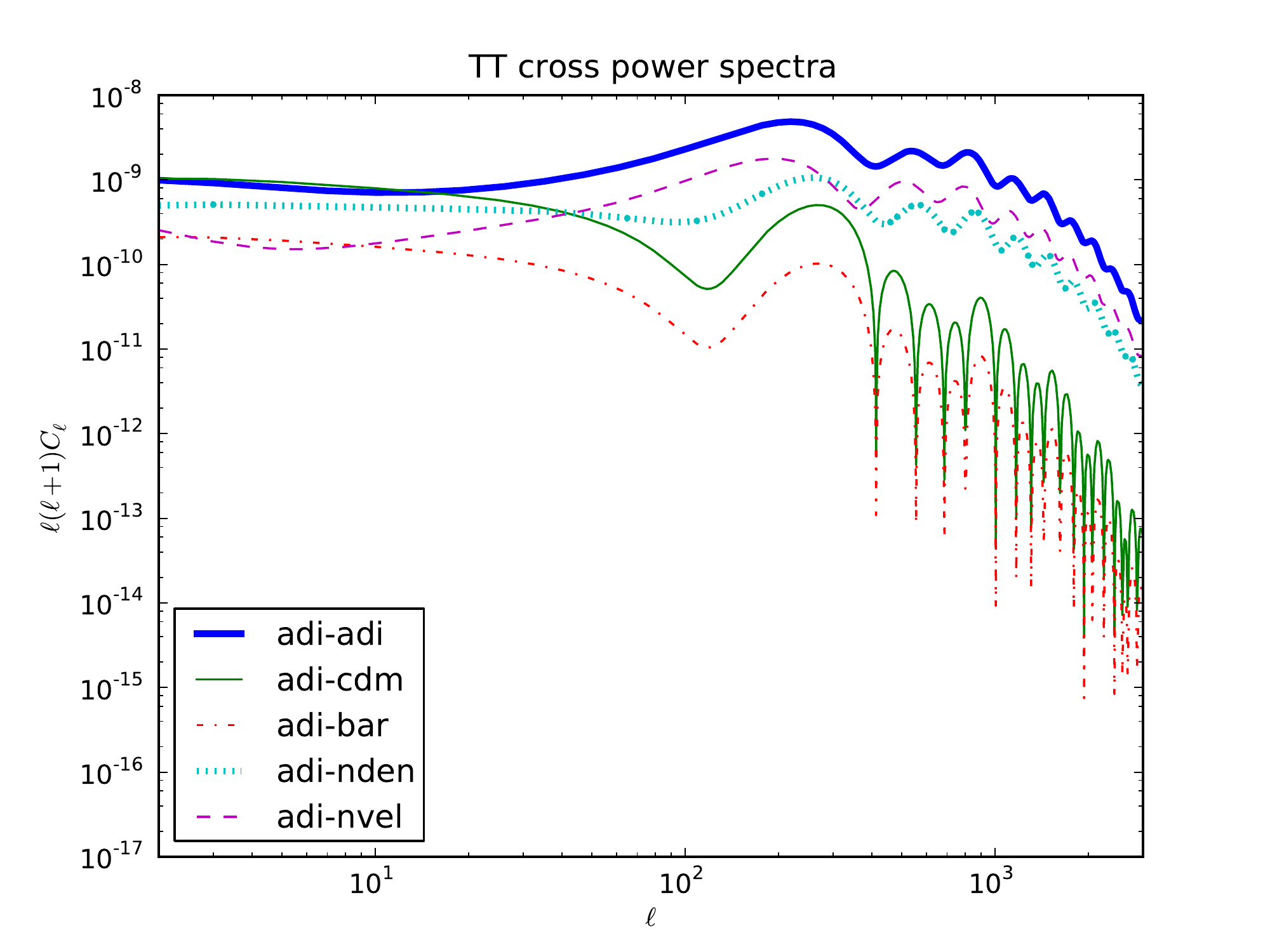}
\includegraphics[width=0.49\textwidth, clip=true]{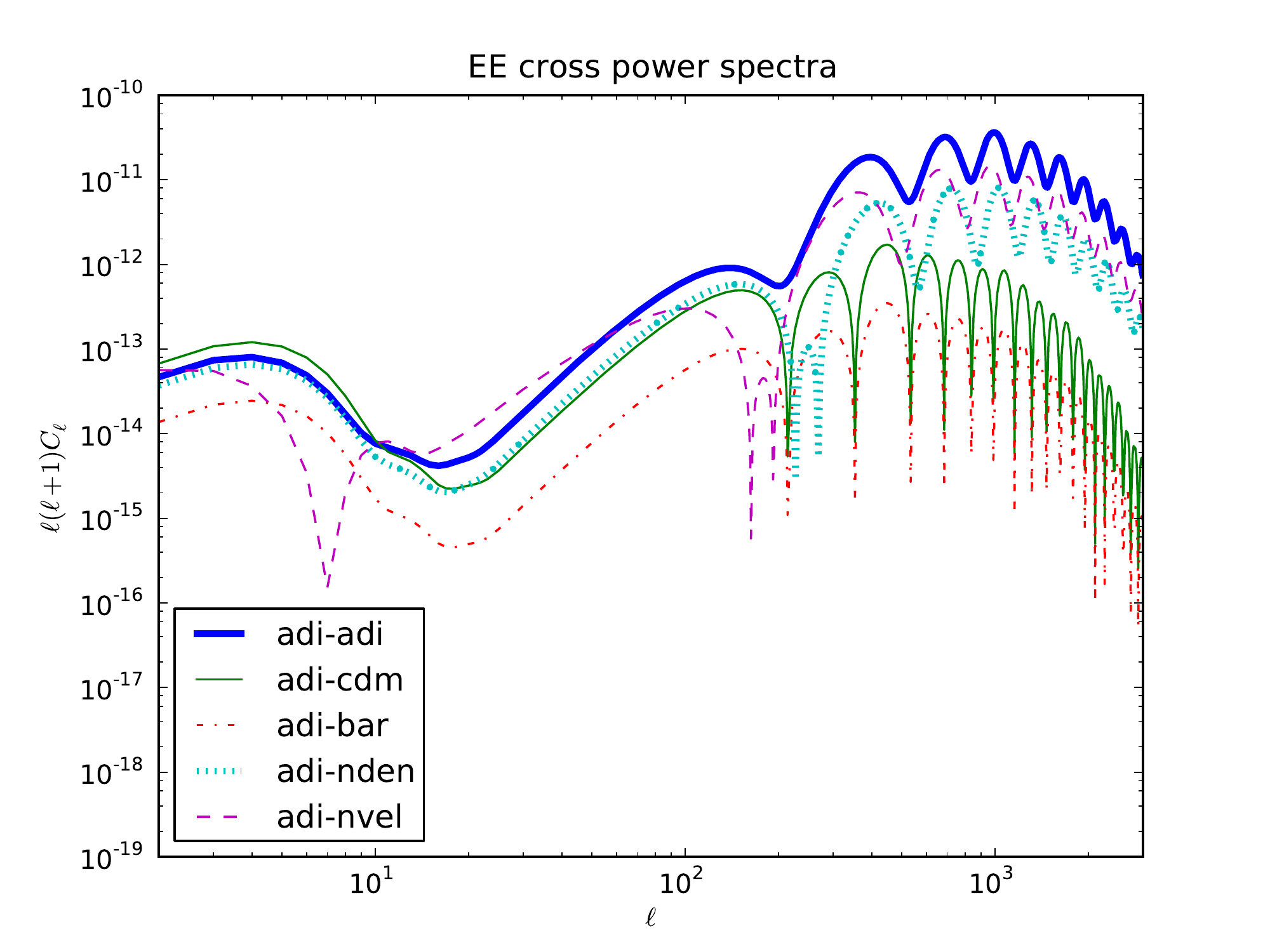}
\caption{The isocurvature cross power spectra (multiplied by $l(l+1)$ and
in absolute value) with one component fixed to be the adiabatic one.
The figure on the left shows temperature (TT), the one on the right 
polarization (EE).}
\label{cross_spectra_fig}
\end{figure}

\begin{figure}
\centering
\includegraphics[width=0.49\textwidth, clip=true]{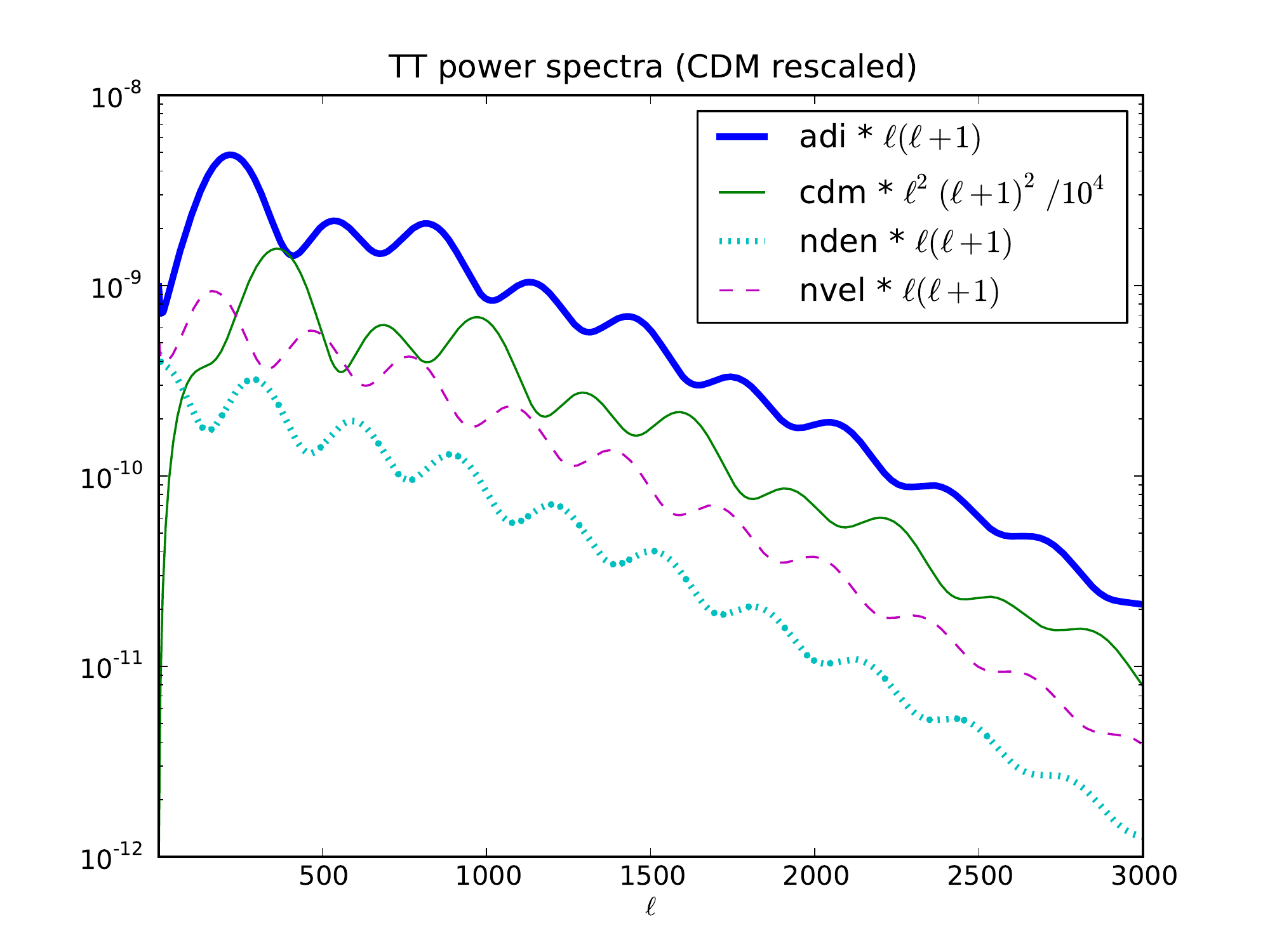}
\includegraphics[width=0.49\textwidth, clip=true]{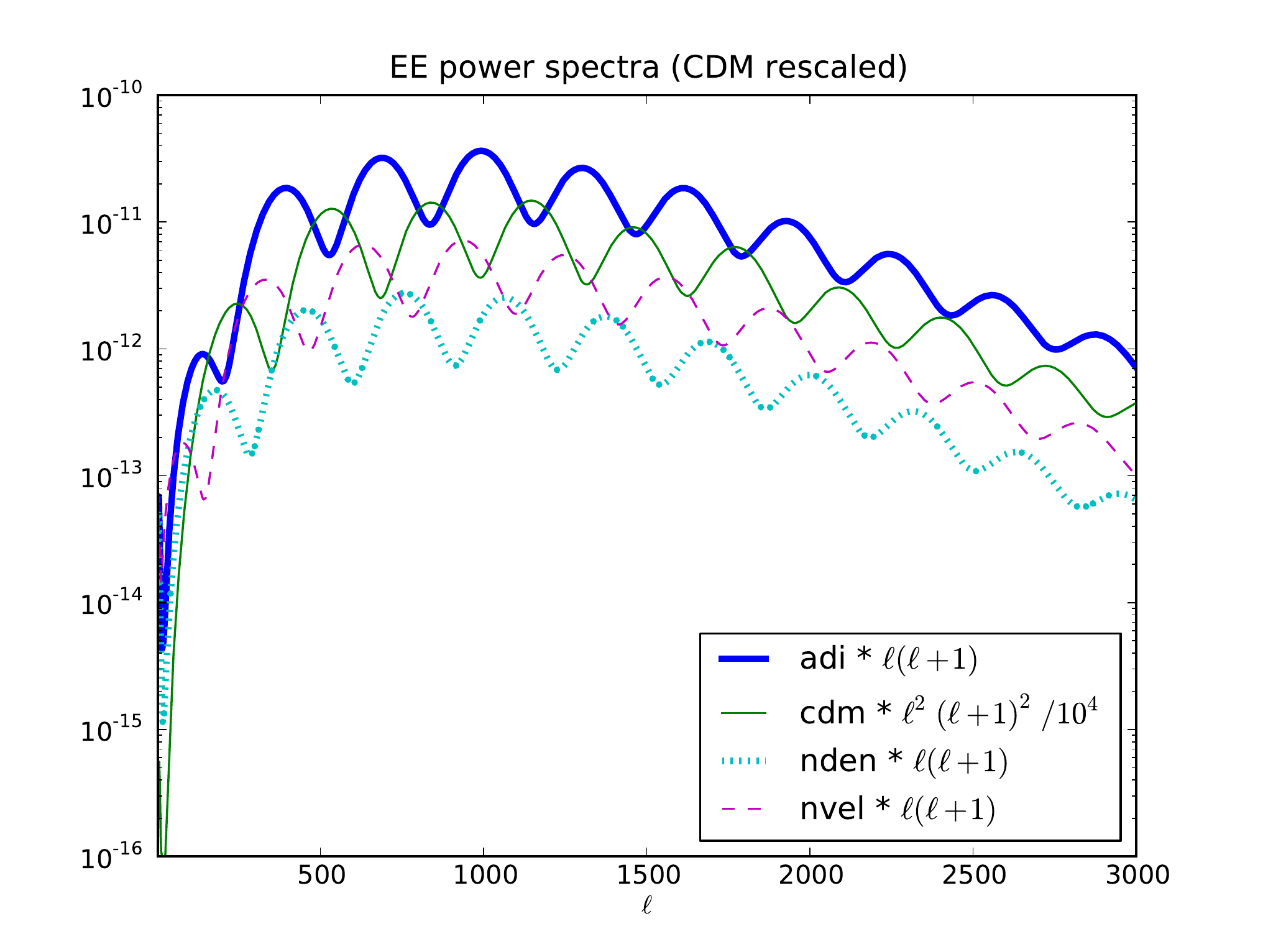}
\caption{Same angular power spectra for the temperature and the polarization as in Fig.~\ref{spectra_fig}, except that the horizontal axis is now linear, and that the CDM isocurvature spectrum is multiplied by $10^{-4} l^2(l+1)^2$ instead of $l(l+1)$.}
\label{spectra_rescaled}
\end{figure}

All this is illustrated in Fig.~\ref{spectra_fig} and \ref{cross_spectra_fig}, 
where we have plotted the angular power spectra for all the various modes 
(Fig.~\ref{spectra_fig}) and  the isocurvature cross power spectra where one of the 
components is adiabatic (Fig.~\ref{cross_spectra_fig}), assuming the same primordial power spectrum for all. 
As one can see from these figures, the CDM (and baryon) isocurvature mode
decreases much faster with $l$ than the other modes. In fact it turns out that
if one multiplies the CDM isocurvature power spectrum by $l^2(l+1)^2$ 
instead of $l(l+1)$, it falls off roughly in the same way as the other modes 
at large $l$, as illustrated in Fig.~\ref{spectra_rescaled}. This figure also 
nicely shows the relative phases of the acoustic peaks for the different modes.

When the ``primordial'' perturbation is a superposition of several modes, 
the multipole coefficients depend on a linear combination of the ``primordial'' modes, 
\beq
\label{a_lm}
a_{lm}=4\pi (-i)^l \int \frac{d^3\k}{(2\pi)^3} \left(\sum_I X^I(\k) \, g^I_l(k)\right) Y^*_{lm}(\hat\k).
\eeq
(Here we have once again omitted the polarization indices, as we will do 
in most of the equations of the paper, in order to improve readability.)
As a result,  the total angular power spectrum is now given by
\beq
\label{C_l}
C_l=\langle a_{lm}a_{lm}^*\rangle=\sum_{I,J}\frac{2}{\pi}\int_0^\infty dk\, k^2 g_l^I(k) g_l^J(k) P_{IJ}(k).
\eeq
We infer from CMB observations that the ``primordial'' perturbation is mainly of the adiabatic type. However, this does not preclude the presence, in addition to the adiabatic mode, of an isocurvature component, with a smaller amplitude. Precise measurement of the CMB fluctuations could lead to a detection
of such an extra component, or at least put constraints on its amplitude. 
For example, constraints on the CDM isocurvature to adiabatic ratio,
\begin{equation}
\label{eq:alpha}
\alpha = \frac{{\cal P}_{S_c}}{{\cal P}_{\zeta}} ,
\end{equation}
based on the WMAP7+BAO+SN data, have been published for the uncorrelated and fully correlated cases (the impact of isocurvature perturbations on the observable power spectrum  indeed depends  on the correlation between adiabatic and isocurvature perturbations, as illustrated in~\cite{Langlois:2000ar}). 
In terms of  the parameter $a\equiv \alpha/(1+\alpha)$, the limits 
given in  \cite{Komatsu:2010fb} 
are\footnote{Our notation differs from that of \cite{Komatsu:2010fb}: our $a$ corresponds to their $\alpha$ and our fully {\sl correlated} limit corresponds to their fully {\sl anti-correlated} limit, because their definition of the correlation has the opposite sign 
(see also \cite{Komatsu:2008hk} for a more detailed discussion).}
\beq
a_0<0.064\quad (95 \% {\rm CL}), \qquad a_{1}< 0.0037 \quad (95 \% {\rm CL})\,,
\eeq
respectively for the uncorrelated case  and for the fully correlated case.

\section{Generalized angular bispectra}
In this section, we turn to non-Gaussianities, including both adiabatic and isocurvature modes.
 
\subsection{Reduced and angular-averaged bispectra}
The angular bispectrum corresponds to the three-point function of the multipole coefficients:
\beq
B_{\l_1 \l_2 \l_3}^{m_1m_2m_3} \equiv \langle a_{\l_1 m_1} a_{\l_2 m_2} a_{\l_3 m_3}\rangle\,.
\eeq
Substituting the  expression (\ref{a_lm}) into the angular bispectrum, one can write it in the form
\beq
\label{B_lm}
B_{\l_1 \l_2 \l_3}^{m_1m_2m_3} ={\cal G}^{m_1m_2m_3}_{l_1l_2l_3}b_{l_1l_2l_3}\,,
\eeq
where the first, purely geometrical,  factor is the  Gaunt integral
\beq
{\cal G}^{m_1m_2m_3}_{l_1l_2l_3}\equiv \int d^2\n\,  Y_{l_1m_1}(\n)\,  Y_{l_2m_2}(\n)\,  Y_{l_3m_3}(\n)\,.
\eeq
The second factor, usually called the {\it reduced} bispectrum, is given by 
\begin{eqnarray}
\label{reduced_bispectrum}
b_{l_1l_2 l_3} & = & \sum_{I,J,K}  \left(\frac{2}{\pi}\right)^3\int \left(\prod_{i=1}^3k_i^2 dk_i\right)  \ g^I_{l_1}(k_1) g^J_{l_2}(k_2) g^K_{l_3}(k_3) 
B^{IJK}(k_1,k_2, k_3) \nonumber\\
&& \times
\int_0^\infty r^2 dr j_{l_1}(k_1r) j_{l_2}(k_2 r) j_{l_3}(k_3 r)\,, 
\end{eqnarray}
which depends on  the bispectra of the primordial $X^I$:
\beq
\label{B_IJK}
\langle X^{I}(\k_1) X^J(\k_2) X^{K}(\k_3) \rangle \equiv  (2 \pi)^3 \delta (\Sigma_i \k_i) B^{IJK}(k_1, k_2, k_3)\,.
 \eeq
 The reduced bispectrum (\ref{reduced_bispectrum}) is  the sum of several contributions, corresponding to different values of  the indices $I$, $J$ and $K$ that vary over the range of  modes included in the primordial perturbations. This expression thus generalizes the purely adiabatic expression given in \cite{Komatsu:2001rj}. 
 
It is also useful to define the angle-averaged  bispectrum
 \beq
 B_{\l_1 \l_2 \l_3} \equiv \sum_{m_1, m_2, m_3} 
  \left(
\begin{array}{ccc}
\l_1 & \l_2 & \l_3 \cr
m_1 & m_2 & m_3
\end{array}
\right)
 B_{\l_1 \l_2 \l_3}^{m_1 m_2 m_3} =\sqrt{\frac{(2\l_1+1)(2\l_2+1)(2\l_3+1)}{4\pi}}  \left(
\begin{array}{ccc}
\l_1 & \l_2 & \l_3 \cr
0 & 0 & 0
\end{array}
\right)b_{\l_1 \l_2 \l_3}\,,
 \eeq
 where the second relation is obtained by substituting (\ref{B_lm}) and by using the identity
 \beq
 \sum_{m_1, m_2, m_3} 
  \left(
\begin{array}{ccc}
\l_1 & \l_2 & \l_3 \cr
m_1 & m_2 & m_3
\end{array}
\right)
{\cal G}^{m_1m_2m_3}_{l_1l_2l_3}=\sqrt{\frac{(2\l_1+1)(2\l_2+1)(2\l_3+1)}{4\pi}}  \left(
\begin{array}{ccc}
\l_1 & \l_2 & \l_3 \cr
0 & 0 & 0
\end{array}
\right)
\,.
\eeq

 \subsection{Non-Gaussianities of local type}
 
To proceed further, one must make some assumption about the functional dependence of the 
bispectra $B^{IJK}(k_1, k_2, k_3)$ in Fourier space. This corresponds to the so-called 
``shape'' of the bispectrum~\cite{Babich:2004gb}, 
which has been discussed at length 
 in the literature
in the  purely adiabatic case where the $B^{IJK}$ reduce to the single bispectrum $B^{\zeta\zeta\zeta}$. 
In the present work, we consider the simplest form of non-Gaussianity, namely  the local shape. 
In the purely adiabatic case, it is defined by 
\beq
\zeta(\bx)=\zeta_L(\bx)+\frac{3}{5}f_{\rm NL}
\left(\zeta_L(\bx)^2-\langle \zeta_L\rangle^2\right)\,,
\eeq
in physical space, where the factor $3/5$ appears because $f_{\rm NL}$ was originally defined with respect to the gravitational potential $\Phi=(3/5)\zeta$, instead of $\zeta$. 
The subscript $L$ here denotes the linear part of the perturbation, which is assumed to be Gaussian.

In Fourier space, this leads to a bispectrum that depends quadratically on the power spectrum:
\beq
\label{B_adiab}
B^{\zeta\zeta\zeta}= \tf_{\rm NL} \left[ P_\zeta(k_2)P_\zeta(k_3)+P_\zeta(k_3)P_\zeta(k_1)+P_\zeta(k_1)P_\zeta(k_2)\right], \qquad \tf_{\rm NL}\equiv\frac65 f_{\rm NL}\,.
\eeq
In the present context where we assume the presence of an isocurvature mode in addition to the dominant adiabatic mode, the simplest extension of (\ref{B_adiab}) is to assume that all the 
generalized  bispectra $B^{IJK}(k_1, k_2, k_3)$ can be written  as the sum of terms  quadratic in the adiabatic power spectrum (note that this implicitly assumes that the power spectrum of the isocurvature mode 
and the isocurvature cross power spectrum,
 if nonvanishing, 
 have the same spectral dependence as the adiabatic one). However, in contrast with (\ref{B_adiab}) where all terms share the same coefficient, as a consequence of the invariance of the bispectrum under 
the exchange of momenta, this is no longer the case for the generalized bispectra when the  indices $I$, $J$ and $K$ are not identical. What  the definition (\ref{B_IJK}) implies  is simply that the bispectra are left unchanged under  the {\it simultaneous} change of two indices and the corresponding momenta (e.g.\ $I$ and $J$, $k_1$ and $k_2$). 
This leads to the decomposition
\begin{eqnarray}
\label{bispectrum_local}
 B^{IJK}(k_1, k_2, k_3)=
 \tf_{\rm NL}^{I, JK}  P_\zeta(k_2) P_\zeta(k_3) 
 +\tf_{\rm NL}^{J, KI}  P_\zeta(k_1) P_\zeta(k_3)
+\tf_{\rm NL}^{K, IJ}   P_\zeta(k_1)P_\zeta(k_2)\,, 
  \end{eqnarray}
  where the coefficients $\tf_{\rm NL}^{I, JK}$ must satisfy the condition 
\beq
\label{f_sym}
\tf_{\rm NL}^{I, JK} =\tf_{\rm NL}^{I, KJ} \,.
\eeq
To keep track of  this symmetry, we  separate the first index from the last two indices with a comma.

\subsection{Link with multiple-field inflation}
It is instructive to show that our definition of generalized local non-Gaussianity is the natural outcome of a generic model of multiple-field inflation. 
Indeed, allowing for several light degrees of freedom during inflation, one can relate, in a very generic way, 
the ``primordial'' perturbations $X^I$ (defined during the standard radiation era)  to the fluctuations of light primordial fields $\phi^a$, generated  at Hubble crossing during inflation, so that one can write, up to second order, 
\beq
\label{X_I}
X^I= N^I_a\,  \delta\phi^a+\frac12 N^{I}_{ab}\,  \delta\phi^a \delta\phi^b + \dots
\eeq
where the $\delta\phi^a$ can usually be treated as  independent quasi-Gaussian fluctuations, i.e. 
\beq
\langle \delta\phi^a (\k) \, \delta\phi^b (\k')\rangle=
 (2\pi)^3 \, \delta^{ab}P_{\delta\phi}(k) \, \delta(\k + \k')\,, \qquad P_{\delta\phi}(k)=2\pi^2k^{-3}\left(\frac{H_*}{2\pi}\right)^2\,,
 \eeq
 where  a star denotes Hubble crossing time. The relation (\ref{X_I}) is very general, and all the details of the inflationary model are embodied by the coefficients $N_a^I$ and $N_{ab}^I$.

 Substituting (\ref{X_I}) into (\ref{B_IJK}) and using Wick's theorem, one finds that the bispectra $B_{IJK}$ can be expressed  in  the form
\beq
\label{bispectrum_fields}
 B^{IJK}(k_1, k_2, k_3)=
 \lambda^{I, JK}  P_{\delta\phi}(k_2) P_{\delta\phi}(k_3) 
 +\lambda^{J, KI}  P_{\delta\phi}(k_1) P_{\delta\phi}(k_3)
+\lambda^{K, IJ}   P_{\delta\phi}(k_1)P_{\delta\phi}(k_2)\,, 
\eeq
with the coefficients 
\beq
\label{lambda}
\lambda^{I, JK} \equiv \delta^{ac}\delta^{bd}N^I_{ab} N^J_{c} N^K_{d}
\eeq
 (the summation over scalar field indices $a$, $b$, $c$ and $d$  is implicit), which are  symmetric under the interchange  of the last two indices, by construction. Since the 
 adiabatic power spectrum is given by
 \beq
 P_\zeta=(\delta^{ab} N_a^\zeta N_b^\zeta) P_{\delta\phi} \equiv A P_{\delta\phi},
 \eeq
 one obtains finally (\ref{bispectrum_local}) with
 \beq
 \label{f_NL}
\tf_{\rm NL}^{I,JK}= \lambda_{NL}^{I, JK} /A^2\,,
\eeq
where it is implicitly assumed that the coefficients $N^I_a$ are weakly time dependent so that the scale dependence of $A^2$ can be neglected.  One can notice that the first index is related to the second-order terms in the decomposition (\ref{X_I}), while  the last two indices come  from   the first-order terms. 

Except in the last section devoted to a specific class of early Universe models, all our considerations will simply follow from our assumption (\ref{bispectrum_local}) and thus apply to any model leading to this local form, whether based on inflation or not.

\subsection{Decomposition of the angular bispectrum}
After substitution of (\ref{bispectrum_local}) into (\ref{reduced_bispectrum}), the reduced bispectrum can 
finally 
be written as 
\beq
\label{b_I}
b_{l_1l_2 l_3}=   \sum_{I,J,K}\tf_{\rm NL}^{I,JK}b_{l_1l_2 l_3}^{I,JK},
\eeq
where each contribution is of the form\footnote{We use the standard notation: $(l_1 l_2 l_3)\equiv [l_1l_2l_3+ 5\,  {\rm perms}]/3!$.}
\begin{eqnarray}
\label{b_IJK}
b_{l_1l_2 l_3}^{I,JK}= 3   \int_0^\infty r^2 dr \, \alpha^I_{(l_1}(r)\beta^{J}_{l_2}(r)\beta^{K}_{l_3)}(r),
\end{eqnarray}
with   
\bea
\label{alpha}
\alpha^I_{\l}(r)&\equiv& \frac{2}{\pi} \int k^2 dk\,   j_\l(kr) \, g^I_{\l}(k),
\\
\label{beta}
\beta^{I}_{\l}(r)&\equiv& \frac{2}{\pi}  \int k^2 dk \,  j_\l(kr) \, g^I_{\l}(k)\,  P_\zeta(k)\,.
\eea
While we have omitted the polarization indices, the reader should keep in mind
that each transfer function carries, in addition to the isocurvature index,
a polarization index, and hence the same is true for $\alpha_l$ and $\beta_l$.
As a consequence, the bispectrum has three
 polarization indices that we do not
show.

\begin{figure}
\centering
\includegraphics[width=0.49\textwidth, clip=true]{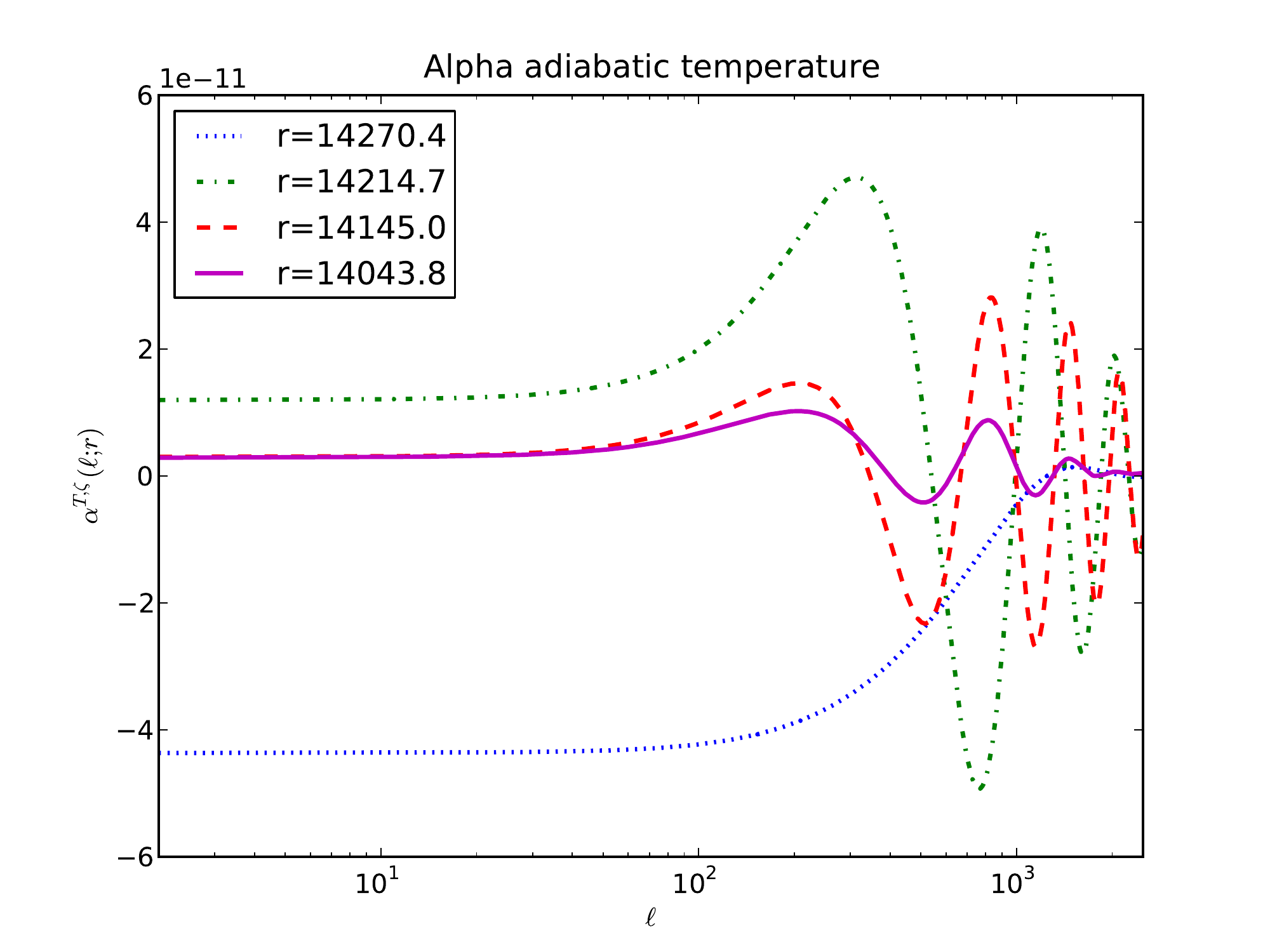}
\includegraphics[width=0.49\textwidth, clip=true]{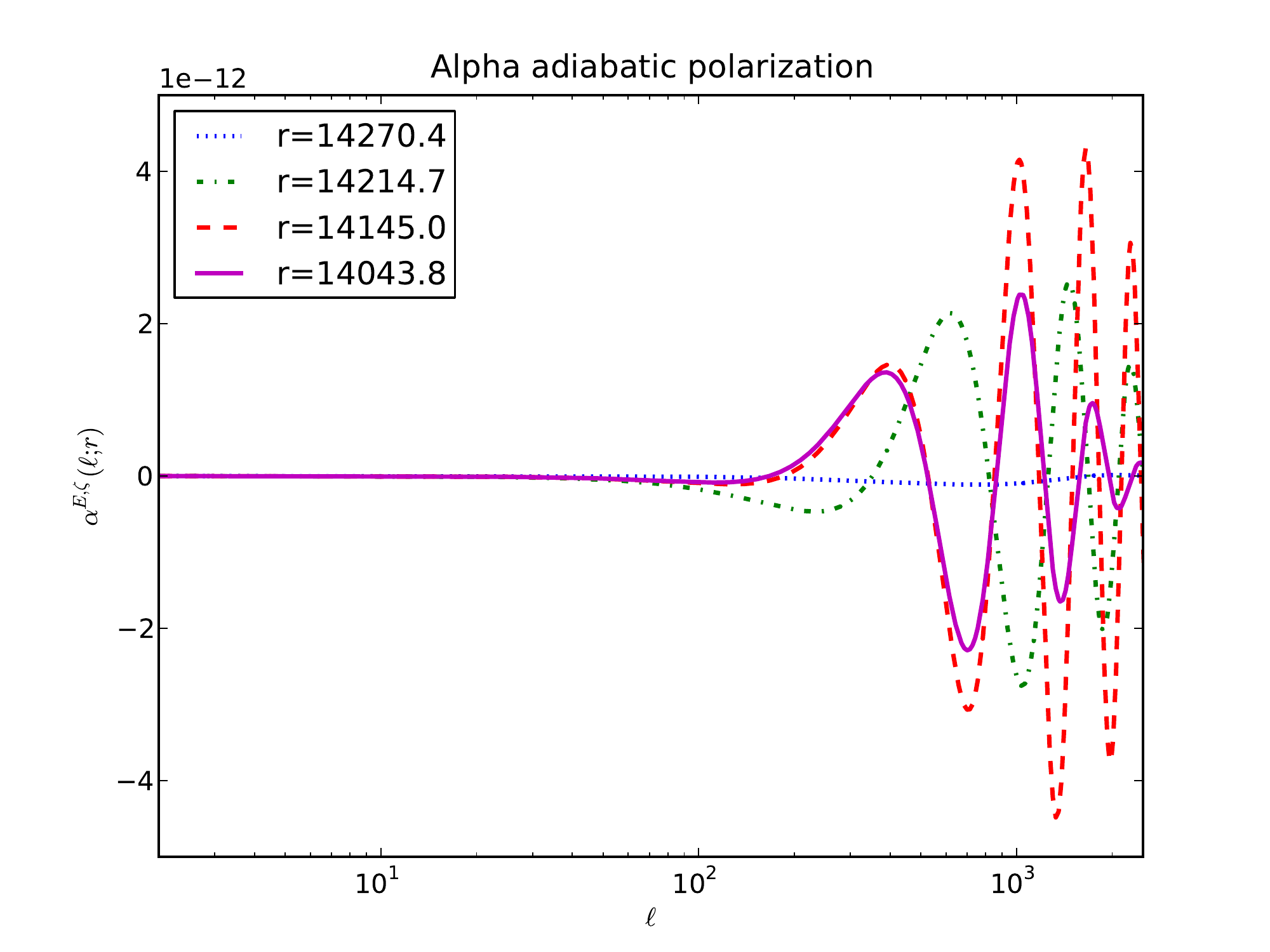}
\caption{The adiabatic $\alpha^{\zeta}_\l(r)$ as a function of $\l$ for 
temperature (left) and polarization (right). It has been evaluated at
four different values of $r$: 14043.8, 14145.0, 14214.7, and 14270.4.}
\label{alpha_a}
\end{figure}

\begin{figure}
\centering
\includegraphics[width=0.49\textwidth, clip=true]{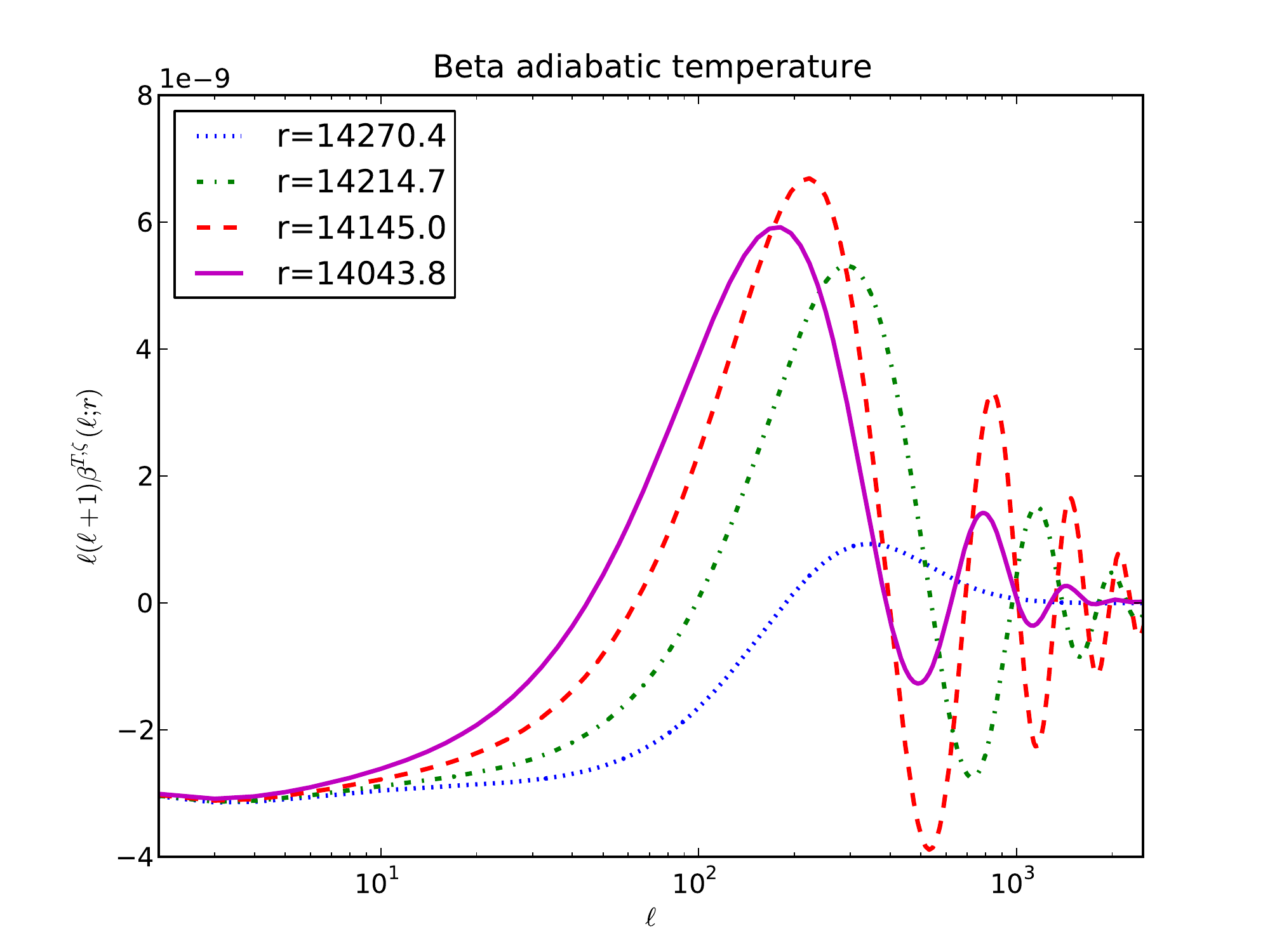}
\includegraphics[width=0.49\textwidth, clip=true]{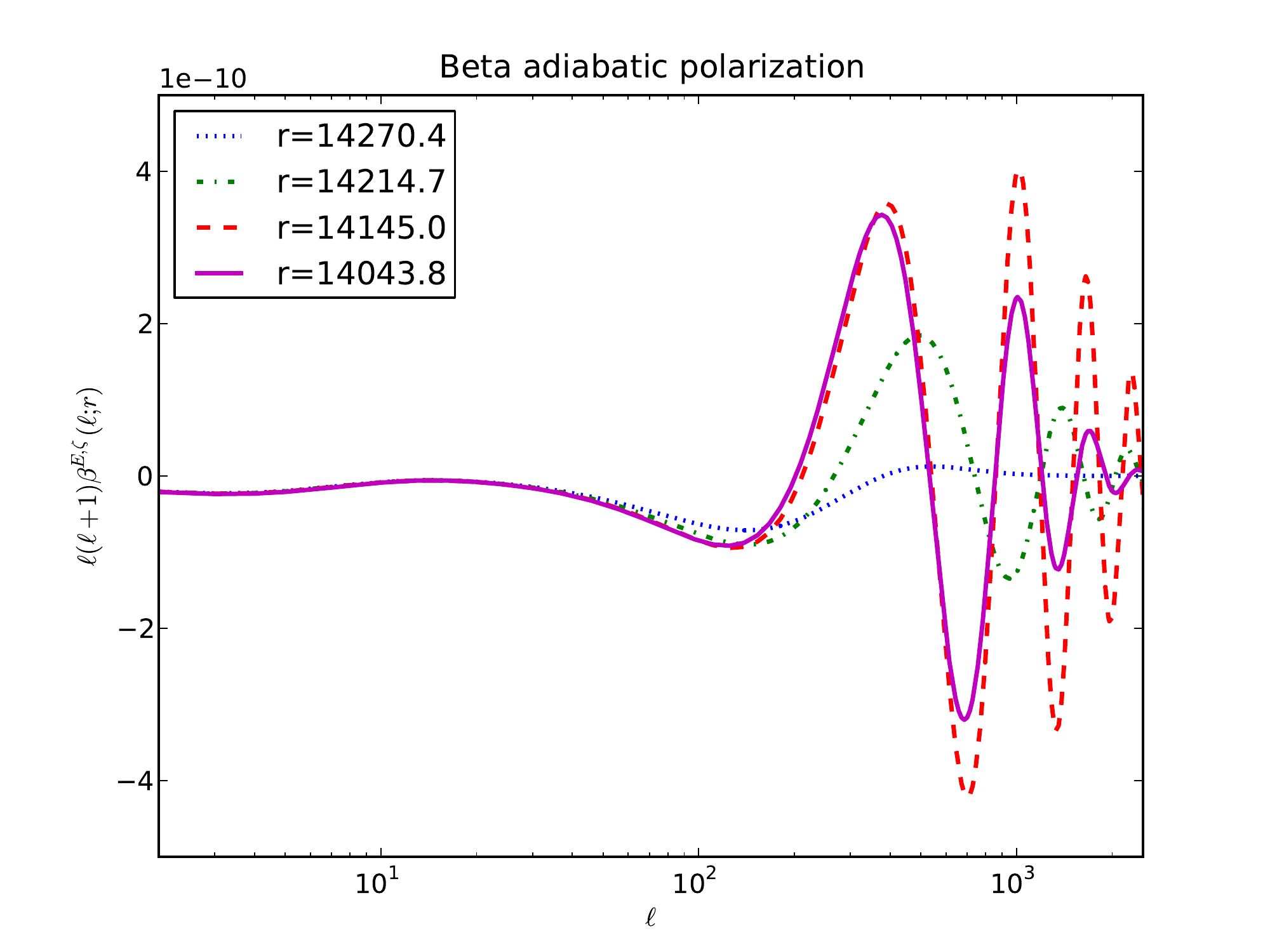}
\caption{The adiabatic $\l(\l+1)\beta^\zeta_\l(r)$ as a function of $\l$ for 
temperature (left) and polarization (right) (for the same values of  
$r$ as in Fig.~\ref{alpha_a}).}
\label{beta_a}
\end{figure}

The purely adiabatic bispectrum, usually the only one considered, can be expressed as 
\begin{eqnarray}
b^{\zeta,\zeta\zeta}_{l_1l_2 l_3}&=&  3 \int_0^\infty r^2 dr \alpha^\zeta_{(l_1}(r)\beta^{\zeta}_{l_2}(r)\beta^{\zeta}_{l_3)}(r)
\cr
&=&   \int_0^\infty r^2 dr \left[\alpha^\zeta_{l_1}(r)\beta^{\zeta}_{l_2}(r)\beta^\zeta_{l_3}(r)+\alpha^\zeta_{l_2}(r)\beta^\zeta_{l_3}(r)\beta^\zeta_{l_1}(r)+\alpha^\zeta_{l_3}(r)\beta^\zeta_{l_1}(r)\beta^\zeta_{l_2}(r)\right].
\label{b_adiab}
\end{eqnarray}
The functions $\alpha^\zeta_{\l}(r)$ and $\beta^\zeta_\l(r)$ are plotted respectively in Fig.~\ref{alpha_a} and Fig.~\ref{beta_a}. We have considered both the temperature and polarization transfer functions. 
The radial distance $r$ can
be expressed as the speed of light times the difference in conformal time
between now and the time in the past we consider. In the figures we have chosen 
some sample values around the time of last scattering, which corresponds to
$r \approx 14100$~Mpc using the WMAP7-only best-fit parameters.

Since we consider local non-Gaussianity, the main contribution to the
bispectrum comes from the squeezed limit, i.e.\ one of the multipole
numbers is much smaller than the other two. To simplify the analysis,
let us assume that $l_1 \ll l_2 = l_3 \equiv l$. One finds that, in
this limit, the integrand in (\ref{b_adiab}) is dominated by (twice)
$\alpha^\zeta_{l}(r)\beta^\zeta_{l}(r)\beta^\zeta_{l_1}(r)$, whereas
the first term is negligible: 
one can see from Fig.~\ref{alpha_a} and \ref{beta_a} that $\beta^\zeta_{l_1}(r)$
is much larger (in absolute value) than $\alpha^\zeta_{l_1}(r)$.
This is true both for temperature and polarization (one should also keep
in mind that the $\beta$'s have been multiplied by $l(l+1)$ in the plots,
which makes them look much larger at large $l$). 

\begin{figure}
\centering
\includegraphics[width=0.49\textwidth, clip=true]{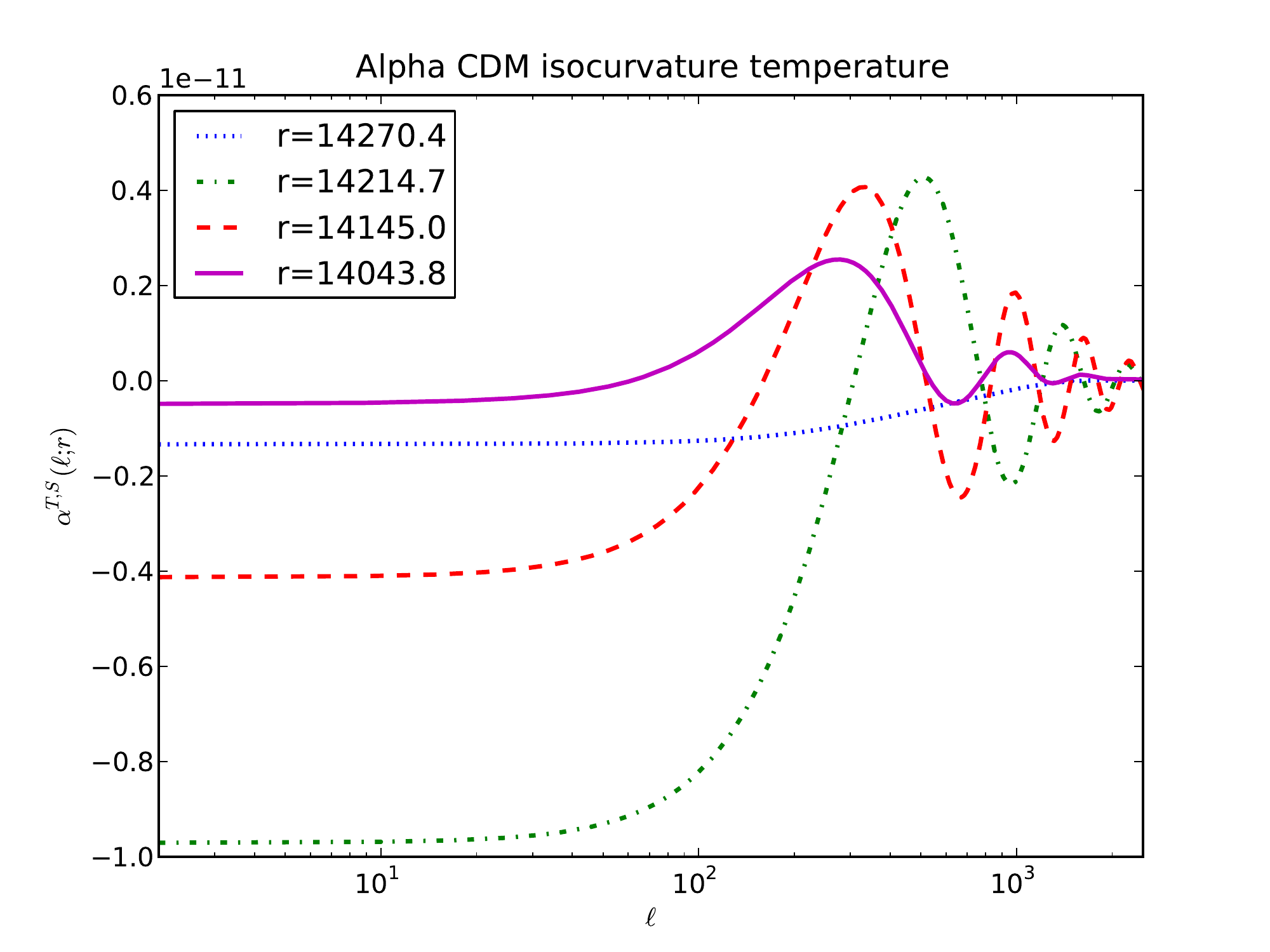}
\includegraphics[width=0.49\textwidth, clip=true]{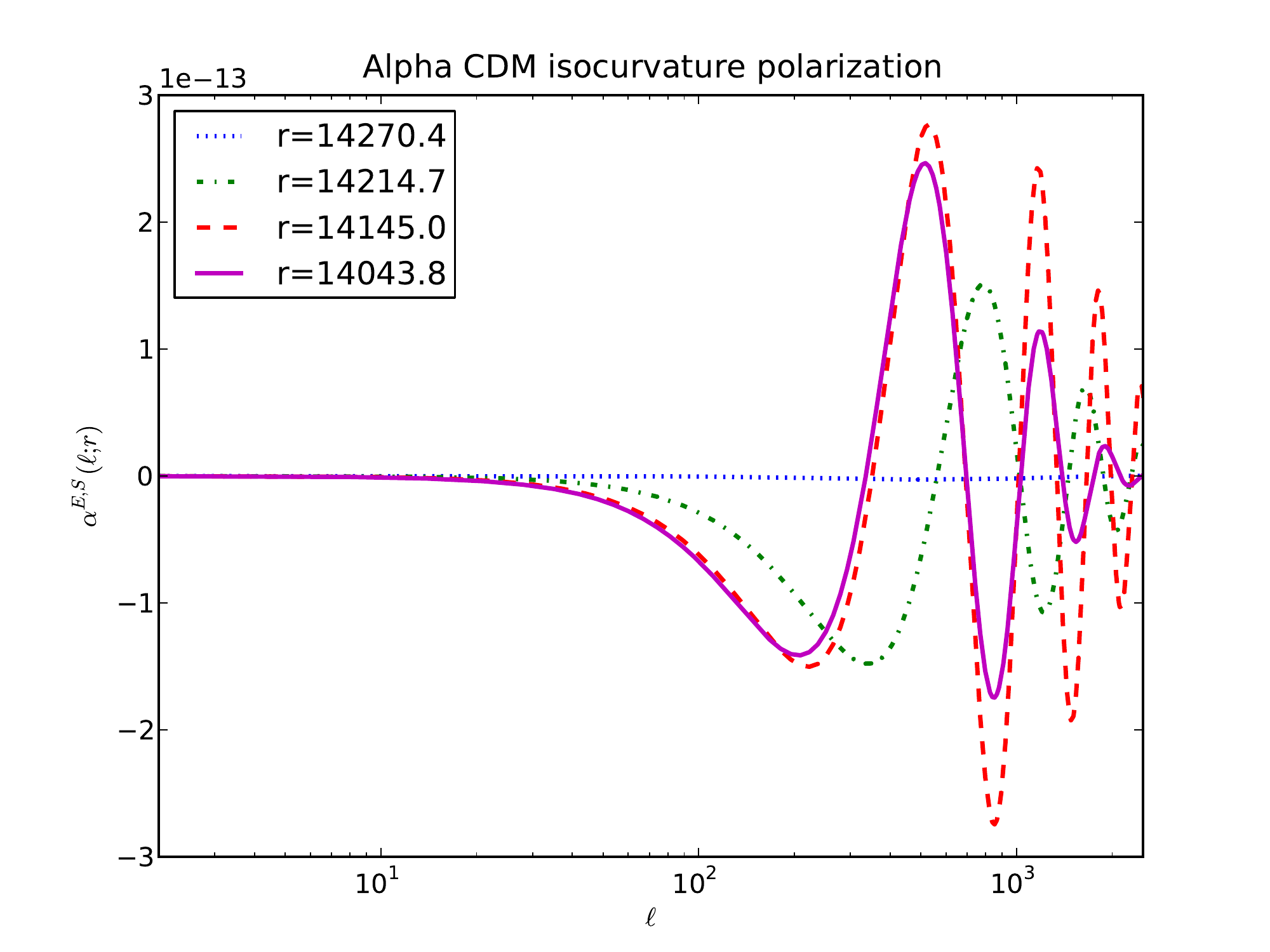}
\caption{The CDM isocurvature $\alpha^{S_c}_\l(r)$ as a function of $\l$ 
for temperature (left) and polarization (right) (for the same values of  
$r$ as in Fig.~\ref{alpha_a}).}
\label{alpha_cdm}
\end{figure}

\begin{figure}
\centering
\includegraphics[width=0.49\textwidth, clip=true]{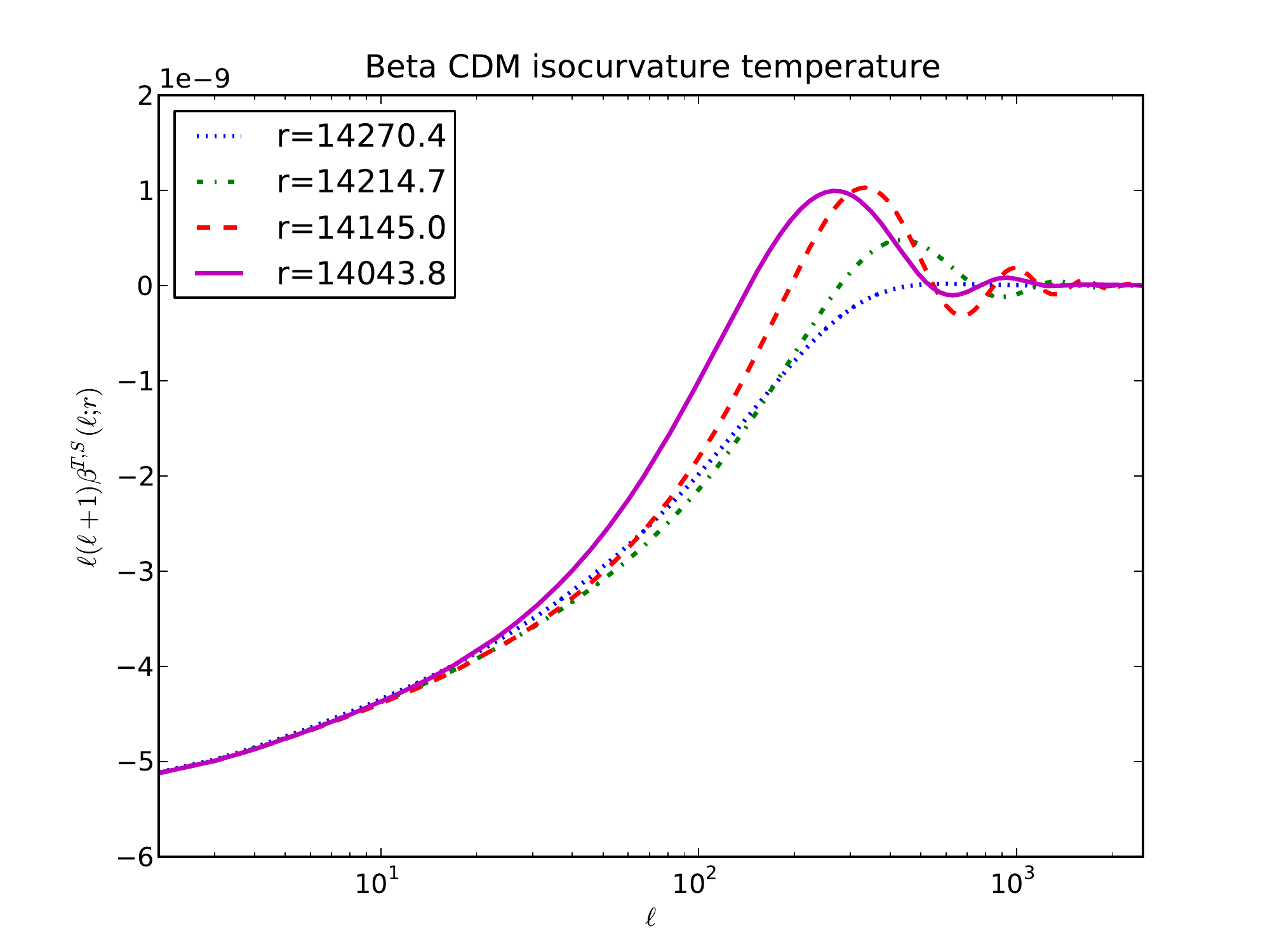}
\includegraphics[width=0.49\textwidth, clip=true]{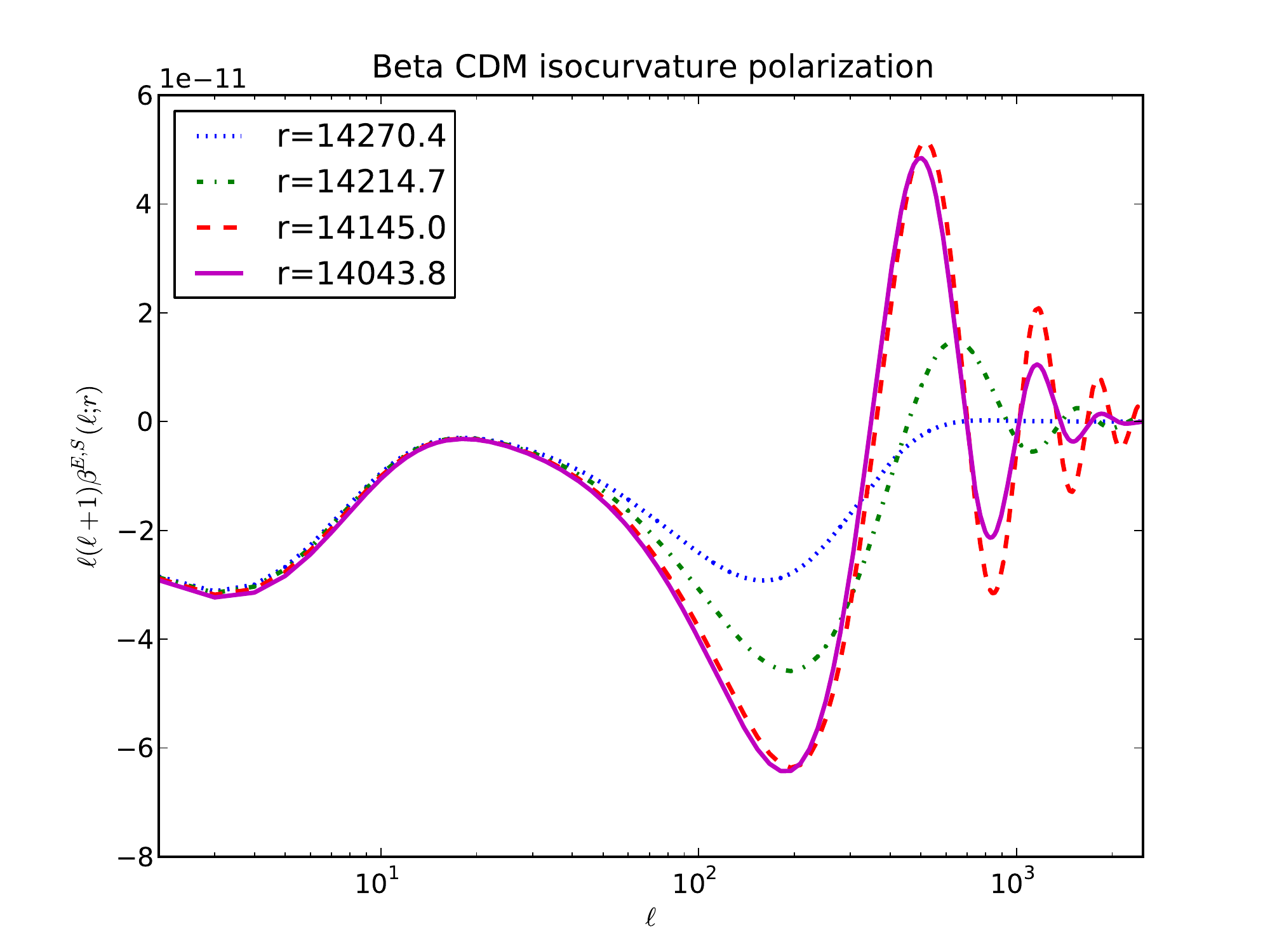}
\caption{The CDM isocurvature $\l(\l+1)\beta^{S_c}_\l(r)$ as a function of $\l$ 
for temperature (left) and polarization (right) (for the same values of  
$r$ as in Fig.~\ref{alpha_a}).}
\label{beta_cdm}
\end{figure}

 \begin{figure}
\centering
\includegraphics[width=0.49\textwidth, clip=true]{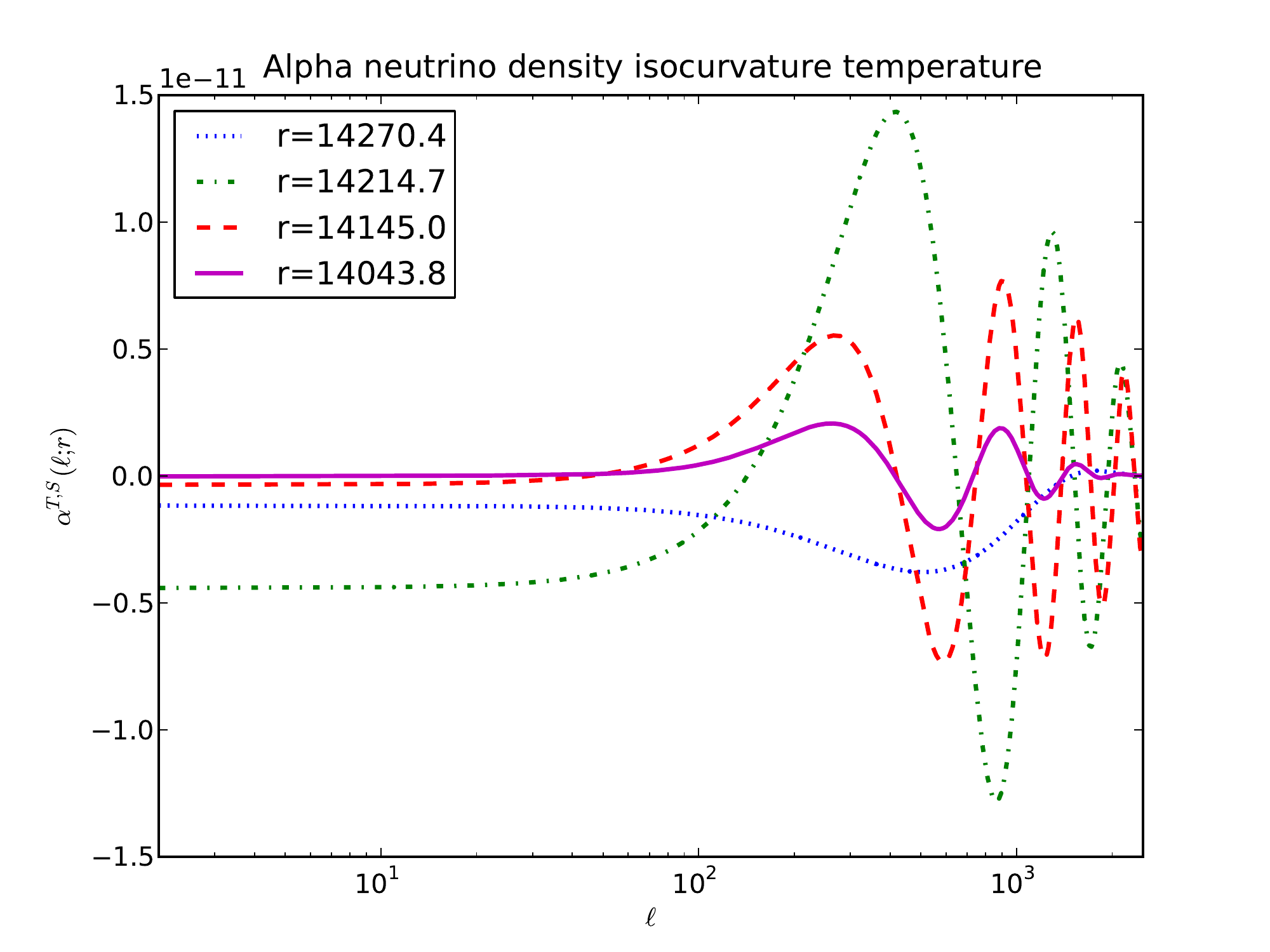}
\includegraphics[width=0.49\textwidth, clip=true]{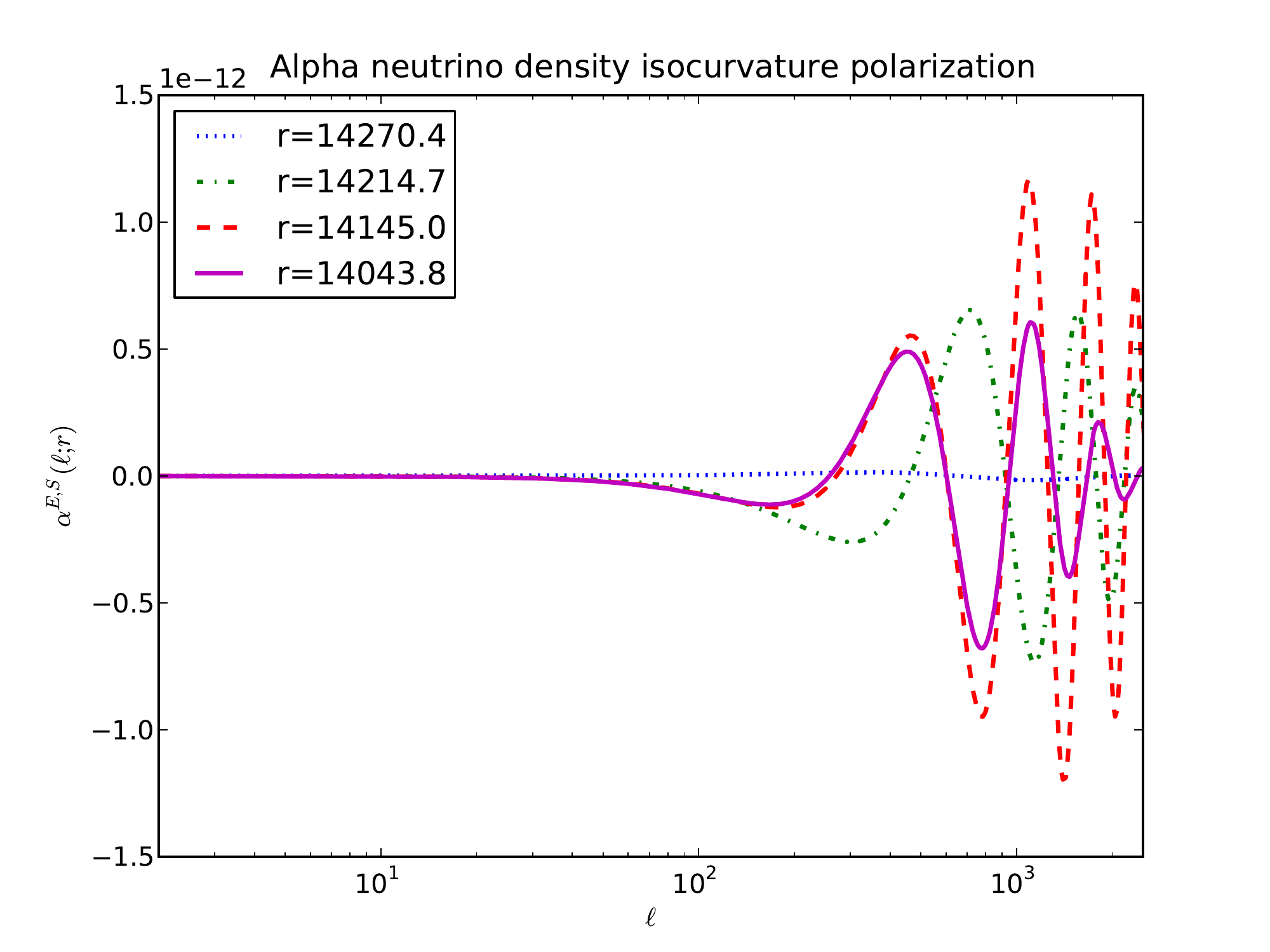}
\caption{The neutrino density isocurvature $\alpha^{S_{\nu d}}_\l(r)$ as a 
function of $\l$ for temperature (left) and polarization (right) (for the 
same values of $r$ as in Fig.~\ref{alpha_a}).}
\label{alpha_nd}
\end{figure}

\begin{figure}
\centering
\includegraphics[width=0.49\textwidth, clip=true]{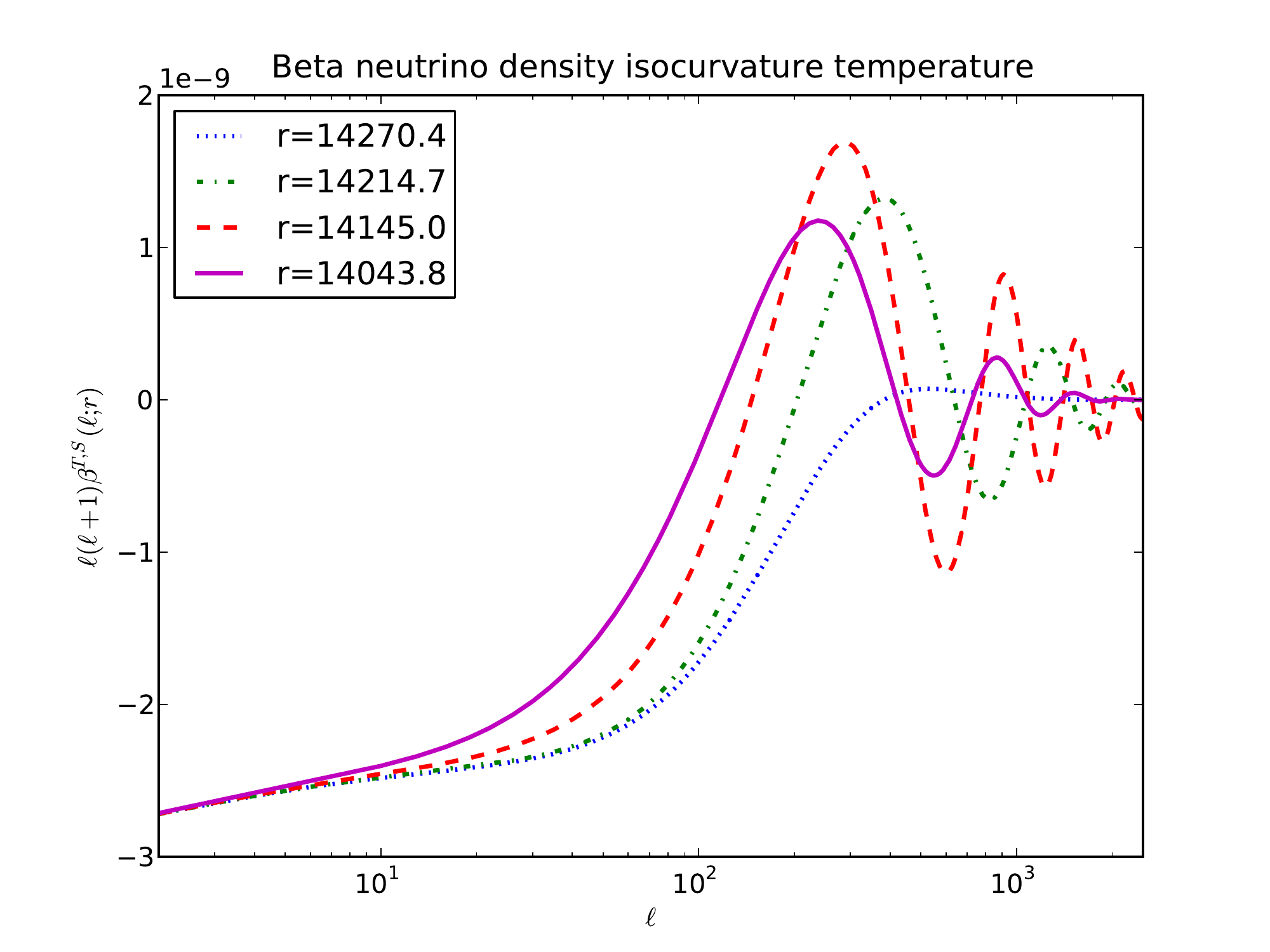}
\includegraphics[width=0.49\textwidth, clip=true]{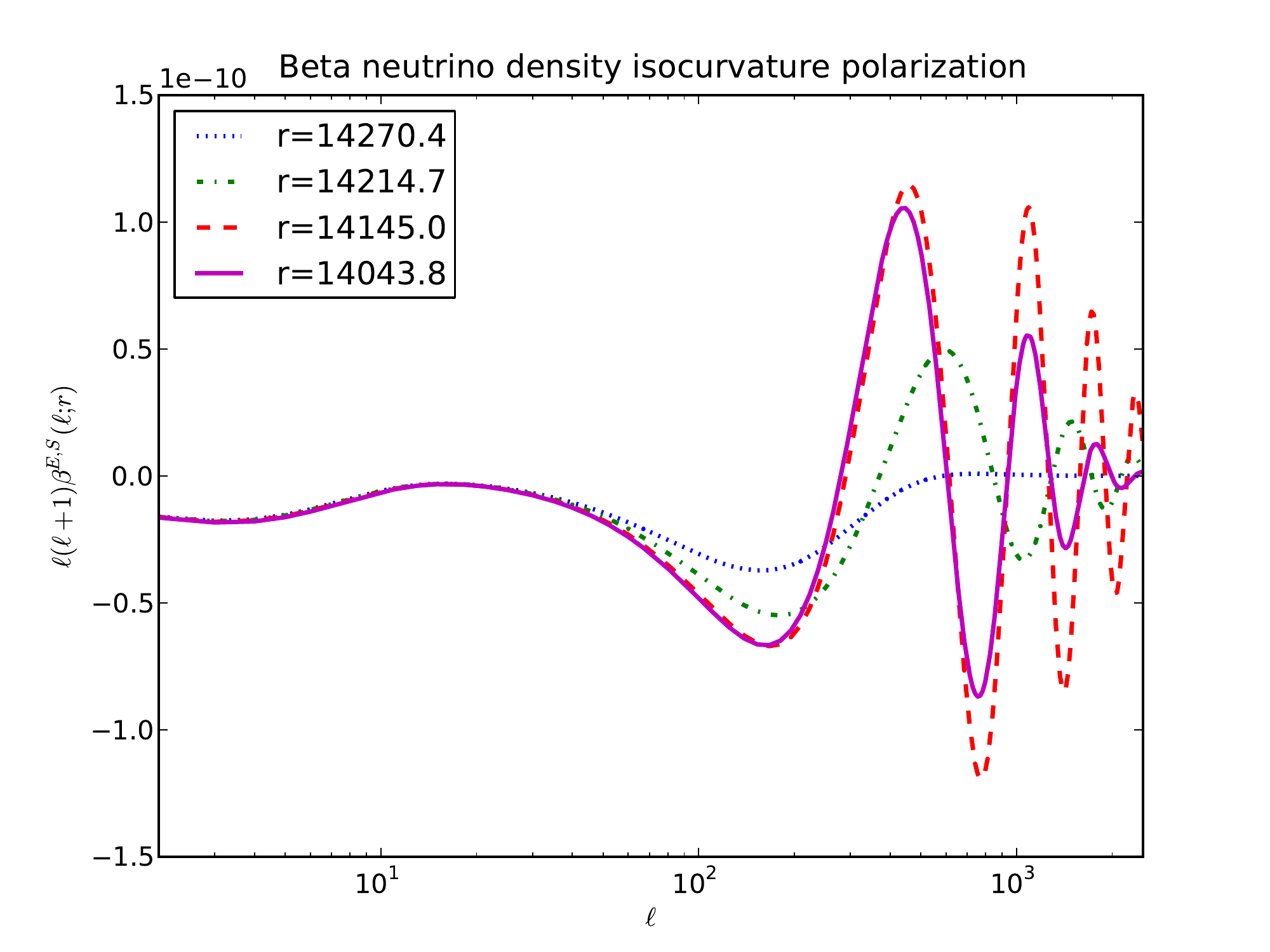}
\caption{The neutrino density isocurvature $\l(\l+1)\beta^{S_{\nu d}}_\l(r)$ 
as a function of $\l$ for temperature (left) and polarization (right) 
(for the same values of $r$ as in Fig.~\ref{alpha_a}).}
\label{beta_nd}
\end{figure}

 \begin{figure}
\centering
\includegraphics[width=0.49\textwidth, clip=true]{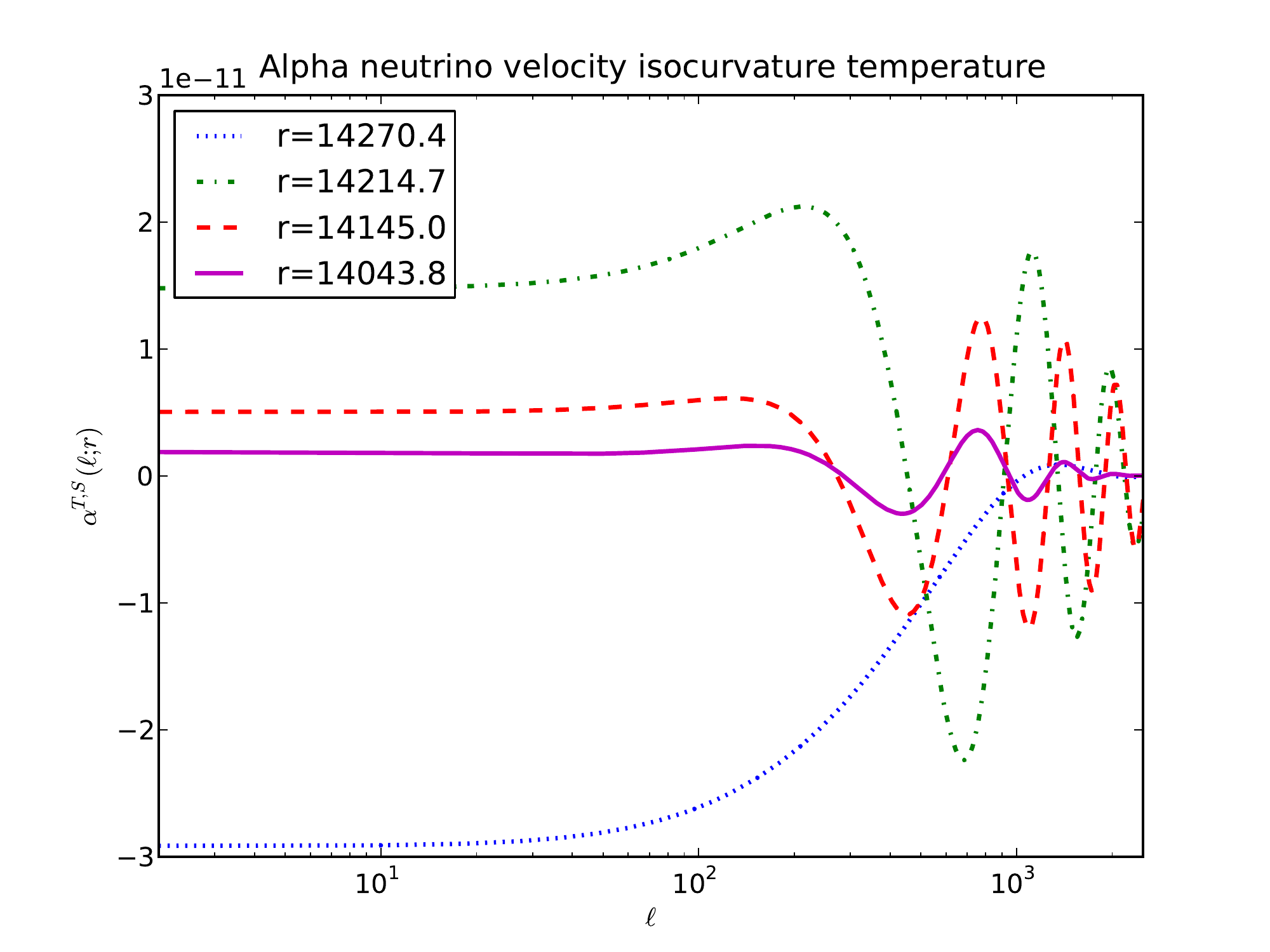}
\includegraphics[width=0.49\textwidth, clip=true]{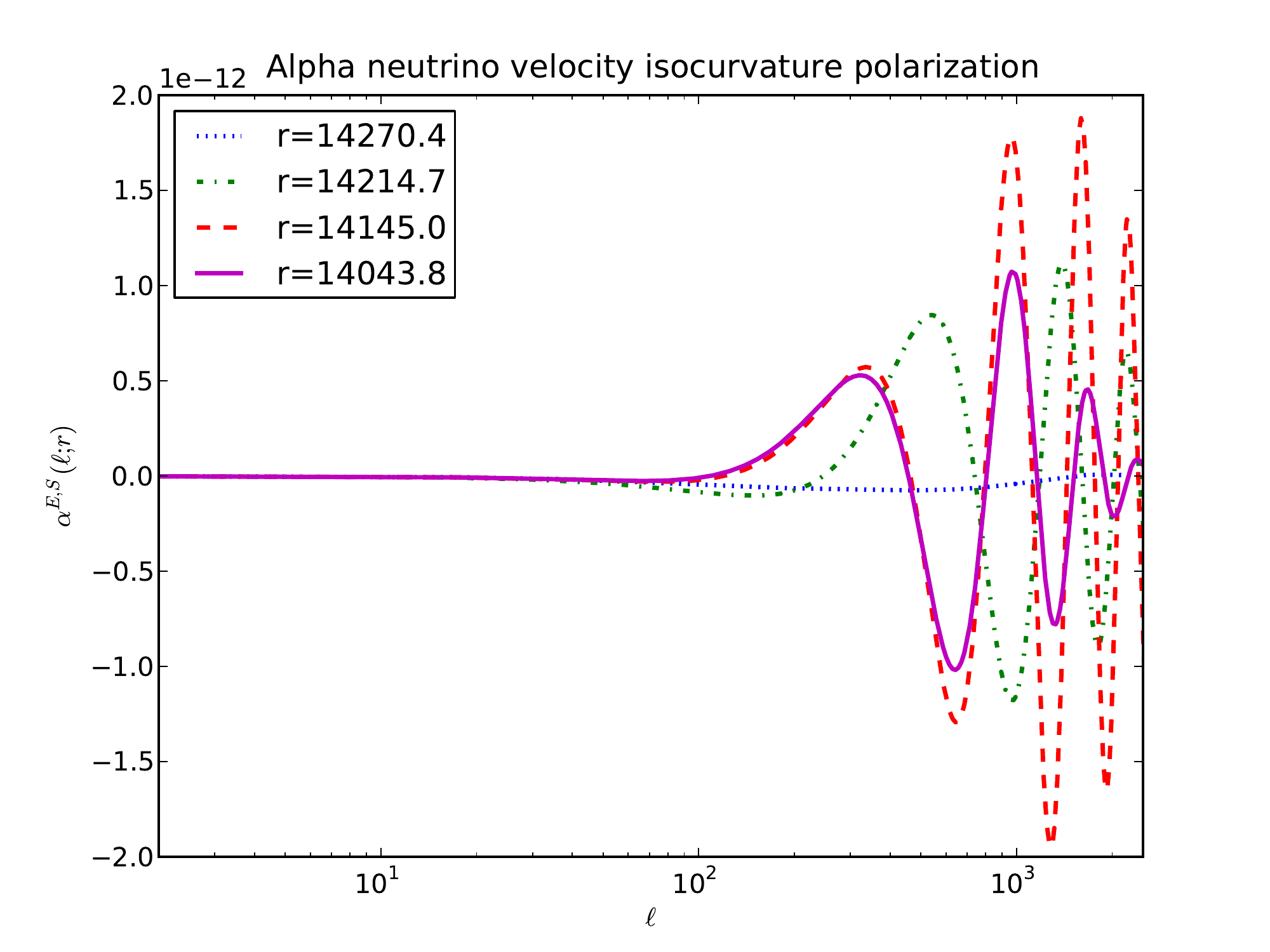}
\caption{The neutrino velocity isocurvature $\alpha^{S_{\nu v}}_\l(r)$ as a 
function of $\l$ for temperature (left) and polarization (right) (for the 
same values of $r$ as in Fig.~\ref{alpha_a}).}
\label{alpha_nv}
\end{figure}

\begin{figure}
\centering
\includegraphics[width=0.49\textwidth, clip=true]{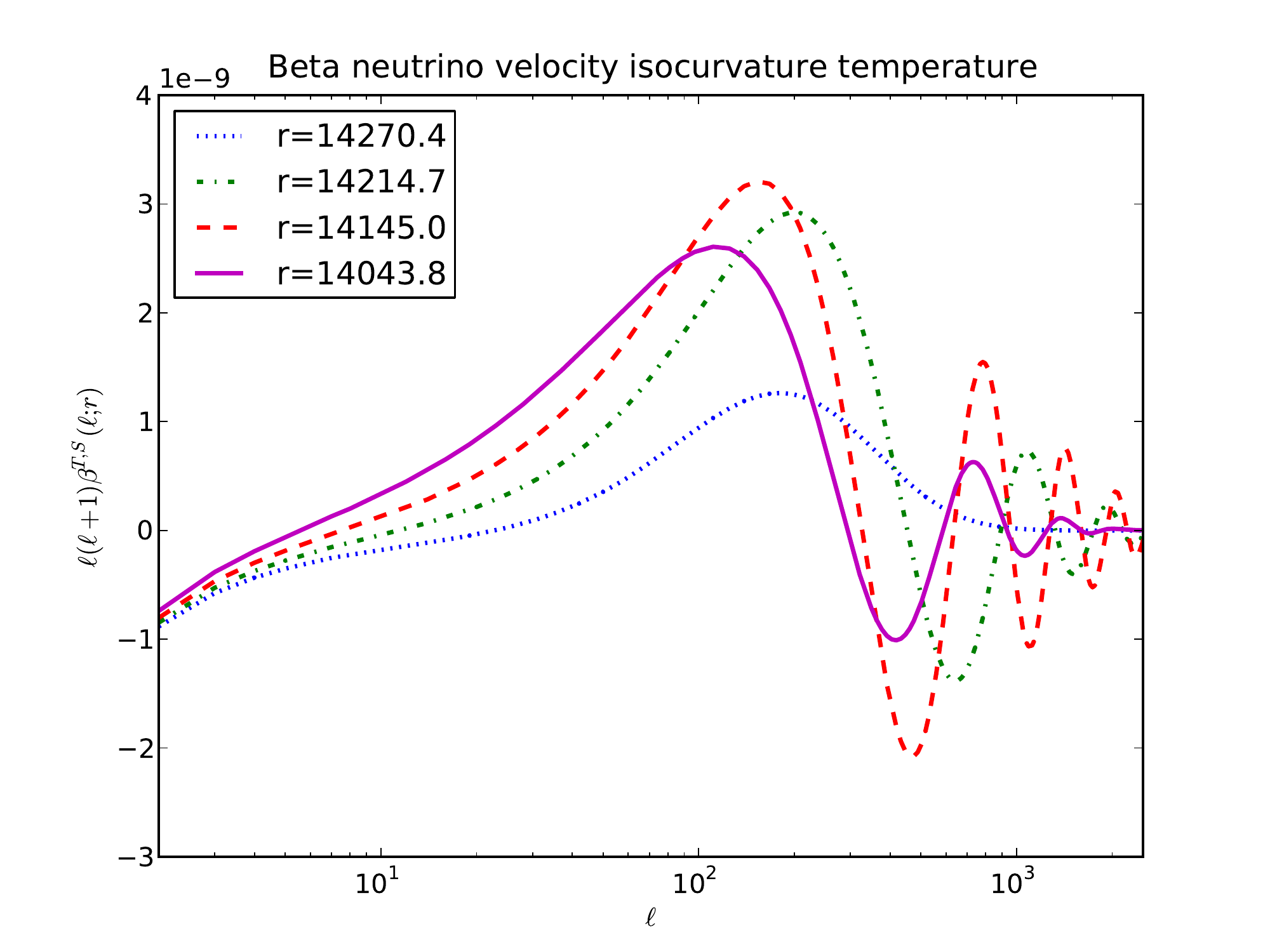}
\includegraphics[width=0.49\textwidth, clip=true]{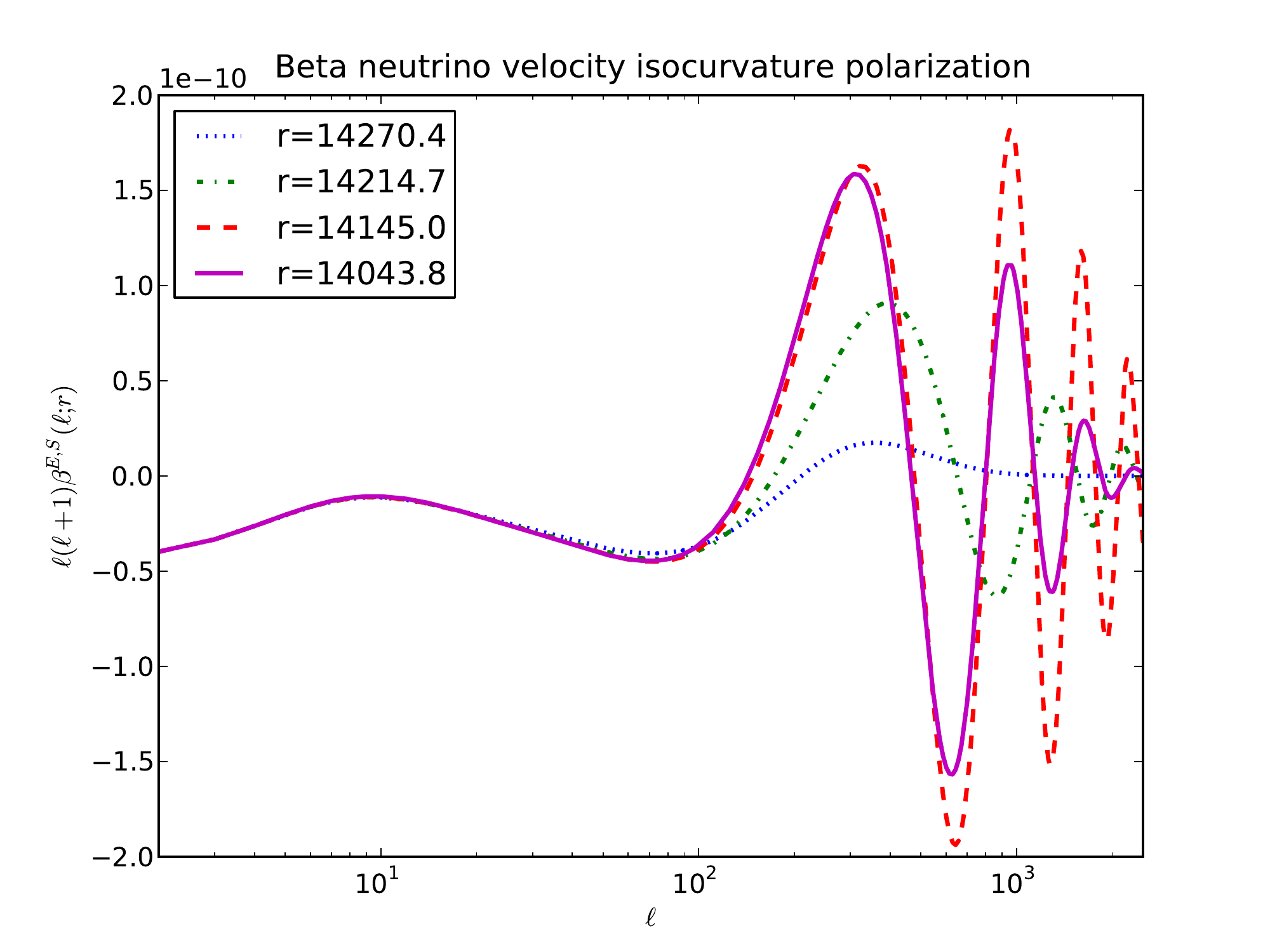}
\caption{The neutrino velocity isocurvature $\l(\l+1)\beta^{S_{\nu v}}_\l(r)$ 
as a function of $\l$ for temperature (left) and polarization (right) 
(for the same values of $r$ as in Fig.~\ref{alpha_a}).}
\label{beta_nv}
\end{figure}

The purely isocurvature bispectrum has exactly the same structure as (\ref{b_adiab}), but 
with the functions $\alpha^S_{\l}(r)$ and $\beta^S_\l(r)$ replacing $\alpha^\zeta_{\l}(r)$ and $\beta^\zeta_\l(r)$. Moreover, the shapes of $\alpha^S_{\l}(r)$ and $\beta^S_\l(r)$ depend on the type of isocurvature mode:
Fig.~\ref{alpha_cdm} and Fig.~\ref{beta_cdm} correspond to the CDM isocurvature mode, Fig.~\ref{alpha_nd} and Fig.~\ref{beta_nd}  to the neutrino density isocurvature mode and Fig.~\ref{alpha_nv} and Fig.~\ref{beta_nv}  to the neutrino velocity isocurvature mode.
The functions $\alpha^S_{\l}(r)$ and $\beta^S_\l(r)$ for  the baryon 
isocurvature mode can be deduced from the CDM isocurvature functions by a simple  rescaling, according to (\ref{omega_bc}):
\beq 
\alpha^{S_b}_{\l}(r)=\omega_{bc} \, \alpha^{S_c}_{\l}(r), \qquad 
\beta^{S_b}_{\l}(r)=\omega_{bc}\,  \beta^{S_c}_{\l}(r)\,.
\eeq

The other bispectra depend on a mixing of the adiabatic and isocurvature functions. For example, one finds 
\begin{eqnarray}
b^{\zeta, \zeta S}_{l_1l_2 l_3}+  b^{\zeta, S \zeta}_{l_1l_2 l_3}&=& 6 \int_0^\infty r^2 dr \alpha^\zeta_{(l_1}(r)\beta^{\zeta}_{l_2}(r)\beta^{S}_{l_3)}(r)
\cr
&=&  \int_0^\infty r^2 dr \left\{\alpha^\zeta_{l_1}(r)
\left[\beta^{\zeta}_{l_2}(r)\beta^{S}_{l_3}(r)+\beta^{\zeta}_{l_3}(r)\beta^{S}_{l_2}(r)\right]
\right.
\\
&&
\left.
+
\alpha^\zeta_{l_2}(r)\left[\beta^{\zeta}_{l_3}(r)\beta^{S}_{l_1}(r)+\beta^{\zeta}_{l_1}(r)\beta^{S}_{l_3}(r)\right]
+\alpha^\zeta_{l_3}(r)\left[ \beta^{\zeta}_{l_1}(r)\beta^{S}_{l_2}(r)+ \beta^{\zeta}_{l_2}(r)\beta^{S}_{l_1}(r)\right]
\right\}.
\nonumber
\end{eqnarray}
Since $b^{\zeta, \zeta S}_{l_1l_2 l_3}$ and $b^{\zeta, S \zeta}_{l_1l_2 l_3}$ cannot be distinguished, we will always consider the sum of the two, and similarly for $b^{S, \zeta S}_{l_1l_2 l_3}$ and $b^{S, S \zeta}_{l_1l_2 l_3}$.

\begin{figure}
\centering
\includegraphics[width=0.49\textwidth, clip=true]{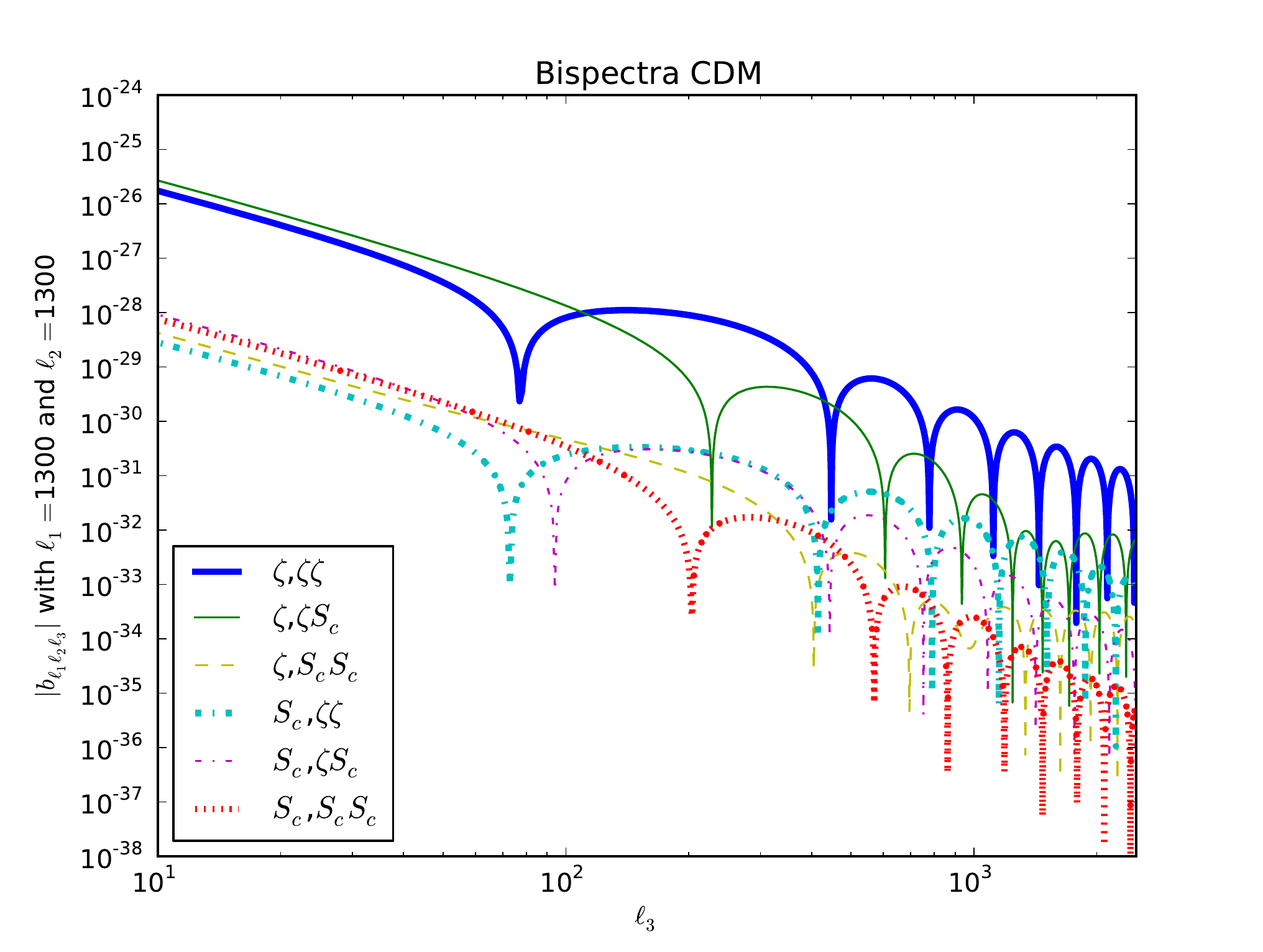}
\includegraphics[width=0.49\textwidth, clip=true]{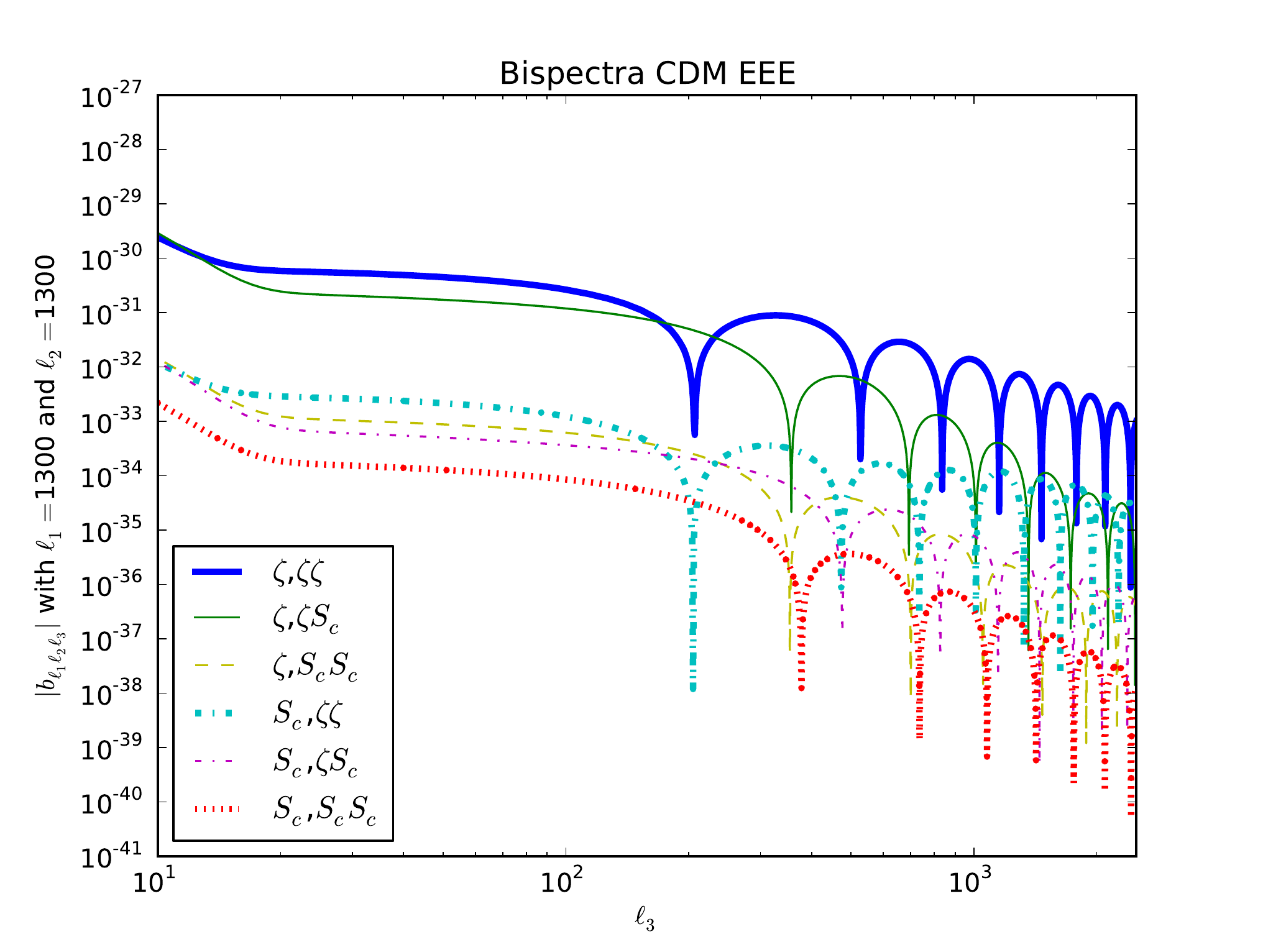}
\caption{Plot of $|b^{I,JK}_{\l_1\l_2\l_3}|$ in the CDM isocurvature case, as functions of $\l_3$, for $\l_1=\l_2=1300$. The figure on the left shows the temperature-only TTT bispectra, the one on the right the pure polarization EEE bispectra.}
\label{fig_bispectra_cdm}
\end{figure}

\begin{figure}
\centering
\includegraphics[width=0.49\textwidth, clip=true]{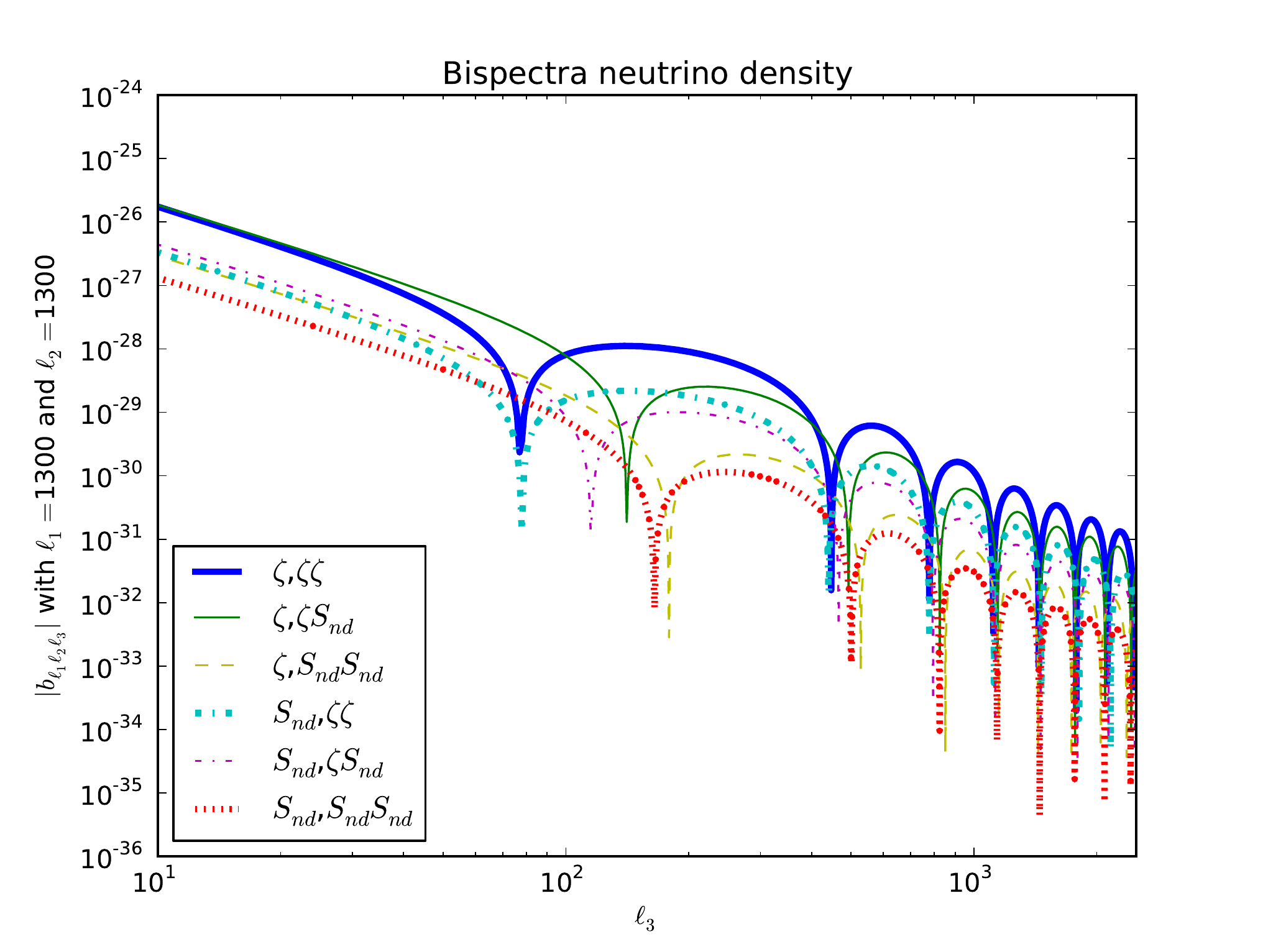}
\includegraphics[width=0.49\textwidth, clip=true]{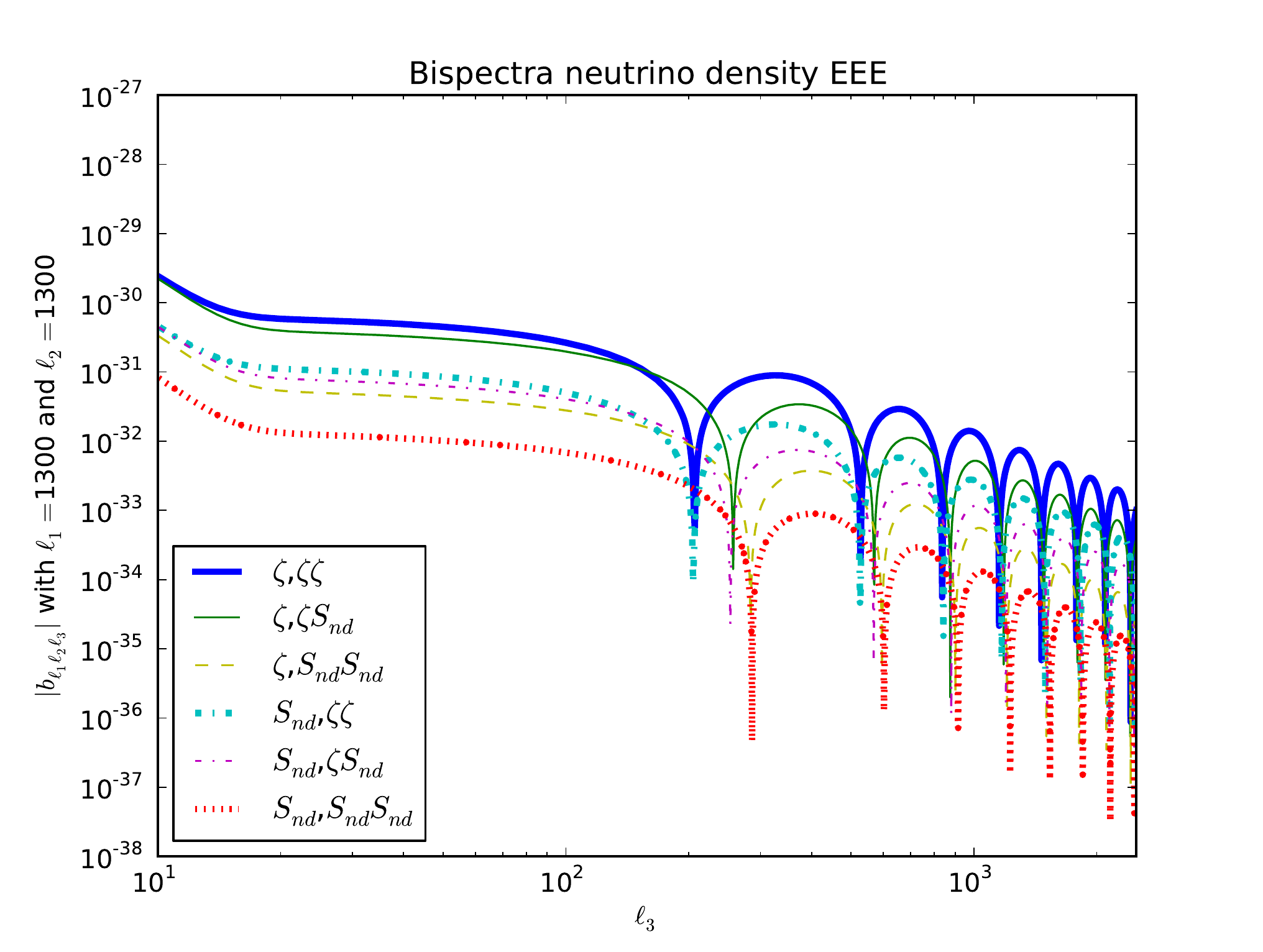}
\caption{Plot of $|b^{I,JK}_{\l_1\l_2\l_3}|$ in the neutrino density isocurvature case, as functions of $\l_3$, for $\l_1=\l_2=1300$. The figure on the left shows the temperature-only TTT bispectra, the one on the right the pure polarization EEE bispectra.}
\label{fig_bispectra_nd}
\end{figure}

\begin{figure}
\centering
\includegraphics[width=0.49\textwidth, clip=true]{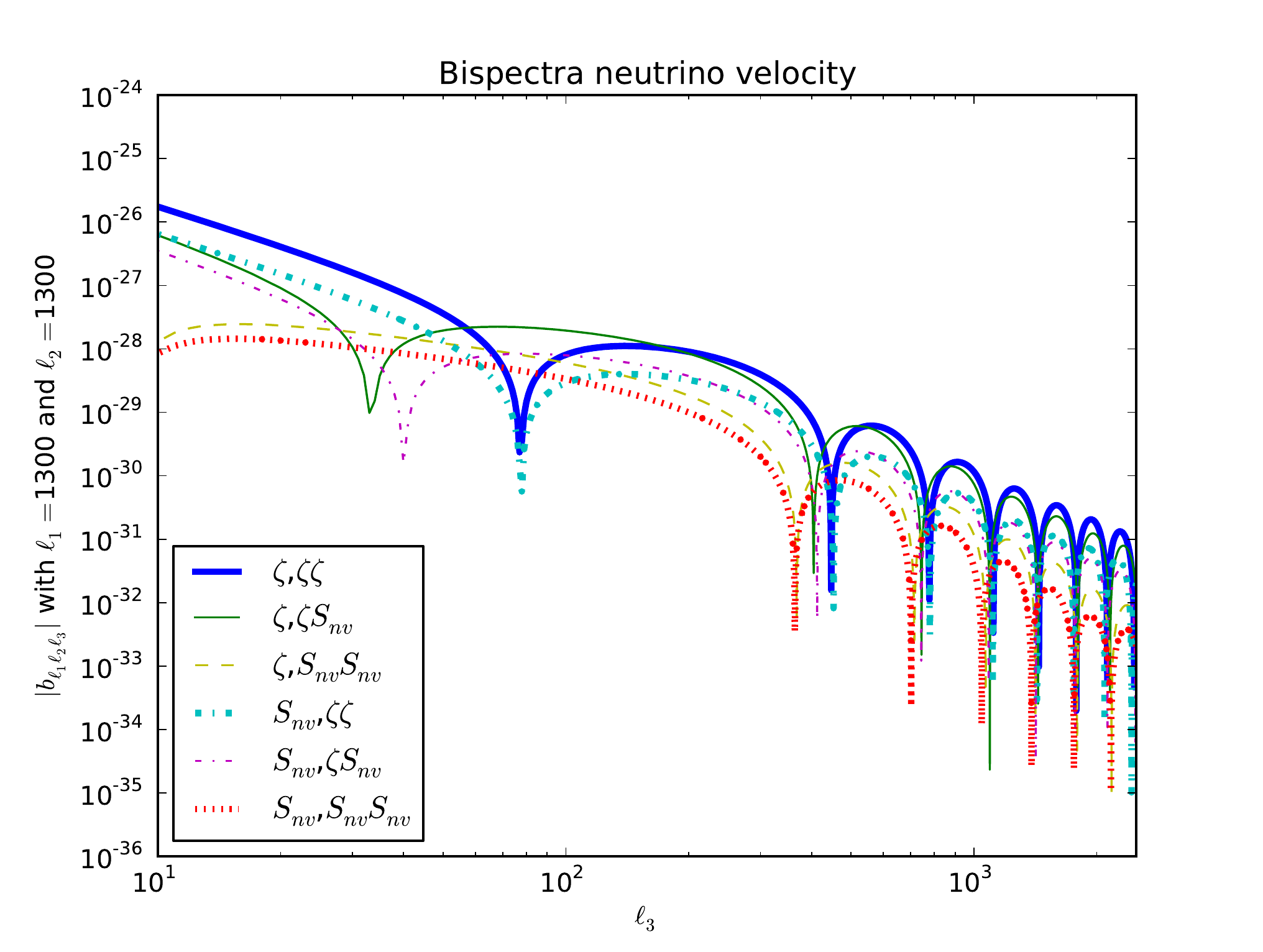}
\includegraphics[width=0.49\textwidth, clip=true]{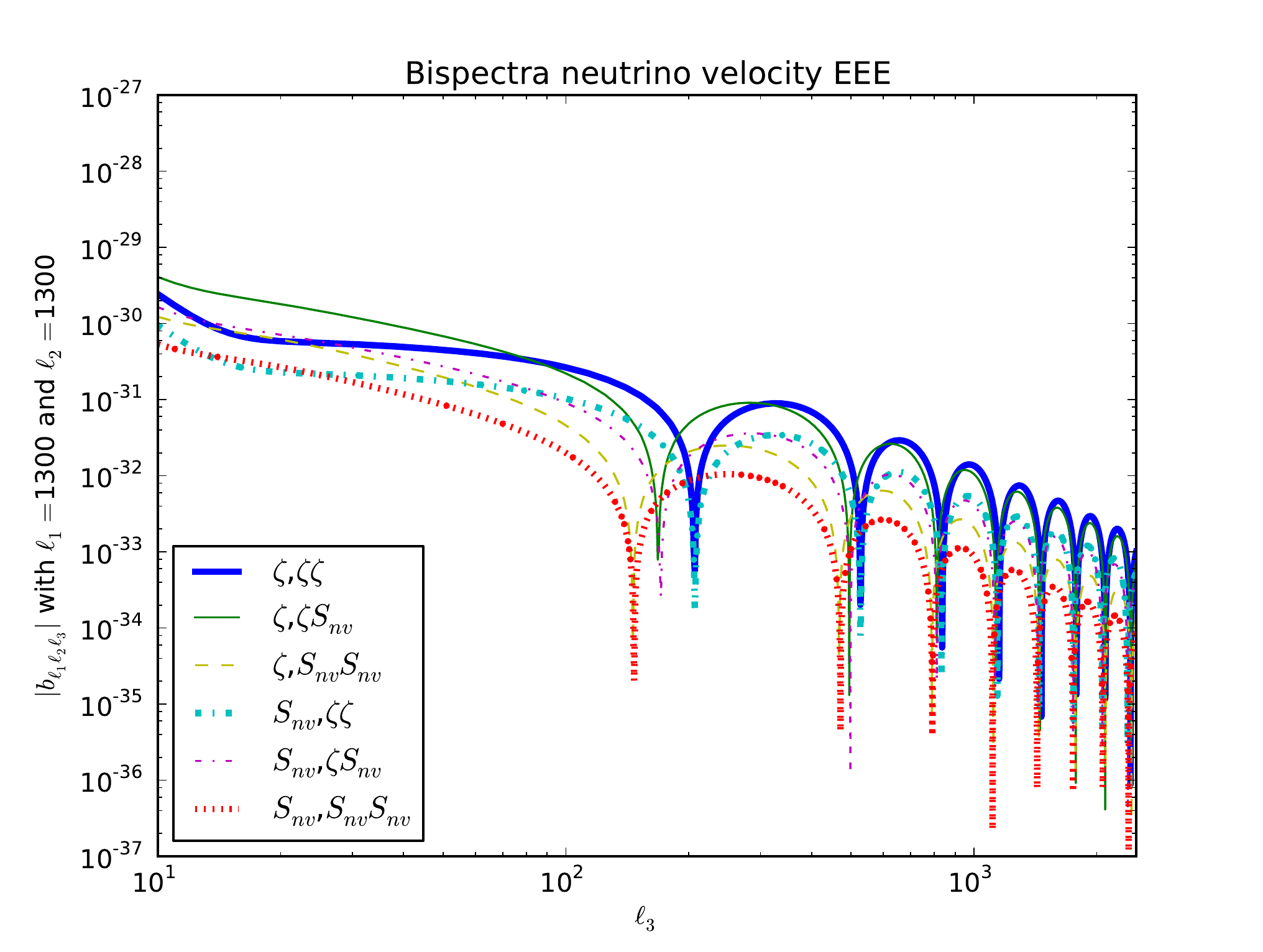}
\caption{Plot of $|b^{I,JK}_{\l_1\l_2\l_3}|$ in the neutrino velocity isocurvature case, as functions of $\l_3$, for $\l_1=\l_2=1300$. The figure on the left shows the temperature-only TTT bispectra, the one on the right the pure polarization EEE bispectra.}
\label{fig_bispectra_nv}
\end{figure}

In summary, after integration over $r$ of these various combinations of $\alpha$ and $\beta$ functions,  we  obtain six independent bispectra, for each type of isocurvature mode. To illustrate the typical angular dependence of these bispectra, we have plotted them as  functions of $l_3$, for fixed values of $l_1$ and $l_2$, respectively in the CDM isocurvature case 
(Fig.~\ref{fig_bispectra_cdm}), in the neutrino density isocurvature case (Fig.~\ref{fig_bispectra_nd}) and in the neutrino velocity case (Fig.~\ref{fig_bispectra_nv}). 
We plot only the pure temperature (TTT) and pure polarization (EEE) bispectra,
but of course one also has all the polarization cross bispectra.
As mentioned before, the $(\zeta, \zeta S)$ curve corresponds to twice $b^{\zeta, \zeta S}_{l_1l_2 l_3}$ since we consider the sum of
$b^{\zeta, \zeta S}_{l_1l_2 l_3}$ and $b^{\zeta, S \zeta}_{l_1l_2 l_3}$,  which cannot be distinguished. The same applies to the $(S, \zeta S)$ curve.
As usual, the bispectra for the baryon isocurvature mode are deduced from the CDM bispectra by the appropriate rescalings. 

\section{Observational prospects}
As we will see in the next section, one can envisage early Universe scenarios that generate significant isocurvature non-Gaussianity, which could dominate the purely adiabatic component, even if the adiabatic mode is dominant in the power spectrum as required by observations. 
 This is why it is important to assess how precisely one can hope to measure and to discriminate the various isocurvature bispectra in the future.
  
  The most general analysis would  require to consider simultaneously all five possible modes, which corresponds to a total of $n^2(n+1)/2=75$ coefficients, taking into account the symmetry 
  (\ref{f_sym}).
  In order to simplify our analysis, we will consider separately the various isocurvature perturbations. In other words, we will assume that the primordial perturbation is the superposition of a dominant  adiabatic mode and of a {\it single} isocurvature mode. In this case, the total bispectrum is characterized by six parameters, which we now denote  $\tf^{(i)}$,  
  \begin{eqnarray}
b_{l_1l_2 l_3}&=&\tf^{\zeta,\zeta\zeta}\,b_{l_1l_2 l_3}^{\zeta,\zeta\zeta}+2\tf^{\zeta,\zeta S}\, b_{l_1l_2 l_3}^{\zeta,\zeta S}+
 \tf^{\zeta,SS}\, b_{l_1l_2 l_3}^{\zeta,SS}+\tf^{S,\zeta\zeta}\, b_{l_1l_2 l_3}^{S,\zeta\zeta}+2\tf^{S,\zeta S}\, b_{l_1l_2 l_3}^{S,\zeta S}+\tf^{S,SS}\, b_{l_1l_2 l_3}^{S,SS}
 \nonumber
 \\
 &=& \sum_{(i)}\tf^{(i)} b_{l_1l_2 l_3}^{(i)}\,,
\end{eqnarray}
where the index $i$ varies between $1$ to $6$, following the order indicated in the upper line. Note that, because of  the factor $2$ in front of  $\tf^{\zeta,\zeta S}$ and 
$\tf^{S,\zeta S}$, we define $b_{l_1l_2 l_3}^{(2)}\equiv 2b_{l_1l_2 l_3}^{\zeta,\zeta  S}$ and $b_{l_1l_2 l_3}^{(5)}\equiv2  b_{l_1l_2 l_3}^{S,\zeta S}$ whereas there is no such factor $2$ for  the other terms.

 \subsection{The Fisher matrix} 
 
 To estimate these six parameters, given some data set, the usual 
procedure is to minimize 
\beq
\chi^2=\left\langle (B^{obs}-\sum_i  \tf^{(i)} B^{(i)}), 
(B^{obs}-\sum_i  \tf^{(i)} B^{(i)})\right\rangle.
\eeq
For an ideal experiment (no noise and no effects due to the beam size) without
polarization, the scalar product is defined  by
\beq
\langle B, B' \rangle\equiv \sum_{l_1 \leq l_2 \leq l_3}
\frac{B_{l_1l_2l_3}B'_{l_1l_2l_3}}{\sigma^2_{l_1l_2l_3}}.
\eeq
The bispectrum variance in that case is given by
\beq
\sigma^2_{l_1l_2l_3}\equiv \langle B^2_{l_1l_2l_3}\rangle
-\langle B_{l_1l_2l_3} \rangle^2\approx
\Delta_{l_1l_2l_3} C_{l_1}C_{l_2} C_{l_3}
\eeq
in the approximation of weak non-Gaussianity,
where  
\beq
\Delta_{l_1l_2l_3} \equiv  1+\delta_{l_1l_2}+\delta_{l_2l_3}+\delta_{l_3l_1}+2\, \delta_{l_1l_2}\delta_{l_2l_3}\,.
\eeq 
The best estimates for the parameters are thus obtained by solving
\beq
\sum_j \langle B^{(i)}, B^{(j)}\rangle \tf^{(j)}
=\langle B^{(i)}, B^{obs}\rangle\, ,
\eeq
while the statistical error on the parameters is deduced from the second-order 
derivatives of $\chi^2$,   which define the Fisher matrix, given in our case by
\beq
F_{ij}\equiv \langle B^{(i)}, B^{(j)}\rangle.
\eeq 
The Fisher matrix is a symmetric matrix, which can be determined by computing
the 21 different scalar products between the six elementary bispectra.

For a real experiment, and if E-polarization is included as well, the above
equations remain valid, except that the definition of the scalar product
has to be replaced by a more complicated expression (see e.g.\ 
\cite{Yadav:2007rk}):
\beq
\langle B^{(i)}, B^{(j)} \rangle \equiv 
\sum_{\alpha\beta\gamma\alpha'\beta'\gamma'}\ 
\sum_{l_1 \leq l_2 \leq l_3} \frac{1}{\Delta_{l_1 l_2 l_3}}
B_{l_1 l_2 l_3}^{(i) \, \alpha\beta\gamma}
({\cal C}_{l_1}^{-1})^{\alpha\alpha'}
({\cal C}_{l_2}^{-1})^{\beta\beta'}
({\cal C}_{l_3}^{-1})^{\gamma\gamma'}
B_{l_1 l_2 l_3}^{(j) \, \alpha'\beta'\gamma'},
\eeq
where $\alpha, \beta, \gamma, \alpha', \beta', \gamma'$ are 
polarization
indices taking the two values $T$ and $E$.
The covariance matrix ${\cal C}_l$ (a matrix in polarization space) is given by
\beq
{\cal C}_l = \left(\begin{array}{cc}
b_l^2 C_l^{TT} + N_l^T & b_l^2 C_l^{TE}\\
b_l^2 C_l^{TE} & b_l^2 C_l^{EE} + N_l^E
\end{array} \right),
\eeq
where $b_l$ is the beam function and $N_l$ the noise power spectrum.
We assumed the same beam function for temperature and polarization detectors,
as well as no correlated noise, but the generalization  is straightforward. In the calculation of the covariance
matrix we only take the adiabatic power spectrum, since from observations
we know that the isocurvature contribution to the power spectrum must be
very small.

For each type of isocurvature mode, we have computed the corresponding Fisher
matrix by extending the numerical code
described in \cite{BucherVanTent} to include isocurvature modes and
E-polarization, according to the expressions presented above. 
We have taken into account the noise
characteristics of the Planck satellite~\cite{Bluebook}, using only
the 100, 143, and 217 GHz channels, combined in quadrature. Our
computation goes up to $l_\mathrm{max} = 2500$ and uses the WMAP-only
7-year best-fit cosmological parameters~\cite{Komatsu:2010fb}.

From the Fisher matrix, one can compute the statistical uncertainty on each of the parameters:
\beq
\label{errors}
\Delta \tf^i=\sqrt{(F^{-1})_{ii}}\,.
\eeq
This takes into account the correlations between the various bispectra. By contrast, if one assumes that the data contain only a single elementary bispectrum, for example the purely adiabatic one, then the corresponding statistical error is 
\beq
\Delta \tf^i=\frac{1}{\sqrt{F_{ii}}} \qquad ({\rm single\    parameter})\,.
\eeq
One can also determine the correlations between any two bispectra:
\beq
\label{correlation_matrix}
{\cal C}_{ij}=\frac{(F^{-1})_{ij}}{\sqrt{(F^{-1})_{ii}(F^{-1})_{jj}}}\,.
\eeq

\subsection{CDM isocurvature mode}
\def\text{}
\begin{table}
\footnotesize
\begin{center}
\begin{tabular}{|cccccc|}
\hline
$(\zeta, \zeta\zeta)$ & $(\zeta, \zeta S)$ & $(\zeta, S S)$ &  $(S,\zeta\zeta)$  & $(S,\zeta S)$ &  $(S,SS)$\\
\hline
$3.9 \,(2.5) \times 10^{\text{-2}}$ & $4.5 \,(3.6) \times 10^{\text{-2}}$ & $2.3 \,(2.1) \times 10^{\text{-4}}$ &
   $2.4\,(1.6) \times 10^{\text{-4}}$ & $6.9 \,(4.3)\times 10^{\text{-4}}$ & $5.3 \,(3.1) \times 10^{\text{-4}}$
\\ 
- & $7.1 \,(6.0) \times 10^{\text{-2}}$ & $5.3 \,(3.8)\times 10^{\text{-4}}$ &
   $3.8\,(2.1)\times 10^{\text{-4}}$ & $11 \,(7.4)\times 10^{\text{-4}}$ & $8.8 \,(5.5)\times 10^{\text{-4}}$
\\ 
- & - & $28 \,(6.4) \times 10^{\text{-5}}$ &
   $16 \,(3.7)\times 10^{\text{-5}}$ & $33 \,(9.5)\times 10^{\text{-5}}$ & $11 \,(5.0)\times 10^{\text{-5}}$
\\ - & - & - & $15 \,(3.0)\times 10^{\text{-5}}$ & $22 \,(5.8)\times 10^{\text{-5}}$ & $7.5 \,(3.2)\times 10^{\text{-5}}$
\\- & - & - & - & $5.1 \,(1.6)\times 10^{\text{-4}}$ & $2.4 \,(1.0)\times 10^{\text{-4}}$
\\- & - & - & - & - & $21 \,(8.3)\times 10^{\text{-5}}$
\\ \hline
\end{tabular}
\caption{Fisher matrix for the CDM isocurvature mode. Only the upper half coefficients are indicated,  since the matrix is symmetric. The value between parentheses corresponds to the Fisher matrix components obtained  without including the polarization.}
\label{table_cdm}
\end{center}
\end{table}

Our results for this mode have already been presented elsewhere \cite{LvT}, but we discuss here  in more detail the peculiarities of the corresponding Fisher matrix, which is given in Table~\ref{table_cdm}.
One can immediately notice the intriguing fact that the coefficients of the upper left \mbox{$2\times 2$} submatrix, corresponding to the purely adiabatic component and the correlated $(\zeta, \zeta S)$ component, are typically  two orders of magnitude larger than all the other coefficients. 
The correlation matrix,  defined in (\ref{correlation_matrix}) and given in Table \ref{correlation_cdm} shows that  the first two bispectra are strongly (anti-)correlated while their correlation with the four other bispectra is weak.

\begin{table}
\begin{center}
\begin{tabular}{|cccccc|}
\hline
$(\zeta, \zeta\zeta)$ & $(\zeta, \zeta S)$ & $(\zeta, S S)$ &  $(S,\zeta\zeta)$  & $(S,\zeta S)$ &  $(S,SS)$\\
\hline

 $1.$ & $-0.84 \,(0.92)$ & $0.18 \,(0.21)$ & $0.003 \,(-0.12)$& $-0.15 \,(0.17)$ & $0.13 \,(0.18)$ \\
 - & $1.$ & $-0.15 \,(0.18)$ & $-0.007 \,(+0.13)$ & $0.13 \,(0.14)$& $-0.16 \,(0.16)$ \\
 - & - & 1. & $-0.069 \,(+0.24)$ & $-0.80 \,(0.97)$ & $0.58 \,(0.92)$ \\
 - & - & - & 1. & $-0.42 \,(0.43)$& $0.29 \,(0.39)$ \\
 - & - & - & - & $1.$ & $-0.82\,(0.98)$ \\
 - & - & - & - & - & 1.
\\ \hline
\end{tabular}
\caption{Correlation matrix for the CDM isocurvature mode. Only the upper half coefficients are indicated,  since the matrix is symmetric. The value between parentheses corresponds to the correlations obtained  without including the polarization (the absence of  sign in the parentheses means that it is unchanged with respect to the value taking into account the polarization).}
\label{correlation_cdm}
\end{center}
\end{table}

From this Fisher matrix, one finds that the  $68$ \% error on the parameters $\tf^i$ is given 
by\footnote{The tiny differences in the 3rd, 5th, and 6th value compared to 
\cite{LvT} are due to small improvements in the computer code.}
\beq
\Delta \tf^i=\sqrt{(F^{-1})_{ii}}=\{9.6, 7.1, 160, 150, 180, 140\}\,.
\label{errors_CDMpol}
\eeq
For ease of readability, we have written $160$ instead of $1.6\cdot 10^2$, etc.,
but we are not claiming more than two digits of significance.
We also remind the reader that in the purely adiabatic case, our 
$\tf^1$, i.e.\ the $(\zeta,\zeta\zeta)$ component, is $6/5$ times the
standard $f_\mathrm{NL}$. 
One sees that the first two uncertainties are typically one order of magnitude  smaller than  the last four. 

It is also interesting to estimate how much the inclusion of the polarization data in the analysis improves the precision of the non-linear parameters. The components of the Fisher matrix when the polarization is not taken into account can be read between the parentheses in Table~\ref{table_cdm}. One notices that whereas the coefficients of the first two lines are reduced by a factor inferior to two, the other coefficients are significantly suppressed when one removes the polarization data. As a consequence, one finds that  
the uncertainties on the parameters without polarization, given by
\beq
\Delta \tf^i=\sqrt{(F^{-1})_{ii}}=\{17, 11, 980, 390, 1060, 700\}\qquad {\rm (no \ polarization)},
\eeq
increase by less than a factor two for the first two parameters, whereas the 
increase is much bigger for the four other ones.

\begin{figure}
\centering
\includegraphics[width=0.6\textwidth, clip=true]{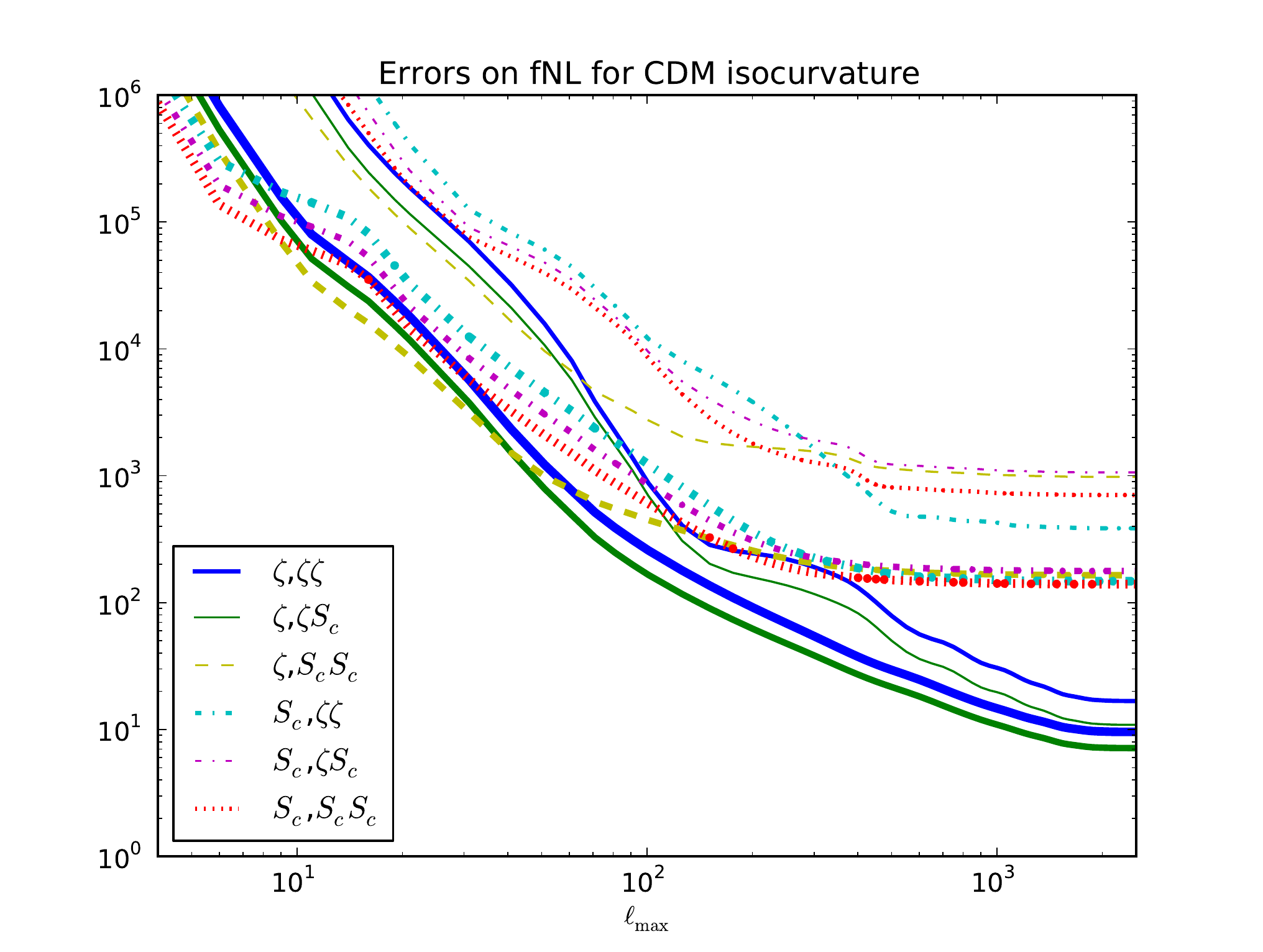}
\caption{Evolution of the $f_{\rm NL}$ parameter uncertainties as one increases 
the cut-off $\l_{\rm max}$, for the CDM isocurvature mode. The six thinner 
curves describe the situation if only temperature data is used, while for the 
six thicker curves both temperature and E-polarization data are included.}
\label{V}
\end{figure}

The evolution of these 
uncertainties as a function of the cut-off $l_\mathrm{max}$ is shown
in Fig.~\ref{V}, both for the case where temperature and E-polarization data 
are used and for the case where only temperature data is included.
One can see that the curves for temperature-only typically look 
bumpier than the curves that include polarization as well.
This can be explained as follows. First,  unlike the power spectrum, the bispectrum is an 
alternating function, so that for certain regions in $l_1 l_2 l_3$ space it is 
zero or close to zero, and the  contribution  to the determination of  $f_\mathrm{NL}$, which is quadratic in the bispectrum, is then negligible in these regions.  Second, as one can
see for example in Fig.~\ref{fig_bispectra_cdm}, the acoustic peaks of the
polarization bispectrum are out of phase with the ones of the temperature
one, so that including polarization neatly fills in the holes 
in $l_1 l_2 l_3$ space and leads to a smoother determination of $f_\mathrm{NL}$,
as first pointed out by \cite{Komatsu:2003iq}.

\begin{figure}
\centering
\includegraphics[width=0.49\textwidth, clip=true]{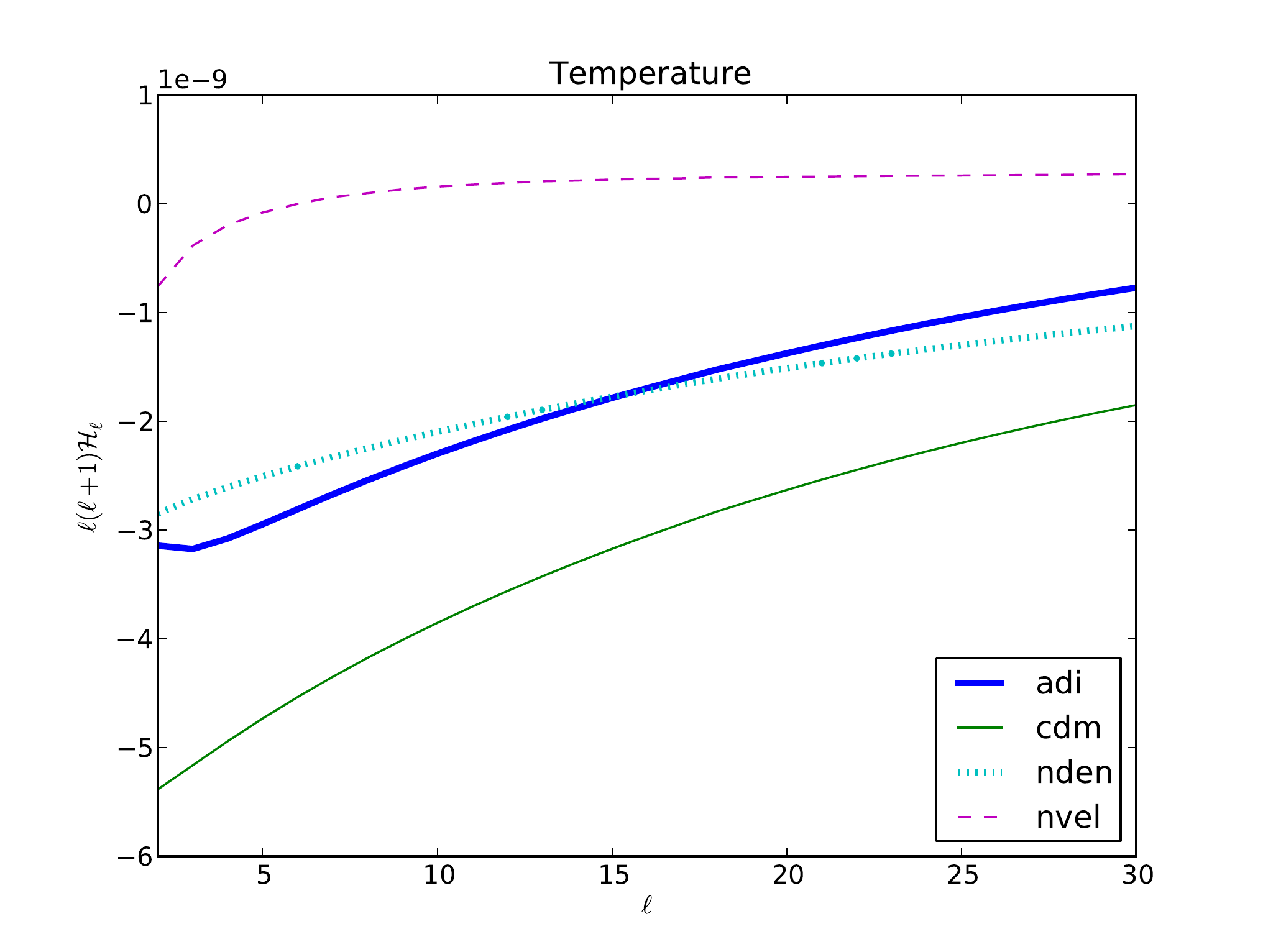}
\includegraphics[width=0.49\textwidth, clip=true]{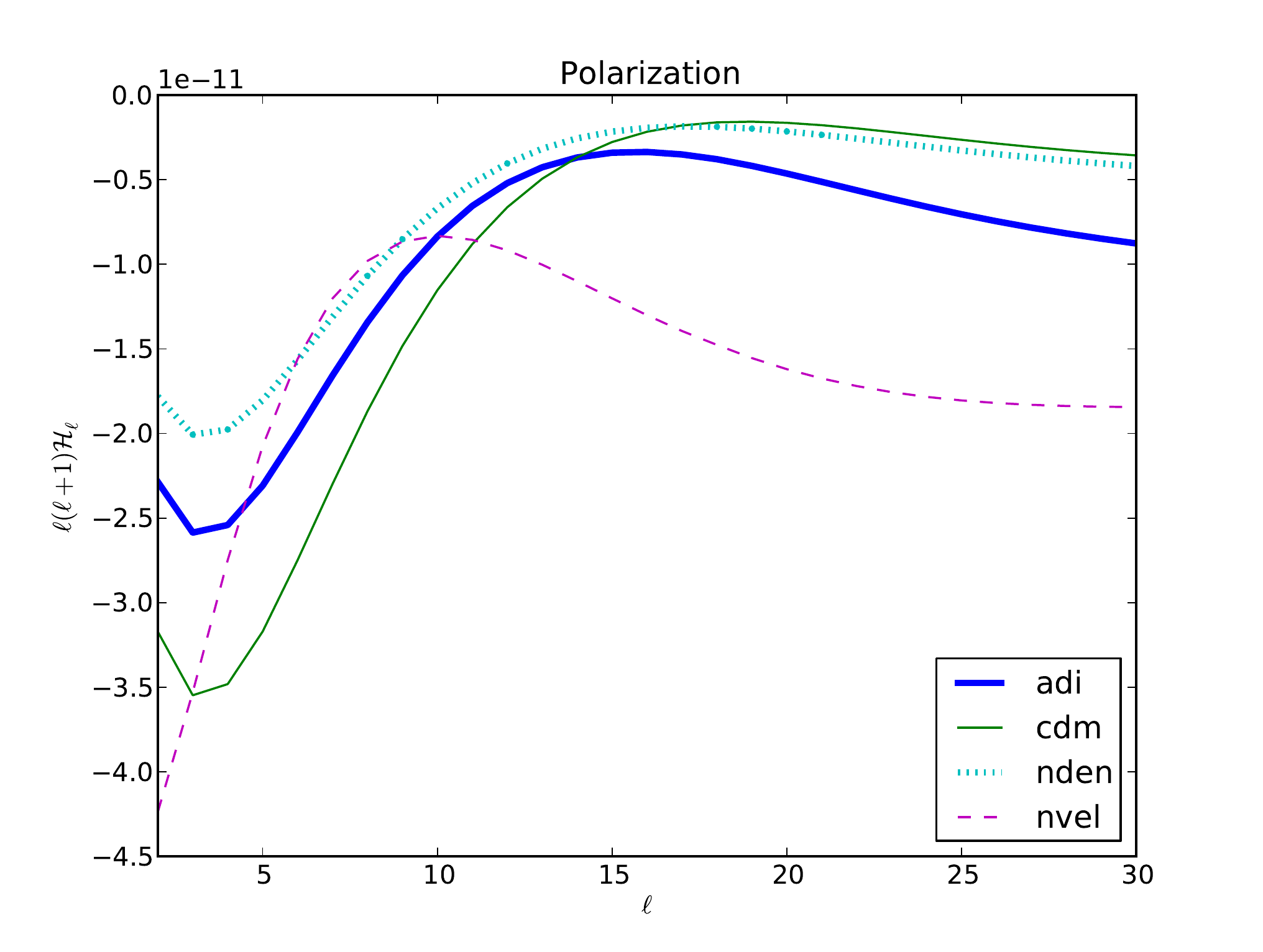}
\caption{The function $l(l+1){\cal H}_K$ plotted as a function of $l \leq 30$ 
for different values of $K$. The figure on the left
shows temperature (T), the one on the right polarization (E).}
\label{HK_fig}
\end{figure}

\begin{figure}
\centering
\includegraphics[width=0.49\textwidth, clip=true]{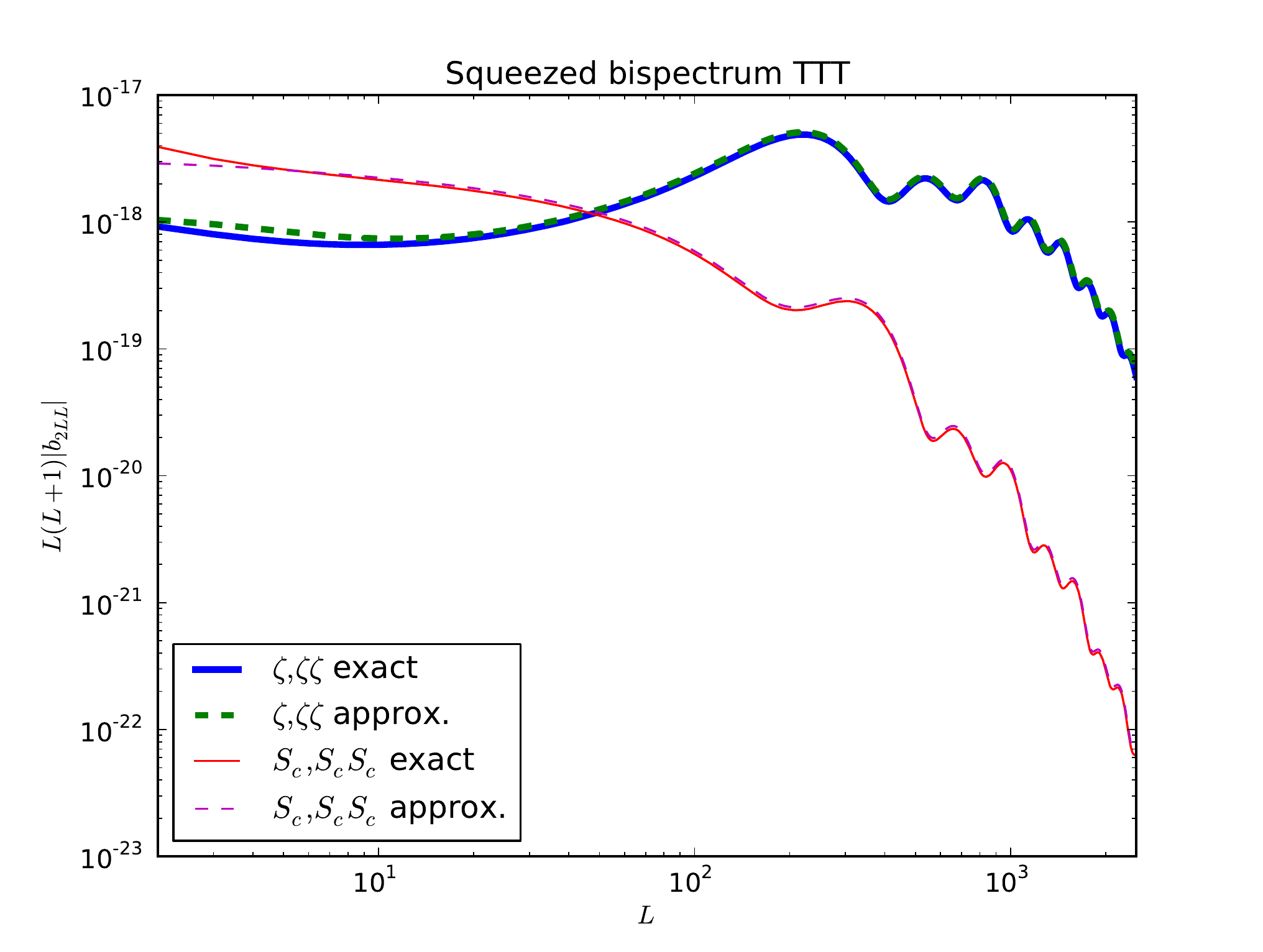}
\includegraphics[width=0.49\textwidth, clip=true]{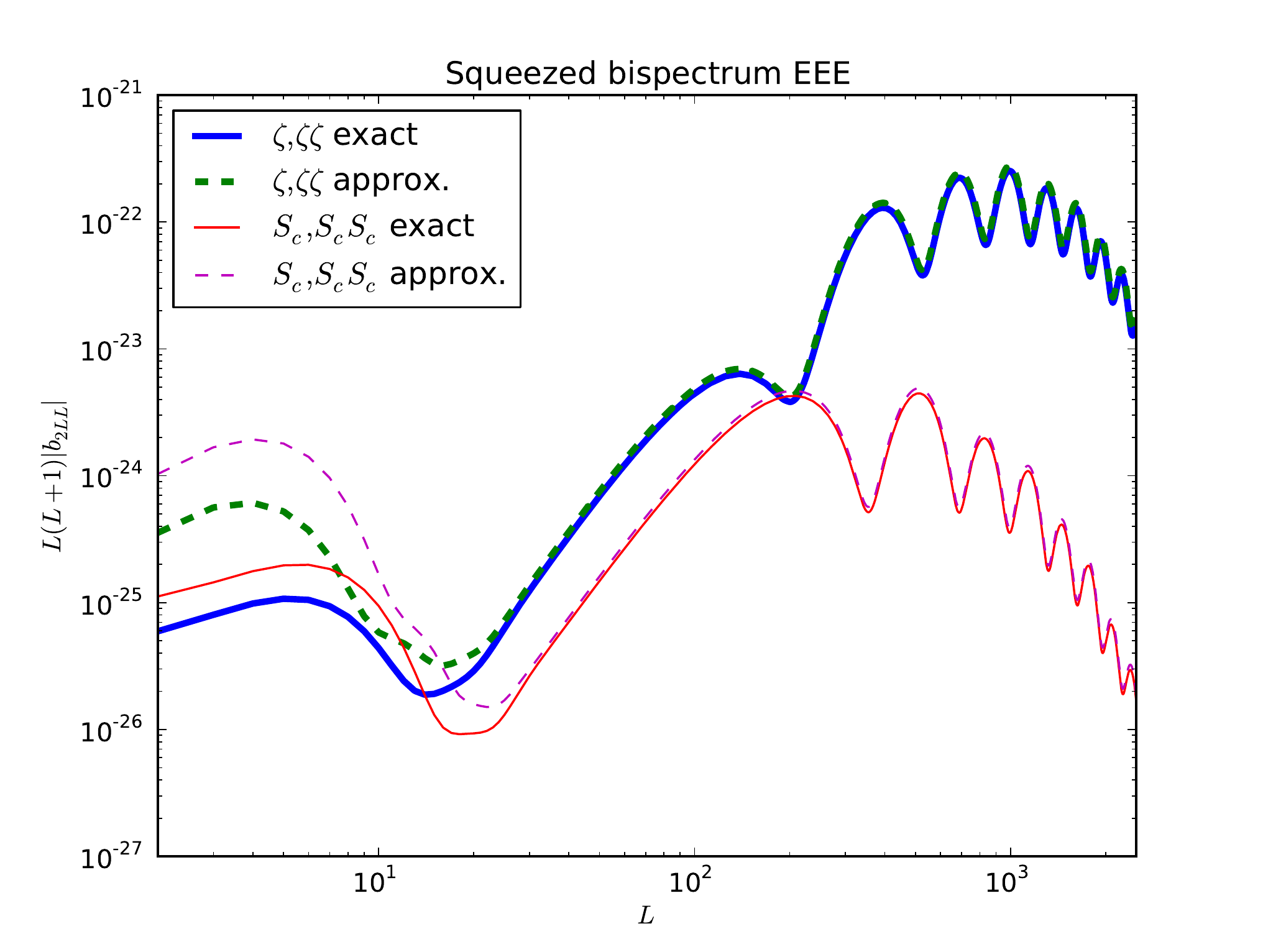}
\caption{The absolute value of the squeezed bispectrum (multiplied by
  $L(L+1)$) with $l_1=2$ and $l_2=l_3\equiv L$ plotted as a function
  of $L$. Both the exact bispectrum and (twice) the approximation
  given in (\ref{G_and_H}) are plotted, for the purely adiabatic mode
  and the purely CDM isocurvature mode. The figure on the left shows
  the temperature-only TTT bispectra, the one on the right the pure
  polarization EEE bispectra. Note that the configuration is only
  squeezed in the case of large $L$; for small $L$ the approximation
  should not be used.}
\label{bispec_approx_fig}
\end{figure}

Our results can be understood by the following analysis in the squeezed limit
(based on \cite{BucherVanTent}), assuming that 
$l_1\equiv \ell\ll l_2\approx l_3\equiv L$. In this limit, 
the bispectra defined in (\ref{b_IJK}) can be decomposed as
\beq
b_{\ell L L}^{I,JK}=    \int_0^\infty r^2 dr \left[\alpha^I_{\ell}(r)\beta^{J}_{L}(r)\beta^{K}_{L}(r)+\alpha^I_{L}(r)\beta^{J}_{\ell}(r)\beta^{K}_{L}(r)+\alpha^I_{L}(r)\beta^{J}_{L}(r)\beta^{K}_{\ell}(r)\right].
\label{squeezed_bispec}
\eeq
The first term is subdominant, since, like the power spectrum, $|\beta_l|$ 
decreases as $1/(l(l+1))$ (or even faster as $1/(l^2(l+1)^2)$ for large $l$), 
as can be seen for example in Fig.~\ref{beta_cdm}. 
The last term, for instance, is explicitly given by
\beq
 \left(\frac2\pi\right)^3 \int k_1^2 dk_1  \, k_2^2 dk_2 \, k_3^2 dk_3 \, g^I_{L}(k_1)g^J_{L}(k_2)g^K_{\l}(k_3)\,  P_\zeta(k_2)P_\zeta(k_3)
\int_0^\infty r^2 dr   j_{L}(k_1r)  j_{L}(k_2r)   j_{\l}(k_3 r) \,,
\eeq
where the last Bessel function  oscillates slowly while the first  two oscillate very rapidly. This leads to a cancellation of the radial integral unless $k_1$ is very close to $k_2$. 
We find that the above expression can thus be approximated by
\bea
&\left(\frac2\pi\right)^3 & \int k_1^2 dk_1 \int k_2^2 dk_2 \int k_3^2 dk_3 \, g^I_{L}(k_1)g^J_{L}(k_2)g^K_{\ell}(k_3) P_\zeta(k_2)P_\zeta(k_3) \frac{5\delta(k_1-k_2)}{k_1 k_2}\int_0^\infty \frac{dr}{r} j_{\ell}(k_3r) 
\cr
&\approx&  \left\{\left(\frac2\pi\right)^2\int k^2 dk \, g^I_{L}(k)g^J_{L}(k) P_\zeta(k) \right\} \left\{ \frac{10}{\pi}\int k_3^2 dk_3 \, g^K_{\ell}(k_3) P_\zeta(k_3) \int_0^\infty \frac{dr}{r} j_{\ell}(k_3r) \right\}
\cr
&\equiv& {\cal G}_{IJ}(L)\, {\cal H}_K(\ell) .
\label{G_and_H}
\eea
The $\delta(k_1-k_2)$ is explained above, but together with the $1/(k_1 k_2)$
also motivated by the closure relation for spherical Bessel functions,
$\int_0^\infty r^2 j_L(k_1 r) j_L (k_2 r) dr = \pi \delta(k_1-k_2)/(2k_1k_2)$.
The $1/r$ follows from a dimensional analysis, and the $5$ has been 
determined heuristically by comparing with the exact bispectrum: the ratio
is approximately $5$ and only weakly dependent on the small $\ell$.
The full squeezed bispectrum (\ref{squeezed_bispec}) is thus approximated by 
\beq
 b_{\ell L L}^{I,JK}\approx {\cal G}_{IJ}(L)\, {\cal H}_K(\ell)+{\cal G}_{IK}(L)\, {\cal H}_J(\ell)\,.
 \eeq
Assuming all primodial power spectra to be equal, the functions ${\cal G}_{IJ}$
are the angular (cross) power spectra plotted in Fig.~\ref{spectra_fig} and
\ref{cross_spectra_fig}. 
The function ${\cal H}_K(\ell)$ is simply an integral over $\beta_\ell^K(r)$:
\beq
{\cal H}_K(\ell) = 5 \int_0^\infty \frac{dr}{r} \, \beta_\ell^K(r)
\eeq
and is shown for small $\ell$ in Fig.~\ref{HK_fig}. The good agreement
of our approximation with the exact squeezed bispectrum is shown in
Fig.~\ref{bispec_approx_fig}.


In the squeezed limit, one thus finds that only the first two elementary bispectra, $(\zeta, \zeta\zeta)$ and $(\zeta, \zeta S)$, depend on ${\cal G}_{\zeta\zeta}(L)$. The four others depend on ${\cal G}_{\zeta S}(L)$ and/or ${\cal G}_{SS}(L)$.
The large $L$ limit of ${\cal G}_{\zeta S}(L)$ and ${\cal G}_{SS}(L)$ are strongly suppressed with respect to ${\cal G}_{\zeta\zeta}(L)$, which explains why the uncertainty on the first two non-Gaussianity parameters can be reduced by probing high multipoles (the bispectrum there is still sufficiently large compared to the noise) 
while the uncertainty on the four other ones saturates as shown in Fig.~\ref{V}.
One can even understand why the curve for the $(\zeta,\zeta S)$ mode is below
the one for $(\zeta,\zeta\zeta)$: their dominant terms both depend on the same 
${\cal G}_{\zeta\zeta}(L)$, but different ${\cal H}(\ell)$, and 
$|{\cal H}_S(\ell)| > |{\cal H}_\zeta(\ell)|$.
Finally, one can understand why including polarization helps much more for
the uncertainty on e.g.\ the $(S,SS)$ mode than for the $(\zeta,\zeta\zeta)$
mode. As one can see from Fig.~\ref{V}, it is in particular in the region
$50 \leq l \leq 200$ that the distance between the two $(S,SS)$ curves increases
compared to the distance between the two $(\zeta,\zeta\zeta)$ curves. A quick
look at Fig.~\ref{spectra_fig} shows that in that region of multipole space
the TT CDM isocurvature power spectrum (i.e.\ ${\cal G}_{SS}(L)$) becomes
very small compared to the TT adiabatic spectrum (i.e.\ 
${\cal G}_{\zeta\zeta}(L)$), but the EE CDM isocurvature spectrum still remains
comparable to the EE adiabatic one.

It is also instructive to  compare (\ref{errors_CDMpol}) with the uncertainties
\beq
\Delta\tf^i= 1/\sqrt{F_{ii}}=\{5.0, 3.7, 60, 82, 44, 69\} \qquad {\mathrm{(single \ parameter)}}
\eeq
  obtained by ignoring the correlations, or, equivalently, by assuming that only one parameter  is nonzero. In particular, the contamination of the purely adiabatic signal by the other shapes induces an increase of the uncertainty, but only by a factor 2, which is rather moderate. 
  
Assuming that the adiabatic and isocurvature modes are {\it uncorrelated} implies that only the purely adiabatic and isocurvature bispectra are relevant. The corresponding  
two-parameter
Fisher matrix, which is the submatrix of $F_{(ij)}$ with entries  $F_{11}$, $F_{66}$ and $F_{16}$,   leads to uncertainties on  $\tf^{(1)}$ and $\tf^{(6)}$ that are almost identical to the corresponding single-parameter errors. 

Finally, let us note that if the observed bispectrum is mainly purely isocurvature with amplitude $\tf^{(6)}$, a naive analysis using 
only the purely adiabatic estimator would lead to an apparent adiabatic coefficient 
\beq
\tf^{(1)}=  \frac{F_{16}}{F_{11}}\, \tf^{(6)}\simeq 10^{-2} \, \tf^{(6)},
\eeq
thus hiding the isocurvature signal with larger amplitude.

\subsection{Baryon isocurvature mode}
The Fisher matrix for the baryon isocurvature mode can be easily deduced from the CDM Fisher matrix. Indeed, as discussed earlier, the CDM and baryon isocurvature transfer functions are identical up to a rescaling by $\omega_{bc}$ introduced in (\ref{omega_bc}).
Consequently, the $\alpha$ and $\beta$ functions are simply rescaled:
\beq
\alpha^{S_b}_{l}(r)=\omega_{bc}\, \alpha^{S_c}_{l}(r),\qquad \beta^{S_b}_{l}(r)=\omega_{bc}\, \beta^{S_c}_{l}(r)\,.
\eeq
The rescaling of the various bispectra will thus depend on the number of $S$ indices, i.e.
\beq
b^{I,JK}_{l_1l_2 l_3}(S_b)=(\omega_{bc})^p \, b^{I,JK}_{l_1l_2 l_3}(S_c),
\eeq
where 
the power $p$ is the number of $S$ among the indices $\{IJK\}$. In summary, 
all coefficients of the baryon isocurvature Fisher matrix can be deduced from 
Table \ref{table_cdm} by using the rescaling 
\beq
F_{ij}^{S_b}=\N_i \, \N_j F_{ij}^{S_c} \quad ({\rm no \, summation}), \qquad \N_i=\{1, \omega_{bc}, \omega_{bc}^2,\omega_{bc},\omega_{bc}^2,\omega_{bc}^3\},
\eeq
where, in our computation, $\omega_{bc}=0.2036$.

The parameter uncertainties can also be deduced from the CDM results via the rescalings $\N_i$: $\Delta \tilde f^i(S_b)=\Delta \tilde f^i(S_c)/\N_i$. One thus obtains:
\beq
\Delta \tilde f^i=
\left\{9.6, 35, 4000, 720, 4300, 16600\right\}\,.
\eeq
Except for the purely adiabatic coefficient, we thus find that the uncertainties on all the other coefficients are significantly larger  than the uncertainties obtained in (\ref{errors_CDMpol}) in the  CDM  case, simply because the elementary bispectra have a smaller amplitude than their CDM counterparts. By contrast, the correlation matrix, which is independent of the normalization of the bispectra, is exactly the same as in the CDM  case. 

\subsection{Neutrino density isocurvature mode}
\def\text{}
\begin{table}
\footnotesize
\begin{center}
\begin{tabular}{|cccccc|}
\hline
$(\zeta, \zeta\zeta)$ & $(\zeta, \zeta S)$ & $(\zeta, S S)$ &  $(S,\zeta\zeta)$  & $(S,\zeta S)$ &  $(S,SS)$\\
\hline
$3.9 \,(2.5) \times 10^{\text{-2}}$ & $3.6 \,(2.6) \times 10^{\text{-2}}$ & $5.6 \,(4.1) \times 10^{\text{-3}}$ &
   $7.9\,(5.1) \times 10^{\text{-3}}$ & $8.8 \,(6.1)\times 10^{\text{-3}}$ & $2.2 \,(1.6) \times 10^{\text{-3}}$
\\ 
- & $3.8 \,(2.9) \times 10^{\text{-2}}$ & $6.3 \,(4.8)\times 10^{\text{-3}}$ &
   $7.4 \,(5.2)\times 10^{\text{-3}}$ & $9.2 \,(6.7)\times 10^{\text{-3}}$ & $2.5 \,(1.8)\times 10^{\text{-3}}$
\\ 
- & - & $11 \,(8.1) \times 10^{\text{-4}}$ &
   $12 \,(8.5)\times 10^{\text{-4}}$ & $1.6 \,(1.1)\times 10^{\text{-3}}$ & $4.4 \,(3.1)\times 10^{\text{-4}}$
\\ - & - & - & $1.8 \,(1.1)\times 10^{\text{-3}}$ & $2.0 \,(1.3)\times 10^{\text{-3}}$ & $5.0 \,(3.2)\times 10^{\text{-4}}$
\\- & - & - & - & $2.5 \,(1.6)\times 10^{\text{-3}}$ & $6.8 \,(4.4)\times 10^{\text{-4}}$
\\- & - & - & - & - & $2.1 \,(1.2)\times 10^{\text{-4}}$
\\ \hline
\end{tabular}
\caption{Fisher matrix for the neutrino density isocurvature mode. Only the upper half coefficients are indicated,  since the matrix is symmetric. The value between parentheses corresponds to the Fisher matrix components obtained  without including the polarization.}
\label{table_nd}
\end{center}
\end{table}
For a neutrino density isocurvature mode, we have obtained the Fisher matrix in Table~\ref{table_nd}. 
Unlike the case of CDM isocurvature, here the difference between the different
entries in the Fisher matrix is smaller, although the coefficients in the 
upper left $2\times 2$ submatrix are still about one order of magnitude 
larger than the others. Also in contrast to the CDM isocurvature case, we see 
that all coefficients increase about equally when polarization is included.

\begin{figure}
\centering
\includegraphics[width=0.6\textwidth, clip=true]{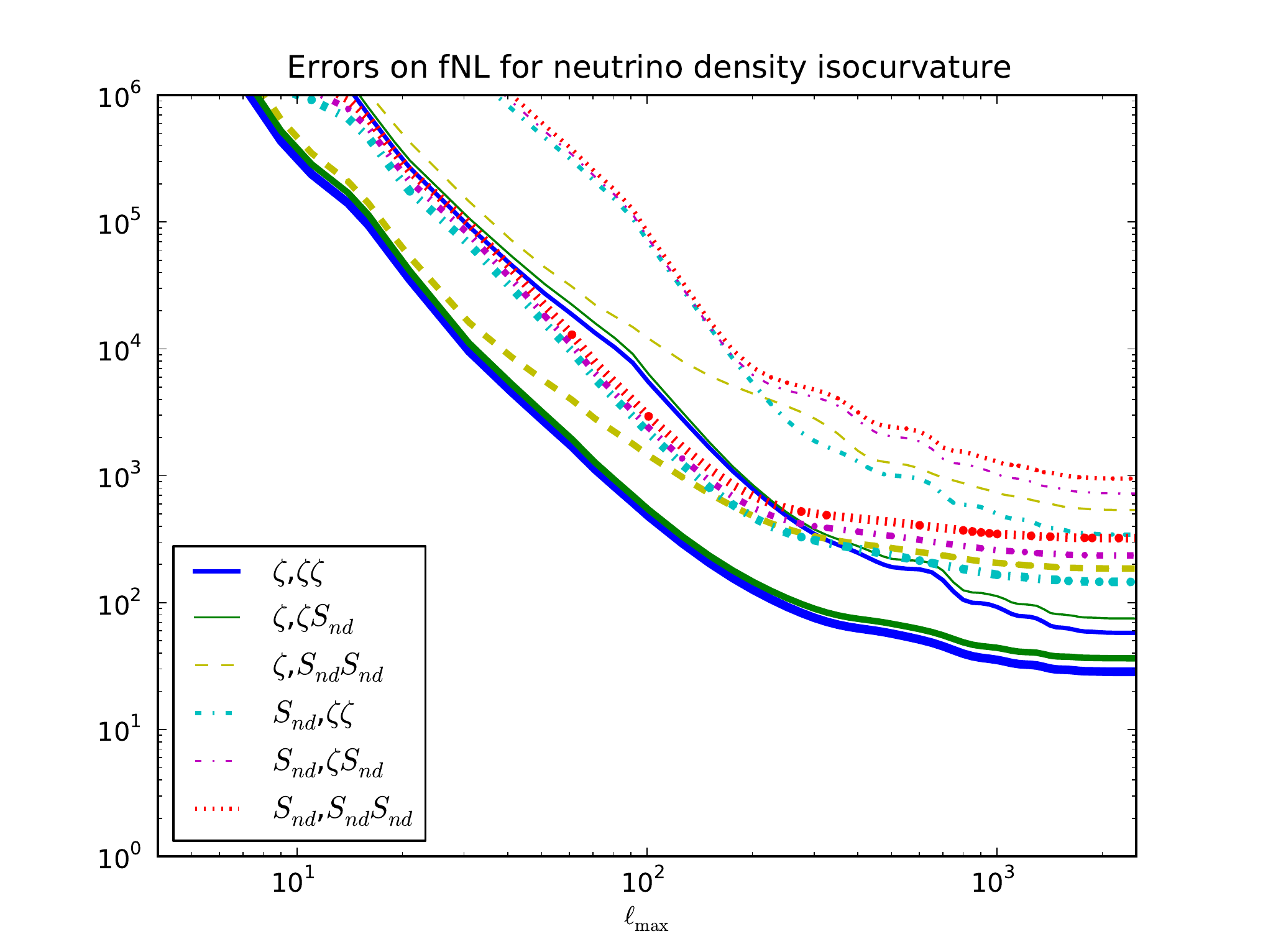}
\caption{Evolution of the $f_{\rm NL}$ parameter uncertainties as one increases 
the cut-off $\l_{\rm max}$, for the neutrino density isocurvature mode. 
The six thinner curves describe the situation if only temperature data is 
used, while for the six thicker curves both temperature and E-polarization 
data are included.}
\label{Va}
\end{figure}

The corresponding uncertainties on the six non-Gaussianity parameters are 
(taking into
account the correlations)\footnote{While we were finalizing our manuscript, 
we became aware of the work \cite{Kawakami:2012ke}, where the authors 
also investigate neutrino density isocurvature non-Gaussianity. Their 
numbers for the uncertainties are very similar to ours 
(note that they use the six non-Gaussianity parameters that we introduced in 
\cite{LvT} but in a different ordering), although they use a different
selection of Planck channels.}
\beq
\Delta \tilde f^i=\{28, 36, 190, 150, 240, 320\}.
\eeq  
When using temperature only, the uncertainties increase to
\beq
\Delta \tf^i=\{58, 75, 540, 340, 720, 950\}\qquad {\rm (no \ polarization)}.
\eeq
The evolution of the uncertainties as a function of $l_\mathrm{max}$ is shown 
in Fig.~\ref{Va}.
As in the CDM case, the $(\zeta,\zeta\zeta)$ and $(\zeta,\zeta S)$ 
non-Gaussianity parameters can be determined more accurately than the other
four, although the difference is not as big as for CDM. Unlike for CDM, all
parameters gain about the same from the inclusion of polarization, and all
uncertainties continue to decrease when higher multipoles are probed,
since the neutrino density isocurvature power spectrum does not decrease
as steeply as the CDM isocurvature one.

The correlation matrix is given in Table~\ref{correlation_nd}.
If one assumes the parameters to be independent, one finds
\beq
\Delta\tf^i= \{5.0, 5.1, 30, 24, 20, 69\} \qquad {\mathrm{(single \ parameter)}}.
\eeq
One sees that the increase of the uncertainties due to the correlations is much
more important here than for CDM, due to the larger correlations between
the various modes.

\begin{table}
\begin{center}
\begin{tabular}{|cccccc|}
\hline
$(\zeta, \zeta\zeta)$ & $(\zeta, \zeta S)$ & $(\zeta, S S)$ &  $(S,\zeta\zeta)$  & $(S,\zeta S)$ &  $(S,SS)$\\
\hline
 $1.$ & $-0.83\,(0.82) $ & $0.42\,(0.31) $ & $-0.73\,(0.67)$& $0.26\,(0.17) $ & $-0.16\,(0.09) $ \\
 - & $1.$ & $-0.58\,(0.45)$ & $0.58\,(0.55)$ & $-0.31\,(0.22)$& $0.25\,(0.20)$  \\
 - & - &  1. & $0.03\,(0.35)$ & $-0.45\,(0.69)$ & $0.19\,(0.41)$ \\
 - & - & - & 1. & $-0.76\,(0.81)$ & $0.58\,(0.64)$ \\
 - & - & - & - & 1. & $-0.84\,(0.86)$ \\
 - & - & - & - & - & 1.
\\ \hline
\end{tabular}
\caption{Correlation matrix for the neutrino density isocurvature mode. Only the upper half coefficients are indicated,  since the matrix is symmetric. The value between parentheses corresponds to the correlations obtained  without including the polarization (the absence of  sign in the parentheses means that it is unchanged with respect to the value taking into account the polarization).}
\label{correlation_nd}
\end{center}
\end{table}

\subsection{Neutrino velocity isocurvature mode}
\def\text{}
\begin{table}
\footnotesize
\begin{center}
\begin{tabular}{|cccccc|}
\hline
$(\zeta, \zeta\zeta)$ & $(\zeta, \zeta S)$ & $(\zeta, S S)$ &  $(S,\zeta\zeta)$  & $(S,\zeta S)$ &  $(S,SS)$\\
\hline
$3.9 \,(2.5) \times 10^{\text{-2}}$ & $3.5 \,(1.1) \times 10^{\text{-2}}$ & $74 \,(5.1) \times 10^{\text{-4}}$ &
   $15\,(9.5) \times 10^{\text{-3}}$ & $16 \,(5.6)\times 10^{\text{-3}}$ & $40 \,(2.6) \times 10^{\text{-4}}$
\\ 
- & $73 \,(7.6) \times 10^{\text{-3}}$ & $23 \,(1.3)\times 10^{\text{-3}}$ &
   $13 \,(4.0)\times 10^{\text{-3}}$ & $30 \,(3.4)\times 10^{\text{-3}}$ & $12 \,(0.65)\times 10^{\text{-3}}$
\\ 
- & - & $80 \,(4.3) \times 10^{\text{-4}}$ &
   $29 \,(1.6)\times 10^{\text{-4}}$ & $96 \,(4.8)\times 10^{\text{-4}}$ & $45 \,(2.1)\times 10^{\text{-4}}$
\\ - & - & - & $5.9 \,(3.6)\times 10^{\text{-3}}$ & $6.3 \,(2.1)\times 10^{\text{-3}}$ & $16 \,(0.88)\times 10^{\text{-4}}$
\\- & - & - & - & $13 \,(1.6)\times 10^{\text{-3}}$ & $54 \,(2.5)\times 10^{\text{-4}}$
\\- & - & - & - & - & $28 \,(1.1)\times 10^{\text{-4}}$
\\ \hline
\end{tabular}
\caption{Fisher matrix for the neutrino velocity isocurvature mode.
Only the upper half coefficients are indicated,  since the matrix is symmetric. The value between parentheses corresponds to the Fisher matrix components obtained  without including the polarization.}
\label{table_nv}
\end{center}
\end{table}

\begin{figure}
\centering
\includegraphics[width=0.6\textwidth, clip=true]{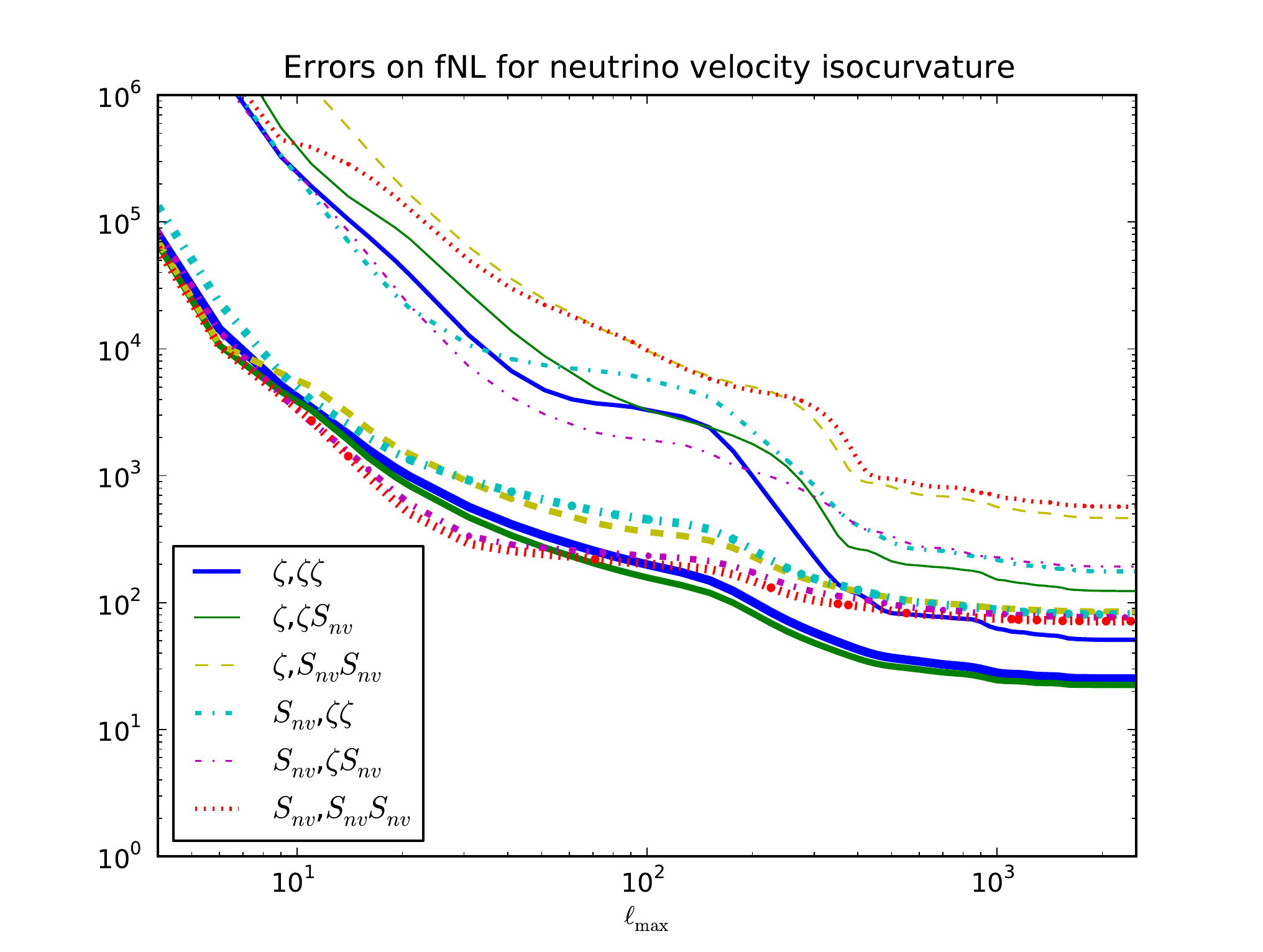}
\caption{Evolution of the $f_{\rm NL}$ parameter uncertainties as one increases 
the cut-off $\l_{\rm max}$, for the neutrino velocity isocurvature mode. 
The six thinner curves describe the situation if only temperature data is 
used, while for the six thicker curves both temperature and E-polarization 
data are included.}
\label{Vb}
\end{figure}

For a neutrino velocity isocurvature mode, we have obtained the Fisher matrix in Table~\ref{table_nv}. One notices that, including polarization, all entries
are roughly of the same order of magnitude, but without polarization, they
vary a lot. 
The corresponding uncertainties on the six non-Gaussianity parameters are 
(taking into account the correlations) 
\beq
\Delta \tilde f^i=\{25, 22, 85, 81, 77, 71\}.
\eeq  
When using temperature only, the uncertainties increase to
\beq
\Delta \tf^i=\{51, 120, 460, 180, 190, 570\}\qquad {\rm (no \ polarization)}.
\eeq
The evolution of the uncertainties as a function of $l_\mathrm{max}$ is shown 
in Fig.~\ref{Vb}.

One sees that in this case the difference between the first two and
the other four uncertainties (when including polarization) is much smaller 
than for CDM or neutrino
density isocurvature (a factor of about 3 compared to a factor of
about 15 in the CDM case). In particular, the latter four can be determined
more accurately than in the case of CDM or neutrino density isocurvature. 
However, the improvement due to
polarization is much more important than in the neutrino density case,
in particular for the $(\zeta,\zeta S)$, the $(\zeta, S S)$ and the
$(S,S S)$ parameters. One can understand why the $(\zeta,\zeta S)$
mode, for example, gains much more from polarization than the
$(\zeta,\zeta\zeta)$ mode with a similar argument as the one presented
for CDM isocurvature. The dominant contributions to these modes both
depend on ${\cal G}_{\zeta\zeta}(L)$ (defined in (\ref{G_and_H})), but
for $(\zeta,\zeta\zeta)$ this is multiplied by ${\cal H}_\zeta(\ell)$
and for $(\zeta,\zeta S)$ by ${\cal H}_S(\ell)$. And as one can see from
Fig.~\ref{HK_fig}, the ratio ${\cal H}_S/{\cal H}_\zeta$ for neutrino
velocity isocurvature increases enormously when one passes from
temperature to polarization (remember that it is the lowest values of
$\ell$ that contribute most to the squeezed configuration).

The correlation matrix is given in Table~\ref{correlation_nv}.
If one assumes the parameters to be independent, one finds
\beq
\Delta\tf^i= \{5.0, 3.7, 11, 13, 8.7, 19\} \qquad {\mathrm{(single \ parameter)}}.
\eeq
Hence one sees that the correlations in the case of neutrino velocity 
isocurvature are more important than for CDM, but less than for neutrino 
density.

\begin{table}
\begin{center}
\begin{tabular}{|cccccc|}
\hline
$(\zeta, \zeta\zeta)$ & $(\zeta, \zeta S)$ & $(\zeta, S S)$ &  $(S,\zeta\zeta)$  & $(S,\zeta S)$ &  $(S,SS)$\\
\hline
 1. & $-0.50\,(0.42)$ & $0.08\,(0.11)$ & $-0.82\,(0.66)$ & $0.34\,(0.34)$ & $-0.13\,(+0.05)$ \\
 - & 1. & $-0.60\,(0.80)$ & $0.25\,(-0.18)$ & $-0.26\,(0.35)$ & $0.36\,(0.39)$ \\
 - & - &   1. & $0.37\,(0.44)$ & $-0.51\,(+0.07)$ & $-0.29\,(0.75)$ \\
 - & - & - & 1. &  $-0.74\,(0.61)$ & $0.25\,(-0.10)$ \\
 - & - & - & - & 1. & $-0.46\,(0.36)$ \\
 - & - & - & - & - & 1.
\\ \hline
\end{tabular}
\caption{Correlation matrix for the neutrino velocity isocurvature mode.
Only the upper half coefficients are indicated,  since the matrix is symmetric. The value between parentheses corresponds to the correlations obtained  without including the polarization (the absence of  sign in the parentheses means that it is unchanged with respect to the value taking into account the polarization).}
\label{correlation_nv}
\end{center}
\end{table}

\section{Constraints on early universe models}

In the previous section, we have studied how to obtain constraints on the six non-linearity coefficient $\tf_{NL}^{(i)}$ without assuming any particular relation between them.
In the context of an early Universe model, or in a class of models, one can go further and use the results of the previous section to  obtain some constraints on the parameters of the model itself.

\subsection{General analysis}

As we have seen earlier, isocurvature perturbations require the existence of at least two degrees of freedom in the early Universe. So, for simplicity, let us focus on models with two scalar fields, $\phi$ and $\sigma$, such that isocurvature perturbations are generated only by the fluctuations 
of $\sigma$ and all non-linearities are also dominated by their contribution from $\sigma$. This means that we have
\beq
\label{ZS}
\zeta=N_\phi^\zeta\delta\phi+N_\sigma^\zeta \delta\sigma+\frac12 N_{\sigma\sigma}^\zeta \delta\sigma^2\,, \qquad 
S=N_\sigma^S \delta\sigma+\frac12 N_{\sigma\sigma}^S \delta\sigma^2\,.
\eeq

Using the general expressions (\ref{lambda}-\ref{f_NL}), one easily finds that the coefficients are interdependent and can be written in the form
\begin{eqnarray}
\tf_{NL}^{I, \zeta\zeta}&=& \mu_I \, \Xi^2,\\
\tf_{NL}^{I, \zeta S}&=& \varepsilon_{_{\zeta S}}\, \mu_I \, \alpha^{1/2} \,  \Xi^{3/2},\\
\tf_{NL}^{I, S S}&=& \mu_I \,\alpha \, \Xi,
\end{eqnarray}
where we have introduced the contribution of $\sigma$ in the adiabatic power spectrum,
\beq
\Xi\equiv \frac{(N^{\zeta}_\curv)^2}{(N^{\zeta}_\phi)^2+(N^{\zeta}_\curv)^2}\,,
\eeq
and the isocurvature to adiabatic ratio,
\beq
\alpha\equiv \frac{(N^{S}_\curv)^2}{(N^{\zeta}_\phi)^2+(N^{\zeta}_\curv)^2}\,.
\eeq
Note that the extraction of an isocurvature component in the power spectrum would fix $\alpha$ while,  in principle, $\Xi$ could be determined from observations by measuring both the bispectrum and the trispectrum coefficients, since 
they satisfy consistency relations~\cite{Langlois:2010fe} similar to the purely adiabatic relation $\tau_{NL}=\tf_{NL}^2/\Xi$.

The two coefficients 
\beq
\mu_I=\frac{N^I_{\curv\curv}}{(N^{\zeta}_\curv)^2}\qquad (I=\zeta, S)
\eeq
fully characterize the non-Gaussianity of the adiabatic and isocurvature perturbations, respectively, while $\varepsilon_{_{\zeta S}}$ denotes the relative sign of $\zeta$ and $S$: $\varepsilon_{_{\zeta S}}=1$ if they have the same sign, $\varepsilon_{_{\zeta S}}=-1$ otherwise.
Interestingly, $\tf_{NL}^{I, \zeta\zeta}$ and $\tf_{NL}^{I, SS}$ share the same sign as $\mu_I$, whereas $\tf_{NL}^{I, \zeta S}$ can have a different sign. The hierarchy between the coefficients $\tf_{NL}^{I, JK}$, $I$ being fixed, depends on the relative amplitude of $\alpha$ and $\Xi$: $\tf_{NL}^{I, \zeta\zeta}$ dominates if $\Xi\gg \alpha$, whereas $\tf_{NL}^{I, SS}$ dominates if $\Xi\ll \alpha$. 

For given values of $\alpha$ and $\Xi$, the uncertainties on the two parameters $\mu_\zeta$ and $\mu_S$ are determined from the ``projected'' Fisher matrix
\beq
{\cal F}_{IJ}=\sum_{i,j}{\cal \pi}_{(i)I} \, F_{ij} \, {\cal \pi}_{(j)J}
\eeq
with
\beq
{\cal \pi}_{(i)\zeta}=\{\Xi^2, \varepsilon_{_{\zeta S}} \alpha^{1/2}\Xi^{3/2}, \alpha\Xi, 0,0,0\},\quad 
{\cal \pi}_{(i)S}=\{0,0,0, \Xi^2, \varepsilon_{_{\zeta S}}\alpha^{1/2}\Xi^{3/2}, \alpha\Xi\}\,.
\eeq
From this $2\times 2$ Fisher matrix, one can easily deduce the expected uncertainties on the two parameters $\mu_\zeta$ and $\mu_S$, by using the analog of (\ref{errors}).

\subsection{Illustrative example}
In  \cite{Langlois:2011zz,LvT} we have studied a class of models which produces perturbations of the above type. In these models, $\sigma$ is a curvaton which decays into radiation and CDM. Since part of the CDM can have been produced before the decay of the curvaton, one can introduce as a parameter the fraction of CDM created by the decay as
\beq
\f\equiv\frac{\gamma_c\,  \Omega_\curv}{\Omega_c+\gamma_c\Omega_\curv},
\eeq 
where the $\Omega$'s represent the relative abundances just before the decay
and $\gamma_c$ is the fraction of the curvaton energy density transferred 
into CDM.
The second relevant parameter,
\beq
r\equiv \frac{3 (1-\gamma_c)\,  \Omega_\curv}{(1-\gamma_c\, \Omega_\curv)(4-\Omega_\curv)}\,,
\eeq
quantifies the transfer between the pre-decay and post-decay perturbations~\cite{Langlois:2011zz} (one finds $\zeta_\gamma^\mathrm{after}=(1-r)\zeta_\gamma^\mathrm{before}$ at the linear level).

As shown in \cite{Langlois:2011zz,LvT}, one can derive the ``primordial'' perturbations $\zeta$ and $S$ as expansions, up to second order, in terms of $\delta\curv$, which yield the coefficients $N^I_{\curv}$ and $N^I_{\curv\curv}$. Using these results and assuming $r\ll1$, one obtains 
\beq
\mu_\zeta=\frac{3}{2r}, \qquad 
\mu_S=\frac{9}{2 r^2}\left(\f(1-2\f)-r\right)\,,
\eeq
while $\varepsilon_{_{\zeta S}}= {\rm sgn}(\f-r)$ and
\beq
\alpha= 9\left(1-\frac{\f}{ r}\right)^2 \,\Xi \, .
\eeq
In the regime $\f\ll r \ll 1$, one finds $\mu_\zeta=3/(2r)$, 
$\mu_S=-9/(2r)$ and $\alpha=9\Xi$, with 
$\varepsilon_{_{\zeta S}}=-1$ and the amplitudes of non-Gaussianities depend only on the parameter $r$. 
By contrast, in the opposite regime $r \ll f_c  \ll 1$, $\mu_\zeta=3/(2r)$, 
$\mu_S=9\f/(2 r^2)$ 
and $\alpha= 9\, \Xi (\f/r)^2$. The coefficients $\tf_{NL}^{S, IJ}$, which depend on $\mu_S$, are thus enhanced with respect to the coefficients $\tf_{NL}^{\zeta, IJ}$ in the latter case.

\section{Conclusion}
We have systematically investigated the angular bispectra generated by initial conditions that combine the usual adiabatic mode with an isocurvature mode, assuming local non-Gaussianity. We have studied successively the four types of isocurvature modes, namely cold dark matter (CDM), 
baryon, neutrino density and neutrino velocity isocurvature modes. In each case, the total bispectrum can be decomposed into six elementary bispectra and we have estimated the expected uncertainties on the corresponding coefficients, which are extensions of the usual purely adiabatic $f_\mathrm{NL}$ parameter, in the context of the forthcoming Planck data of the cosmic microwave background radiation (CMB). As we showed, the results
for baryon isocurvature can be obtained from a simple rescaling of the
CDM isocurvature results, but the others are distinct.

In the squeezed limit, where one multipole $l$ is much smaller than the other
two (which then have to be almost equal due to the triangle inequality), we have shown that  the six elementary bispectra  factorize as a function of the small $l$ times the power
spectrum as a function of the large $l$.  Since the  squeezed limit components dominate the bispectrum for {\it local} non-Gaussianity, we have been able,  
using this factorization,  to give simple explanations for the various interesting results that we have observed.

By enlarging the space of initial conditions, one obviously expects a larger uncertainty on the purely adiabatic $f_\mathrm{NL}$ coefficient. Interestingly, this uncertainty is increased only by a factor $2$ for CDM and baryon isocurvature modes, whereas it increases by a factor $6$ in the neutrino density isocurvature case and by a factor $5$ in the neutrino velocity isocurvature case.
This can be explained by the fact that the CDM isocurvature power spectrum 
decreases much faster with $l$ than the adiabatic and neutrino isocurvature
ones. As we have shown, this means that the uncertainties on $\tf^{\zeta,\zeta\zeta}$ 
and $\tf^{\zeta, \zeta S}$ in the case of CDM isocurvature continue to improve
as one increases the number of available multipoles, while the other four
saturate at a much lower $l$. As a consequence the first two can be
determined much more accurately than the other four, and are only weakly 
correlated with them. This small correlation also means that it is 
important to look at the data with the full $f_\mathrm{NL}$ estimator, and
not just the adiabatic one, as a large CDM isocurvature non-Gaussianity could
be hiding behind a small adiabatic signal.

We have  shown that the E-polarization often plays a crucial role in reducing the uncertainties. In the CDM isocurvature case, polarization improves slightly the precision on the coefficients $\tf^{\zeta,\zeta\zeta}$ and $\tf^{\zeta, \zeta S}$, but the precision of the other four coefficients improves by a factor of order five.
Polarization is also very important for some of the parameters in the
neutrino velocity case, the uncertainty on the purely 
isocurvature $\tf^{S,SS}$, for example, improving by a factor 8 when
polarization is included. Again we were able to explain these results
from the behaviour of the power spectra, using the factorization of the
squeezed bispectrum.

The decomposition of the bispectrum into six elementary bispectra and
the CMB constraints on the six $\tf^{(i)}$ parameters do not depend on any
assumptions about the specific model of the early Universe. We only assumed
that (possibly correlated) primordial adiabatic and isocurvature modes are 
produced, with a primordial bispectrum of local type and with power spectra 
that all have the same shape. Note that neither of these assumptions appears
to be essential; they were only made for simplicity. In future work we will
look into the possibility of relaxing these assumptions.

If, however, one does consider an explicit early Universe model, there are
often relations between the six parameters, and an observational detection
or constraint could then be used to check for such a relation and put
constraints on the parameters of the model. 
We have discussed a general class
of models with two scalar fields, where only one of the fields generates
both the isocurvature perturbations and all non-linearities. 
We have also
considered a specific implementation of this general model, where the
two fields are an inflaton and a curvaton. In this model, a  CDM isocurvature mode
is produced and the six $\tf^{(i)}$ parameters only depend on two model parameters. In some ranges of the model parameters,
the isocurvature mode is subdominant in the power spectrum but provides observable  non-Gaussianity that can dominate the usual adiabatic non-Gaussianity. Looking for these new angular shapes in  the CMB data would thus provide interesting information on the very early Universe.

\section*{Acknowledgements}
The original numerical bispectrum code, which was extended here to include isocurvature modes, was developed by M. Bucher and BvT. We also
acknowledge the use of CAMB (http://camb.info/). D.L. is partially supported by the
ANR grant ÒSTR-COSMOÓ, ANR-09-BLAN-0157.

\end{document}